\documentclass[11pt]{article}
\usepackage{amssymb,latexsym, amsmath}
\newcommand{\bfi}{\bfseries\itshape}
\newcommand{\singlespace}{\baselineskip 4.2333mm \parskip 4.2333mm} 
\makeatletter
\@addtoreset{figure}{section}
\def\thefigure{\thesection.\@arabic\c@figure}
\def\fps@figure{h, t}
\@addtoreset{table}{bsection}
\def\thetable{\thesection.\@arabic\c@table}
\def\fps@table{h, t}
\@addtoreset{equation}{section}

\makeatother
\def\mathbb#1{\mathbf{ #1}}
\def\mathfrak#1{\mathbf{ #1}}

\def\setmarsing-ok{
\textwidth 6.5in
\textheight 8.5in
\oddsidemargin 0in
\topmargin -0.5in}

\setmarsing-ok

\begin{document}

\title{{\bf Euler-Poincar\'e Dynamics of \\
Perfect Complex Fluids}}

\author{
Darryl D. Holm
\\Theoretical Division and Center for Nonlinear Studies
\\Los Alamos National
Laboratory, MS B284
\\ Los Alamos, NM 87545\\
{\footnotesize dholm@lanl.gov}\\
}
\date{August 1, 2000}

\maketitle
\normalsize
\vspace{-0.1in}

\begin{abstract}
Lagrangian reduction by stages is used to derive the Euler-Poincar\'e
equations for the nondissipative coupled motion and micromotion of 
complex fluids. We mainly treat perfect complex fluids (PCFs) whose order
parameters are {\sl continuous material variables}. These order parameters may
be regarded geometrically either as objects in a vector space, or as coset
spaces of Lie symmetry groups with respect to subgroups that leave these
objects invariant. Examples include liquid crystals, superfluids, Yang-Mills
magnetofluids and spin-glasses. A Lie-Poisson Hamiltonian formulation of the
dynamics for perfect complex fluids is obtained by Legendre transforming
the Euler-Poincar\'e formulation. These dynamics are also derived by using the
Clebsch approach. In the Hamiltonian and Lagrangian formulations of perfect
complex fluid dynamics Lie algebras containing two-cocycles arise as a
characteristic feature. 

After discussing these geometrical formulations of the dynamics of perfect
complex fluids, we give an example of how to introduce defects into the order
parameter as imperfections (e.g., vortices) that carry their
own momentum. The defects may move relative to the Lagrangian fluid
material and thereby produce additional reactive forces and stresses.

\end{abstract}

\clearpage
\tableofcontents

\section{Introduction}\label{intro}

\paragraph{Definitions.}
The hydrodynamic motion of a complex fluid depends on variables called {\bfi
order parameters} that describe the macroscopic variations of the internal
structure of the fluid parcels. These macroscopic variations may form
observable patterns, as seen, for example, via the gradients of optical
scattering properties in liquid crystals arising due to the spatially varying
orientations of their molecules, as discussed in, e.g., {Chandrasekhar
[1992]} and {de Gennes and Prost [1993]}. This {\bfi micro-order} of a
complex fluid is described by an auxiliary macroscopic continuum field of
geometrical objects associated with each fluid element and taking values in a
vector space (or a manifold) called the {\bfi order parameter space}. The
canonical example is the description of the local directional asymmetries of
nematic liquid crystal molecules by a spatially and temporally varying
macroscopic continuum field of unsigned unit vectors called ``directors,''
see, e.g., {Chandrasekhar [1992]} and {de Gennes and Prost [1993]}.

Thus, the presence of micro-order breaks the symmetry group $\mathcal{O}$
of the uniform fluid state to a subgroup $\mathcal{P}$. This {\bfi spontaneous
symmetry breaking} occurs in every phase transition of condensed matter into an
ordered state. The symmetry subgroup $\mathcal{P}\subset\mathcal{O}$ that
remains is the isotropy group of whatever geometrical object appears (e.g., a
vector, a spin, a director, etc.) when the symmetries of the uniform fluid
state are broken to produce the micro-order. That is, the remaining symmetry
subgroup $\mathcal{P}$ is the symmetry group of the micro-order. Equivalently,
the order parameter may also be regarded as taking its values in the coset
space $\mathcal{C}=\mathcal{O}/\mathcal{P}$ and its space and time variations
may be represented by a space and time dependent curve in the Lie symmetry
group $\mathcal{O}$ through the action of $\mathcal{O}$ on its coset space
$\mathcal{C}$. Thus, associating a geometrical object in a vector space (say,
a director) with an order parameter is a way of visualizing the coset
space $\mathcal{C}$, see, e.g., {Mermin [1979]} and Volovick [1992] for
physical examples. The symmetry group $\mathcal{O}$ of the original uniform
fluid state is called the {\bfi order parameter group}, or the {\bfi broken
symmetry}. The order parameter dynamics that generates a curve parameterized
by space and time in the broken symmetry group for a complex fluid is called
its {\bfi micromotion}, although it refers to continuum properties at the
coarse-grained macroscopic scale. 

Spatial and temporal variations in the micro-order are measured relative to a
{\bfi reference configuration}. Let $\mathcal{O}$ act transitively from the
right on a manifold $M$. Suppose the subgroup $\mathcal{P}\subset\mathcal{O}$
leaves invariant an arbitrarily chosen reference point $m_0\in M$, i.e.,
$m_0p=m_0\,,\forall p\in\mathcal{P}$. The reference point $m_0$ then
corresponds to the coset
$[e]=e\mathcal{P}=\mathcal{P}$ of $\mathcal{O}/\mathcal{P}$, where $e$ is the
identity element of $\mathcal{O}$. For another choice of reference point, say 
$m'_0=m_0h$ with $h\in\mathcal{O}$, the isotropy subgroup
becomes the conjugate subgroup
$\mathcal{P}\rightarrow\mathcal{P}'=h^{-1}\mathcal{P}h$ and the corresponding
coset space $\mathcal{O}/\mathcal{P}'$ is, thus, isomorphic to the original
coset space $\mathcal{O}/\mathcal{P}$. Hence, the conjugacy equivalence
classes of the order parameter coset spaces account for the arbitrariness in
the choice of reference point $m_0$. One may think equally well of the
order parameter group as acting either on a manifold $M$ with ``origin''
$m_0$, or on a coset space $\mathcal{O}/\mathcal{P}$, where
$\mathcal{P}$ is the stabilizer of the reference point $m_0$. 

For example, let $\mathcal{O}$ be the group of proper orthogonal
transformations $SO(3)$ acting transitively on directors in $\mathbb{R}^3$
(unit vectors with ends identified) and choose $m_0$ to be the vertical
director in an arbitrary reference frame; so $m_0$ is invariant under the
$O(2)$ group of rotations around the vertical axis in that frame and
reflections across its horizontal plane.  This example applies to
cylindrically symmetric nematic liquid crystal properties. The isotropy
group $\mathcal{P}$ is $O(2)$, and the liquid crystal director may be
represented in the coset space $SO(3)/O(2)$, which is isomorphic to
$S^{\,2}/Z_2$, the unit sphere $S^{\,2}$ with diametrically opposite points
identified, i.e., the real projective plane $RP^2$. Thus, after a reference
configuration has been chosen, the micro-order of nematic liquid crystals may
be represented equivalently as a space and time dependent curve in the broken
symmetry group $\mathcal{O}$ acting on either the order parameter manifold
$M=RP^2$, or on the coset space $SO(3)/O(2)$ of the broken symmetry $SO(3)$.

In addition to its micromotion, the {\bfi motion} of a complex fluid also
possesses the usual properties of classical fluid dynamics. In particular, the
motion of a complex fluid involves the advection of thermodynamic state
variables such as heat and mass, regarded as fluid properties taking values in
a vector space $V^\ast$.  The complete dynamical equations for complex fluids
must describe both their motion and their micromotion. In general, these two
types of motion will be nonlinearly coupled to each other. This coupling
typically arises because properties of both types of motion appear
in the stress tensor governing the total momentum.

This definition of complex fluid motion in terms of the continuum dynamics of
both its order parameter and its usual fluid properties encompasses a wide
range of models for complex fluid motion, including binary fluids, multi-phase
fluids, polymeric materials, spin glasses, various other types of magnetic
materials, superfluids and, of course, liquid crystals. The order parameter
for each of these models provides a continuum (i.e., coarse-grained)
description of the complex fluid's internal degrees of freedom, or
micro-order. Here we will discuss mainly the case in which these order
parameters are {\bfi continuous material variables}, that is, they are
continuous functions carried along with the fluid parcels.%
%
\footnote{In some cases, order parameters are determined
from constraint relations that are Eulerian in nature, e.g., the volume
fraction in two-phase interpenetrating flow as in 
{Holm and Kupershmidt [1986]}. This case will not be discussed here.}
Such media shall be called {\bfi perfect complex fluids}. This name
(abbreviated PCF) is chosen to contrast with the perfect simple fluid (which
has no internal micro-degrees of freedom) and to provide a geometrical basis
for treating defects as imperfections in the order parameter field that can
propagate, or move relative to the material labels. The relative motion of
defects through the medium generally introduces additional reactive forces and
sources of dissipation. We believe the dynamics of defects in complex fluids
is best approached after first discussing the dynamics of perfect complex
fluids, which is interesting in its own right. Here we shall concentrate on
deriving the nonlinear dynamical equations for the ideal continuum motion and
micromotion of perfect complex fluids.

Once the equations for the nonlinear dynamics of their ideal (nondissipative)
motion and micro-motion are established, dissipative processes must be included
for most physical applications of complex fluids. By tradition, this is
accomplished phenomenologically in these models, by introducing kinetic
coefficients, such as viscosity, mobility, thermal diffusivity, etc., so as to
obey the requirements of the Clausius-Duhem relation that the entropy
production rate be positive when the dynamics of {\it all thermodynamic
variables} (including order parameters) are included, as in
{Dunn and Serrin [1985]}, {Hohenberg and Halperin [1977]}. Here 
we shall ignore dissipation entirely, trusting that it can be added later by
using the standard phenomenological methods. 

For the case that the order parameter group $\mathcal{O}$ is the proper
orthogonal group $SO(3)$, a geometrical approach to complex fluids  
exists as part of the rational theory of continuum dynamics for materials with
orientational internal degrees of freedom, such as liquid crystals.  Rational
theories of such complex media began with E. and F. {Cosserat [1909]}. The
Cosserat theories were recapitulated at various times by many different
people. See, e.g., {Kleman [1983]} and {Eringen [1997]} for descriptions of
recent developments and proposed applications of this approach for treating,
e.g., liquid crystal dynamics in the tradition of the rational theory of
continuum media. 

The present paper starts with the example of the Ericksen-Leslie theory of
nematic liquid crystals and develops the geometrical framework for continuum
theories of perfect complex fluids. In this geometrical framework, the motion
and micromotion are nonlinearly and self-consistently coupled to one another by
the composite actions of the diffeomorphisms and the order parameter group. The
micromotion follows a curve in the order parameter group depending on
time and material coordinate, and the motion is a time-dependent curve in the
group of diffeomorphisms, which acts on the material coordinates of the fluid
parcels to carry them from their reference configuration to their current
positions. A feedback develops between the composite motion and micromotion,
because the stress tensor affecting the velocity of the diffeomorphisms
depends on the gradient of the order parameter. The mathematical basis for our
development is the method of {\bfi Lagrangian reduction by stages}, due to
{Cendra, Marsden and Ratiu [1999]}. As we shall see, obtaining the
Euler-Poincar\'e equations for perfect complex fluids requires two stages of
Lagrangian reduction, first by the order parameter group and then by the
diffeomorphism group.

\paragraph{The main results in this paper.}  The  {\bfi Euler-Poincar\'e
approach} provides a unified framework for modeling the dynamics of perfect 
complex fluids that preserves and extends the mathematical structure inherent
in the dynamics of classical fluids and liquid crystals in the Eulerian
description. This paper provides detailed derivations and discusses
applications of the Euler-Lagrange equations, the Lagrange-Poincar\'e
equations, the Euler-Poincar\'e equations, the Clebsch equations and the
Lie-Poisson Hamiltonian structure of perfect complex fluids. 
 The action principles for these equations appear at their various
 stages of transformation under reduction by symmetries. 
 The theme of this paper is the reduction of degrees of freedom by 
 transforming to variables in the action principles that are 
 invariant under the symmetries of the Lagrangian. 
 (This is the Lagrangian version of Poisson reduction 
 on the Hamiltonian side.) The transformation is done in stages, not all at
 once, for the additional perspective we hope it brings and to illustrate how
 Lagrangian reduction can be iterated in condensed matter applications. It may
 also be possible to impose nonholonomic constraints by stages, by
 transforming to variables that either respect the constraints, or appear in
 their specifications.

The main new results for liquid crystals and PCFs in this
paper are:
\begin{enumerate}

\item Four action principles for PCF dynamics and their associated motion and
micromotion equations at the various stages of reduction -- the Euler-Lagrange
equations, the Lagrange-Poincar\'e equations, the Euler-Poincar\'e equations,
and the Clebsch equations.

\item The canonical and Lie-Poisson Hamiltonian formulations of these
equations, their Lie algebraic structures and the Poisson maps between them.

\item The momentum conservation laws and Kelvin-Noether circulation theorems
for these equations.

\item The reduced equations for one-dimensional dependence on either space or
time, and the relation of these reduced equations to the Euler-Poisson
equations for the dynamics of generalized tops.

\item A strategy for composing the dynamics of defects in a complex fluid with
its underlying PCF dynamics.

\end{enumerate}

\paragraph{Outline.}  In Section \ref{liqxtal}, we develop these results for
the motion and micromotion in the example of nematic liquid crystals, in forms
that shall parallel the results of the general theory derived in the following
section. The Euler-Lagrange equations from Hamilton's principle specialize to
the Ericksen-Leslie equations, upon making the appropriate choices of the
kinetic and potential energies. The Lagrange-Poincar\'e equations and
Euler-Poincar\'e equations that follow from applying two successive stages of
Lagrangian reduction of Hamilton's principle with respect to its symmetries
provide geometrical variants and generalizations of the Ericksen-Leslie
equations.

In Section \ref{Ham-Princ-Lag-Red}, we perform two successive stages of
Lagrangian symmetry reduction to derive first the Lagrange-Poincar\'e
equations and then the Euler-Poincar\'e equations for PCFs with an
arbitrary order parameter group. These Euler-Poincar\'e equations are 
Legendre-transformed to their Lie-Poisson Hamiltonian form. Their
derivation by the Clebsch approach is also given. 

Finally, in Section \ref{strat-defect} we discuss a strategy of including
defect dynamics into the theory of PCF dynamics. This strategy introduces an
additional set of fluid variables that describe the motion of the defects
relative to the material coordinates of the PCF. We shall illustrate this
strategy in the example of rotating superfluid $^4$He, in which $U(1)$ is the
broken symmetry and the defects are quantum vortices.

\section{The example of liquid crystals}\label{liqxtal}

\subsection{Background for liquid crystals}

We begin with a hands-on example that illustrates the utility of the
ideas we shall develop in this paper. Liquid crystals provide a ubiquitous
application that embodies these ideas and supplies a guide for developing them.

Liquid crystals are the prototype for complex fluids. For extensive reviews,
see {Chandrasekhar [1992]} and {de Gennes and Prost [1993]}. An orientational
order parameter for a molecule of arbitrary shape is given in 
{Chandrasekhar [1992]}, p.40, as
\begin{equation} \label{orient-order-param}
{\rm S} = \frac{1}{2} \Big\langle  3\,\chi\otimes \chi 
- Id \otimes  Id \Big\rangle
\,,
\end{equation}
where $\langle\,\cdot\,\rangle$ is a statistical average and $\chi\in SO(3)$
is a rotation that specifies the local molecular orientation relative to a
fixed reference frame. Thus, in index notation,
\begin{equation} \label{orient-order-param-index}
{\rm S}_{klKL} 
=
 \frac{1}{2} \Big\langle  3\,\chi_{kK}\,\chi_{lL} 
- 
 \delta_{kl}\delta_{KL} \Big\rangle
\,.
\end{equation}
This order parameter is traceless in both pairs of its indices, since
$\chi^T=\chi^{-1}$. 

For cylindrically shaped molecules, (e.g., nematics, cholesterics or smectics%
\footnote{Smectics form layers, so besides their director orientation they have
an additional order parameter for their broken translational symmetry.})
we may choose the 3-axis, say, as the reference axis of symmetry,
i.e., choose
$K=3=L$ and set 
\begin{equation} \label{director-def}
n_k \equiv \chi_{k\,3} 
\,,
\quad\hbox{so that}\quad
|\mathbf{n}|^2 = \chi_{3k}\chi_{k3} = 1
\,,
\end{equation}
and S becomes,
\begin{equation} \label{stat-director-order-param}
{\rm S}_{kl33} 
\equiv
{\rm S}_{kl} 
=
 \frac{1}{2} \Big\langle  3\,n_k\,n_l 
- 
 \delta_{kl} \Big\rangle
\,.
\end{equation}
Note: The order parameter S does not distinguish between
$\mathbf{n}$ and $-\mathbf{n}$. Physically, S may be regarded as the 
quadrupole moment of a local molecular charge distribution. For a clear
description of the use of this order parameter in assessing nematic
order-disorder phase transitions using the modern theory of critical
phenomena, see {Lammert et al. [1995]}. Instead of
considering such phase transitions, here we shall be interested in the
continuum dynamics associated with the interaction of this order parameter
with material deformations.

In passing to a continuum mechanics description, one replaces statistical
averages by a local space and time dependent unit vector, or
``director'' $\mathbf{n}(\mathbf{x},t)$. Then, the continuum order parameter
$S$ corresponding to the statistical quantity S in
(\ref{stat-director-order-param}) is the symmetric traceless tensor,
\begin{equation} \label{director-order-param}
S_{kl} (\mathbf{n})
\equiv
 \frac{1}{2} \Big(  3\,n_k\,n_l 
- 
 \delta_{kl} \Big)
\,,
\end{equation}
which satisfies $S(\mathbf{n}) = S(-\mathbf{n})$ and admits $\mathbf{n}$ as an
eigenvector, $S\cdot\mathbf{n} = \mathbf{n}$. The hydrodynamic tensor order
parameter $S$ represents the deviation from isotropy of any convenient tensor
property of the medium. For example, the residual dielectric and diamagnetic
energy densities of a nematic liquid crystal due to anisotropy may be
expressed in terms of the tensor order parameter $S$ as,
\begin{equation} \label{dielect/diamag-ergs}
\frac{\Delta\epsilon}{2}\, \mathbf{E}\cdot S(\mathbf{n})\cdot\mathbf{E}
\quad\hbox{and}\quad
\frac{\Delta\mu}{2}\, \mathbf{B}\cdot S(\mathbf{n})\cdot\mathbf{B}
\,,
\end{equation}
for (external) electric and magnetic fields $\mathbf{E}$ and $\mathbf{B}$,
respectively. (For simplicity, we neglect any dependence of the electric and
magnetic polarizabilities of the medium on the gradients $\nabla\mathbf{n}$,
although this possibility is allowed.)

\subsection{Four action principles for perfect liquid crystals}

\paragraph{Liquid crystal action.}
The standard equations for the continuum dynamics of liquid crystals without
defects are the Ericksen-Leslie equations, due to {Ericksen [1960, 1961]}, 
{Leslie [1966, 1968]}, and reviewed, e.g., in {Leslie [1979]}.
These equations express the dynamics of the director $\mathbf{n}$ and may
be derived from an action principle $\delta \mathcal{S}=0$ with action
$\mathcal{S} = \int dt\,L $ in the following class, 
\begin{equation}\label{liqxtal-action-director}
\mathcal{S} = \int dt\int d^3X \
\mathcal{L}(\boldsymbol{\dot{\mathbf{x}}}, J, \mathbf{n},
\boldsymbol{\dot{\mathbf{n}}},\nabla\mathbf{n})
\,,
\end{equation} 
with notation $\boldsymbol{\dot{\mathbf{x}}} =
\partial\mathbf{x}(\mathbf{X},t)/\partial{t}$,
so that overdot denotes material time derivative,
$J={\rm det}(\partial\mathbf{x}/\partial\mathbf{X})$ and 
$\nabla\mathbf{n}(\mathbf{X},t)$ has spatial components given by 
the chain rule expression,
\begin{equation}\label{director-grad}
\nabla_{i\,}\mathbf{n}
=
\bigg[\Big(\frac{\partial\mathbf{x}}
{\partial\mathbf{X}}\Big)^{-1}\bigg]_{i\,A}
\frac{\partial\mathbf{n}}{\partial X_A}
\equiv
\frac{\partial X_A}{\partial x_i}
\,
\frac{\partial }{\partial X_A}\mathbf{n}(\mathbf{X},t)
\equiv
\mathbf{n}_{,i}
\,\;.
\end{equation} 
Note that the coupling between the fluid dynamics $\mathbf{x}(\mathbf{X},t)$
and director dynamics $\mathbf{n}(\mathbf{X},t)$ occurs in the
Lagrangian density $\mathcal{L}$ through the {\bfi inverse deformation
gradient}, $(\partial\mathbf{x}/\partial\mathbf{X})^{-1}$, via the chain rule
expression above for $\nabla_{i\,}\mathbf{n}(\mathbf{X},t)$.

Varying the action $\mathcal{S}$ in the fields $\mathbf{x}$ and
$\mathbf{n}$ at fixed material position $\mathbf{X}$ and time $t$ gives
\begin{eqnarray}
\delta\mathcal{S}
\!\!\!&=&\!\!\!
-\int dt\int d^3X
\bigg\{\delta x_p\bigg[ \bigg(
\frac{\partial \mathcal{L}}{\partial
\dot{x}_p}\bigg)^{\!\!\boldsymbol{\dot{\,}}}
 +
\,J\frac{\partial}{\partial x_p}\frac{\partial \mathcal{L}}{\partial J}
-
J\frac{\partial}{\partial x_m}
\bigg( J^{-1}
\frac{\partial \mathcal{L}}{\partial \mathbf{n}_{,m}}
\boldsymbol{\cdot}\mathbf{n}_{,p}\bigg)
\bigg]
\nonumber\\
&&\hspace{1in}
+\
\delta\mathbf{n}\boldsymbol{\cdot}
\,\Big[\Big(
\frac{\partial \mathcal{L}}
{\partial\boldsymbol{\dot{\mathbf{n}}}}\Big)^{\boldsymbol{\dot{\,}}} 
- 
\frac{\partial \mathcal{L}}{\partial\mathbf{n}}
+
\, J \frac{\partial}{\partial x_m}
\bigg(J^{-1}
\frac{\partial \mathcal{L}}{\partial \mathbf{n}_{,m}}\bigg)
\bigg]
\bigg\}
\,,\label{stat-act-director}
\end{eqnarray}
with natural (homogeneous) conditions, expressing continuity of 
\begin{equation}\label{liqxtal-bc}
\frac{\partial \mathcal{L}}{\partial J}
\quad\hbox{and}\quad
\hat{n}_m\frac{\partial \mathcal{L}}{\partial \mathbf{n}_{,m}}\,,
\end{equation} 
on a boundary, or at a material interface, whose unit normal has spatial
Cartesian components $\hat{n}_m$. (We always sum over repeated indices.)

\paragraph{Action principle \#1 -- Euler-Lagrange equations.}
Stationarity of the action, $\delta\mathcal{S}=0$, i.e., {\bfi Hamilton's
principle}, thus yields the following {\bfi Euler-Lagrange equations} in the
Lagrangian fluid description,
\begin{eqnarray}
&&\delta x_p: \quad
 \bigg(
\frac{\partial \mathcal{L}}{\partial
\dot{x}_p}\bigg)^{\!\!\boldsymbol{\dot{\,}}}
 +
\,J\frac{\partial}{\partial x_p}\frac{\partial \mathcal{L}}{\partial J}
-
J\frac{\partial}{\partial x_m}
\bigg( J^{-1}
\frac{\partial \mathcal{L}}{\partial \mathbf{n}_{,m}}
\boldsymbol{\cdot}\mathbf{n}_{,p}\bigg)
=
0\,,
\label{liqxtal-motion-Lag}
\\
&&\delta\mathbf{n}:\quad
\Big(
\frac{\partial \mathcal{L}}
{\partial\boldsymbol{\dot{\mathbf{n}}}}\Big)^{\boldsymbol{\dot{\,}}} 
- 
\frac{\partial \mathcal{L}}{\partial\mathbf{n}}
+
\, J \frac{\partial}{\partial x_m}
\bigg(J^{-1}
\frac{\partial \mathcal{L}}{\partial \mathbf{n}_{,m}}\bigg)
=
0
\,.
\label{liqxtal-director-Lag}
\end{eqnarray}
The material volume element 
$J={\rm det}(\partial\mathbf{x}/\partial\mathbf{X})$
satisfies an auxiliary kinematic equation, 
obtained from its definition,
\begin{equation} \label{J-dot-eqn}
\dot{J}
= 
J\frac{\partial X_A}{\partial x_i}
\frac{\partial \dot{x}_i}{\partial X_A}
\,.
\end{equation} 
Imposing constant $J$, say $J=1$, gives incompressible flow.
The last term in equation (\ref{liqxtal-motion-Lag}) is the
divergence of the Ericksen stress tensor arising due to the dependence of
the potential energy of the medium on the strain
$\nabla\mathbf{n}$, see   {Ericksen [1960, 1961]}.

With the proper choice of Lagrangian density, namely,
\begin{equation} \label{liqxtal-OF-Lag}
\mathcal{L} 
= 
\frac{\rho}{2}|\boldsymbol{\dot{\mathbf{x}}}|^2 
+
\frac{I}{2}|\boldsymbol{\dot{\mathbf{n}}}|^2 
+
p(J-1)
+
q(|\mathbf{n}|^2-1)
-
F(\mathbf{n}, \nabla\mathbf{n})
\,,
\end{equation} 
where $\rho$ and $I$ are material constants one finds
that the Euler-Lagrange equations (\ref{liqxtal-motion-Lag}) and
(\ref{liqxtal-director-Lag}) produce the {\bfi Ericksen-Leslie equations}
for incompressible liquid crystal flow, {Chandrasekhar [1992]} and {de Gennes
and Prost [1993]},
\begin{eqnarray}\label{Ericksen-Leslie eqns}
\rho\,\ddot{x}_i 
+
\frac{\partial}{\partial x_j}
\bigg( p\,\delta_{ij}
 +
\mathbf{n}_{,i}\boldsymbol\cdot\frac{\partial F}{\partial \mathbf{n}_{,j}}
\bigg)
&=& 0
\,,
\\
I\boldsymbol{\ddot{\mathbf{n}}}
-
2q\mathbf{n} 
+
\mathbf{h} 
= 0
\,,
\quad\hbox{with}\quad
\mathbf{h} 
&=& 
\frac{\partial F}{\partial \mathbf{n}}
-
\frac{\partial}{\partial x_j}
\frac{\partial F}{\partial \mathbf{n}_{,j}}
\,.\label{director-eqn}
\end{eqnarray}
In the Lagrangian density (\ref{liqxtal-OF-Lag}), the Lagrange multipliers
$p$ and $q$ enforce the incompressibility condition $J=1$ and the director
normalization condition $|\mathbf{n}|^2=1$, respectively. The standard choice
for the function $F(\mathbf{n},\nabla\mathbf{n})$ is the Oseen-Z\"ocher-Frank
Helmholtz free energy density, as discussed, e.g., in
{Chandrasekhar [1992]} and {de Gennes and Prost [1993]}, namely,%
\footnote{This notation treats spatial and director components on the same
(Cartesian) basis. Later, we will distinguish between these types of
components, see, e.g., equation (\ref{Ham-matrix-diff-liqxtal-compon}).}
\begin{eqnarray} \label{Oseen-Frank erg}
F(\mathbf{n},\nabla\mathbf{n})
 &=& 
\underbrace{
k_2 (\mathbf{n}\cdot\nabla\times\mathbf{n})
}_{\hbox{{\bfi chirality}}}
\
+\
\underbrace{
\frac{k_{11}}{2} 
(\nabla\cdot\mathbf{n})^2
}_{\hbox{{\bfi splay}}}
\nonumber\\
&+&
\underbrace{
\frac{k_{22}}{2} (\mathbf{n}\cdot\nabla\times\mathbf{n})^2
}_{\hbox{{\bfi twist}}}
\
+\
\underbrace{
\frac{k_{33}}{2} |\mathbf{n}\times\nabla\times\mathbf{n}|^2
}_{\hbox{{\bfi bend}}}
\nonumber\\
&+&
\frac{\Delta\epsilon}{2}\, \mathbf{E}\cdot S(\mathbf{n})\cdot\mathbf{E}
\
+\
\frac{\Delta\mu}{2}\, \mathbf{B}\cdot S(\mathbf{n})\cdot\mathbf{B}
\,,
\end{eqnarray}
with $k_2=0$ for nematics (but nonzero for cholesterics) and
$\mathbf{n}\cdot\nabla\times\mathbf{n}=0$ for smectics. Since smectics
form layers that break translational symmetry, their order parameter group may
be taken as the Euclidean group $E(3)$, or perhaps as $SO(3)\times U(1)$ in
simple cases. We shall concentrate our attention on nematics in this section.
However, the general theory for arbitrary order parameter groups developed in
the next section would also encompass smectics.

\paragraph{Key concepts: Material angular frequency and spatial strain of
rotation.}  The rate of rotation of a director $\mathbf{n}$ in the rest frame
of the fluid material element that carries it is given by
\begin{equation}
\boldsymbol\nu =
\mathbf{n}\times\boldsymbol{\dot{\mathbf{n}}}\,.
\end{equation}
In terms of this {\bfi material angular frequency} $\boldsymbol\nu$ (which is
orthogonal to the director $\mathbf{n}$), the dynamical equation
(\ref{director-eqn}) becomes simply,
\begin{equation}
I\boldsymbol{\dot\nu} 
=
\mathbf{h}\times \mathbf{n}\,.
\end{equation}
Likewise, the amount by which a specified director field
$\mathbf{n}(\mathbf{x})$ rotates under an infinitesimal spatial displacement
from $x_i$ to $x_i + dx_i$ is given by,
\begin{equation}\label{gamma-vector-def}
\boldsymbol\gamma_i = \mathbf{n}\times\nabla_i\,\mathbf{n}
\quad\hbox{or,}\quad
\gamma\cdot dx 
\equiv
\boldsymbol\gamma_i dx_i = \mathbf{n}\times d\mathbf{n}
\,.
\end{equation}
%

\paragraph{Remarks on kinematics.}
An orthogonal matrix with components $\chi_{kL}$ is associated with an
orthonormal frame of unit vectors $(\mathbf{l},\mathbf{m},\mathbf{n})$ whose
components
$(l_k,m_k,n_k)=(\chi_{k1},\chi_{k2},\chi_{k3})$ satisfy
\begin{equation}\label{connex-def}
\epsilon_{jkl}(d\chi\chi^{-1})_{lk}
=
(\mathbf{l}\times d\mathbf{l}
+
\mathbf{m}\times d\mathbf{m}
+
\mathbf{n}\times d\mathbf{n})_j
\,,
\end{equation}
which is $O(3)$ right invariant. Accordingly, the spatial rotational strain,
with components
\begin{equation}\label{spat-rot-strn-def}
(\mathbf{n}\times d\mathbf{n})_j
=
\epsilon_{jkl}(d\chi_{l3}\chi_{k3})
\,,
\end{equation}
and the material angular velocity,
\begin{equation}
(\mathbf{n}\times\boldsymbol{\dot{\mathbf{n}}})_j
=
\epsilon_{jkl}(\dot{\chi}_{l3}\chi_{k3})
\,,
\end{equation}
are each elements of $TS^2/Z_2$, i.e., they are tangent to the sphere
$|\mathbf{n}|^2=1$ and $O(2)$ right invariant under $SO(2)$ rotations about
$\mathbf{n}$ and under reflections
$\mathbf{n}\to-\mathbf{n}$. Also, $\mathbf{n}\times d\mathbf{n}$ and
$\mathbf{n}\times\boldsymbol{\dot{\mathbf{n}}}$ are $O(2)$ right
invariant elements of the Lie algebra, $so(3)\simeq TSO(3)/SO(3)$, so they are
elements of $so(3)/O(2)$. Being orthogonal to $\mathbf{n}$, their evolution
preserves $|\mathbf{n}|^2$. Therefore,
$|\mathbf{n}|^2=1$ may be taken as an initial condition, and $\mathbf{n}$ may
be reconstructed from $\boldsymbol{\dot{\mathbf{n}}}
=
\boldsymbol\nu \times \mathbf{n}
$.\medskip

The {\bfi spatial rotational strain} $\mathbf{n}\times d\mathbf{n}=\gamma\cdot
dx$ satisfies the kinematic relation,
\begin{equation}\label{LPgamma-dot}
\big(\gamma\cdot dx\big)\boldsymbol{\dot{\,}}
=
\big(2\boldsymbol\nu\times\boldsymbol\gamma_i
+
\nabla_i\,\boldsymbol\nu\big)dx_i
\,,
\end{equation}
obtained from its definition. This kinematic relation involves only
$\boldsymbol\gamma_i$, $\boldsymbol\nu$ and $x_i$, which suggests 
transforming variables from
$(\mathbf{n},\boldsymbol{\dot{\mathbf{n}}},\nabla\mathbf{n})$ to
$\boldsymbol\gamma_i$ and $\boldsymbol\nu$. (Note that $\boldsymbol\gamma_i$
and $\boldsymbol\nu$ do not distinguish between $\mathbf{n}$ and
$-\mathbf{n}$.) Each component of $\boldsymbol\gamma_i$ is orthogonal to the
director $\mathbf{n}$. That is, $\mathbf{n}\cdot\boldsymbol\gamma_i = 0$, for
$i=1,2,3$. Using this fact and $|\mathbf{n}|^2=1$ gives the relations,
\begin{equation}\label{gamma-grad-n relns}
\nabla_i\,\mathbf{n} =
-\,\mathbf{n}\times\boldsymbol\gamma_i
\ \ \hbox{and}\ \
\nabla\times\mathbf{n} 
= -\,\mathbf{n}\;{\rm tr}\boldsymbol{\gamma}
+
n^m\,\boldsymbol{\gamma}_m
\,,
\end{equation}
or, equivalently, in index notation,
\begin{equation}\label{gamma-grad-n relns-index}
\nabla_i\,n^a
=
-\,\epsilon^{abc}\,n^b\gamma^c_i
\ \ \hbox{and}\ \
\epsilon^{ijk}\,\nabla_j n^k
= - n^i\;\gamma^m_m
+
n^m\,\gamma^i_m
\,,
\end{equation}
where lower indices are spatial and upper indices denote director components.
Consequently, we have the following useful {\bfi transformation formulas},
\begin{equation}\label{gamma-grad-n lns}\hspace{-.1in}
\nabla_i\,\mathbf{n}\cdot\!\nabla_j\,\mathbf{n}
=
\boldsymbol\gamma_i\cdot\boldsymbol\gamma_j
\,,\
\nabla_i\,\mathbf{n}\times\!\nabla_j\,\mathbf{n}
=
\boldsymbol\gamma_i\times\boldsymbol\gamma_j
\ \ \hbox{and}\ \
\mathbf{n}\cdot\nabla\times\mathbf{n}
=
 - {\rm tr}\boldsymbol{\gamma}
\,.
\end{equation}
%

\paragraph{Connection, curvature,  singularities and topological indices.}
The Eulerian curl of the spatial rotational strain $\boldsymbol\gamma_i$
yields the remarkable relation,
\begin{equation}\label{curl-gamma}
B_{ij}
\equiv
\boldsymbol{\gamma}_{i,j} - \boldsymbol{\gamma}_{j,i} 
+
2\boldsymbol\gamma_i\times\boldsymbol\gamma_j
=
0
\,,
\end{equation}
where $B_{ij}$ vanishes, provided the vector field $\mathbf{n}$ is
continuous. In geometrical terms, vanishing of $B_{ij}$ is the Maurer-Cartan
relation for the flat connection one-form (or, left-invariant Cartan one-form)
$\boldsymbol\gamma_i dx_i = \mathbf{n}\times d\mathbf{n}$, as discussed in,
e.g., {Flanders [1989]}. That is, the curvature two-form $B_{ij}dx_i\wedge
dx_j$ vanishes in the absence of singularities (disclinations) in the director
field $\mathbf{n}(\mathbf{x})$. Thus, $B_{ij}$ may be regarded as the areal
density of disclinations in a nematic liquid crystal.  In Section
\ref{Ham-Princ-Lag-Red}, we shall discuss how to proceed when defects exist
and the disclination density $B_{ij}$ does not vanish.

A second remarkable formula also stems from the
curl$-\boldsymbol\gamma$ equation (\ref{curl-gamma}), 
which implies the relation,
\begin{equation}\label{Hopf-vector}
\frac{\partial \gamma^a_j}{\partial x_i} dx_i\wedge dx_j 
=
\epsilon^{abc} 
\frac{\partial n^b}{\partial x_i}
\frac{\partial n^c}{\partial x_j}dx_i\wedge dx_j
=
\epsilon^{abc} dn^b\wedge dn^c
\,,
\end{equation}
for a continuous director field.
Contracting with $n^a$ and taking
the exterior derivative of 
this relation implies
\begin{equation}\label{d3n}\hspace{-.1in}
\frac{\partial }{\partial x_k}\bigg(\!\!
n^a\frac{\partial \gamma^a_j}{\partial x_i} 
\bigg)
\!dx_i\wedge \!dx_j \wedge dx_k
=
\epsilon^{abc} dn^a\!\wedge dn^b\!\wedge dn^c 
={\rm det}(\!\nabla n)d^3x
={\rm det}(\gamma)d^3x
\,.
\end{equation}
By $|\mathbf{n}|^2=1$, these expressions must vanish,
when the director field has no singularities.
Thus, in the absence of singularities, there is a vector $\mathbf{v}$, for
which
\begin{equation}\label{Hopf-vector-is-a-curl}
\epsilon_{kij} n^a\frac{\partial \gamma^a_j}{\partial x_i}
=
\epsilon_{kij}\epsilon^{abc} n^a
\frac{\partial n^b}{\partial x_i}
\frac{\partial n^c}{\partial x_j}
=
\epsilon_{kij} v_{j,i}
=
({\rm curl}\,\mathbf{v})_k
\,,
\end{equation}
or, equivalently,
\begin{equation}\label{Hopf-vec-curl}
\epsilon^{abc} n^a \nabla n^b \times \nabla n^c
=
{\rm curl}\,\mathbf{v}
\,.
\end{equation}
Up to an inessential multiplicative factor, this formula is the well known
{\bfi Mermin-Ho relation} between three-dimensional superfluid texture
$\mathbf{n}$ and vorticity curl$\,\mathbf{v}$ in dipole-locked $^3He-A$, found
in {Mermin and Ho [1976]}. Since
$\mathbf{n}$ is a unit vector, a simple calculation in spherical polar
coordinates transforms this relation to
\begin{equation}\label{Hopf-vec}
\epsilon^{abc} n^a \nabla n^b \times \nabla n^c
= 
\nabla \phi \times \nabla
{\rm cos}\theta 
=
{\rm curl}\,\mathbf{v}
\,,
\end{equation}
for $\mathbf{n}=({\rm sin}\theta{\rm cos}\phi, {\rm sin}\theta{\rm sin}\phi,
{\rm cos}\theta)^T$ in terms of the polar angle $\theta$ and azimuthal angle
$\phi$ on the sphere. Hence,
\begin{equation}\label{area-elmnt}
{\rm curl}\,\mathbf{v}\cdot d\mathbf{S}
=
\epsilon^{abc} n^a dn^b\wedge dn^c 
= 
d\phi \wedge d{\rm cos}\theta 
\,,
\end{equation}
which is the area element on the sphere. Thus, by Stokes' theorem for a
given closed curve $C$ in $\mathbb{R}^3$ with unit tangent vector
$\mathbf{n}(\mathbf{x})$, the integral 
\begin{equation}\label{write-int}
2\pi {\rm Wr}
\equiv
\oint_C \mathbf{v} \cdot d\mathbf{x}
=
\int\!\!\!\!\int_{S} 
\epsilon^{abc} n^a \gamma^b_k\,dx_k\wedge \gamma^c_l\,dx_l
\,,
\end{equation}
is the (signed) area swept out by $\mathbf{n}$ on a portion of the unit
sphere upon traversing the curve $C$ which is the boundary of the surface $S$
in three dimensions. This is also equal to the {\bfi writhe} of the curve $C$,
see {Fuller [1978]}, as cited in {Kamien [1998]}. Thus, if the field lines of
the three dimensional vector field $\mathbf{n}(\mathbf{x})$ form a closed curve
$C$,  the writhe of this curve is given by the area integral in equation
(\ref{write-int}) over a surface whose boundary is $C$. Equation
(\ref{write-int}) is the second remarkable formula implied by the
curl$-\boldsymbol\gamma$ equation (\ref{curl-gamma}).

Finally, the integral of $\mathbf{v}\cdot{\rm curl}\,\mathbf{v}$ over three
dimensional space gives a number,
\begin{equation}\label{Hopf-index}
N
=
\frac{1}{4\pi}\int_{S^3}
\mathbf{v}\cdot{\rm curl}\,\mathbf{v}\ d^3x
\,,
\end{equation}
which takes integer values and is called the {\bfi Hopf index} or degree of
mapping for the map $\mathbf{n}(\mathbf{x}):S^3\to S^2$. The Hopf degree of
mapping $N$ counts the number of times the map $\mathbf{n}(\mathbf{x}):S^3\to
S^2$ covers the unit sphere, as discussed in, e.g., {Flanders [1989]}.
Equivalently, the Hopf degree of mapping $N$ counts the number of linkages
of the three dimensional divergenceless vector field curl$\,\mathbf{v}$ with
itself. See {Mineev [1980]} for the physical interpretation of this formula in
terms of disclinations in nematic liquid crystal experiments. See also {Trebin
[1982]}, {Kleman [1983]}, {Kleman [1989]} and references therein for additional
discussions of the differential geometry of defects in liquid crystal physics.

The quantity $\epsilon^{abc} n^a \nabla n^b \times \nabla n^c $ in
equation (\ref{Hopf-vec}) appears in many differential geometric contexts in
physics: in the Mermin-Ho relation for the vorticity of superfluid $^3He-A$ in
terms of the ``texture'' $\mathbf{n}$ due to
{Mermin and Ho [1976]};  in the Skyrmion Lagrangian, as discussed in, e.g., 
{Trebin [1982]};  as the $n-$field topological Wess-Zumino term in the $O(3)$
nonlinear sigma model, discussed in, e.g.,  {Yabu and Kuratsuji [1999]},
{Tsurumaru [1999]};  and as the instanton number
density in coset models.  For references to the last topic see
{Coquereaux and Jadcyk [1994]} and for a
discussion of the associated Poisson-Lie models, see {Stern [1999]}. This same
term also produces forces on vortices in $^3He-A(B)$ as found in
{Hall [1985]} and in ferromagnets, discussed in 
{Kuratsuji and Yabu [1998]}. Also, we have seen
that this term also allows the calculation of the ``writhe'' of a
self-avoiding closed loop of, say, DNA, or some other polymer filament with
unit tangent vector $\mathbf{n}$, see, e.g., {Fuller [1978]}, 
{Goldstein et al. [1998]}, {Kamien [1998]}.  See also
{Klapper [1996]} and {Goriely and Tabor [1997]}, for additional recent
discussions of the differential geometric writhe of DNA conformations. 
Finally, we have identified the relation of the Hopf degree of
mapping to the linkage number of the {\bfi writhe flux}, curl$\,\mathbf{v}$.
Perhaps surprisingly, in the case of liquid crystals, it will turn out that
this topological index, the writhe of a fluid loop, and the flux of curvature
$B_{ij}$ through a fluid surface (which measures the presence of singularities
in the director field) are are all three {\bfi created} nonabelian PCF
dynamics. 

\paragraph{Action principle \#2 -- Lagrange-Poincar\'e equations.}
Intrigued by the geometrical properties of the variables $\boldsymbol\gamma_i$
and $\boldsymbol\nu$, we shall seek to express the Ericksen-Leslie equations
solely in terms of these variables, by applying symmetry reduction methods to
Hamilton's principle for liquid crystal dynamics. The trace of the
spatial rotational strain $\boldsymbol\gamma$ gives
\begin{equation}\label{trace-gamma}
\gamma^a_m 
\equiv 
\epsilon^{abc} n^b \nabla_m\, n^c 
\quad\Rightarrow\quad
{\rm tr}\boldsymbol{\gamma} 
\equiv \gamma_m^m
 =
-\,\mathbf{n}\cdot\nabla\times\mathbf{n}
\,,
\end{equation}
which figures in the energies of chirality and twist
in equation (\ref{Oseen-Frank erg}) for $F(\mathbf{n}, \nabla\mathbf{n})$. 

We shall now drop the anisotropic dielectric and diamagnetic terms, which
exert torques on the director angular momentum due to external $\mathbf{E}$
and $\mathbf{B}$ fields given by, 
\begin{equation}\label{EM-torques-def}
\mathbf{h}_{EM}\times \mathbf{n}
=
\frac{3\Delta\epsilon}{2}(\mathbf{E}\cdot\mathbf{n})
\mathbf{E}\times \mathbf{n}
+
\frac{3\Delta\mu}{2}(\mathbf{B}\cdot\mathbf{n})
\mathbf{B}\times \mathbf{n}
\,,
\end{equation}
but which do not contribute in the motion equation.
Upon neglecting these torques, which may always be restored later (see the
Appendix), the remaining Oseen-Frank free energy $F(\mathbf{n},
\nabla\mathbf{n})$ may be expressed (modulo a divergence term that will not
contribute to the Euler-Lagrange equations) entirely as a function of a subset
of the invariants of
$\boldsymbol\gamma$, namely, ${\rm tr}\boldsymbol{\gamma}$, 
${\rm tr}(\boldsymbol{\gamma} + \boldsymbol{\gamma}^T)^2$ and 
${\rm tr}(\boldsymbol{\gamma} - \boldsymbol{\gamma}^T)^2$, 
with the help of the identities below, obtained using $|\mathbf{n}|^2~=~1$. 
In the notation of {Eringen [1997]}, one has
\begin{eqnarray}\label{grad-n-identities}
I_1
&\equiv&
(\nabla\times\mathbf{n})^2
=
[\mathbf{n}\cdot(\nabla\times\mathbf{n})]^2
+
[\mathbf{n}\times(\nabla\times\mathbf{n})]^2
\,,\\
I_2
&\equiv&
n^i_{,j}n^i_{,j}
=
(\nabla\cdot\mathbf{n})^2
+
\Big(\nabla\times\mathbf{n}\Big)^2
+
\Big(n^jn^i_{,j} - n^i n^j_{,j}\Big)_{,i}
\label{I2-div term}
\\
&&\hspace{.45in} =
\Big(\nabla\times\mathbf{n}\Big)^2
+\
n^j_{,i}n^i_{,j}\
\,,
\nonumber\\
I_3
&\equiv&
[(\mathbf{n}\cdot\nabla)\mathbf{n})]^2
=
[\mathbf{n}\times(\nabla\times\mathbf{n})]^2
\,.
\end{eqnarray}
Geometrically, the divergence term 
\begin{equation}\label{div-term}
\Big(n^jn^i_{,j} - n^i n^j_{,j}\Big)_{,i}
=
2\kappa_1\kappa_2
\,,
\end{equation}
appearing in equation (\ref{I2-div term}) for $I_2$ is (twice) the Gaussian
curvature of the surface $\Sigma$ whose normal is $\mathbf{n}(\mathbf{x})$.
This equation follows from Rodrigues' formula
\begin{equation}\label{Rodrigues-form}
d\mathbf{n} = - \, \kappa \, d\mathbf{x}
\,,
\end{equation}
for the principle directions of the surface $\Sigma$. 
Such a surface exists globally, provided
$\mathbf{n}\cdot{\rm curl}\,\mathbf{n}=0$,
see {Weatherburn [1974]}, as cited in {Kleman [1983]}.  Straightforward
calculations give the following linear relations, see Eringen [1997]
\begin{equation}\label{gamma-inv}
{\rm tr}(\boldsymbol\gamma_S^2)
=
-  \,\frac{1}{2}\, I_1
+  I_2
+  \,\frac{1}{2}\, I_3
\,,\quad
{\rm tr}(\boldsymbol\gamma_A^2)
=
-  \,\frac{1}{2}\, (I_1+I_3)
\,,
\end{equation}
where $\boldsymbol\gamma_S$ and $\boldsymbol\gamma_A$ are the symmetric and
antisymmetric parts of $\boldsymbol\gamma$,
\begin{equation}\label{gamma-inv-sym-antisym}
\boldsymbol\gamma_S
=
\frac{1}{2}\,(\boldsymbol{\gamma} + \boldsymbol{\gamma}^T)
\,,\quad\hbox{and}\quad
\boldsymbol\gamma_A
=
\frac{1}{2}\,(\boldsymbol{\gamma} - \boldsymbol{\gamma}^T)
\,.
\end{equation}
Hence, modulo the divergence term in (\ref{I2-div term}), the Oseen-Frank free
energy in (\ref{Oseen-Frank erg}) may be written in terms of the invariants of
$\boldsymbol\gamma$ as $F(\,{\rm tr}\boldsymbol\gamma, {\rm
tr}(\boldsymbol\gamma_S^2), {\rm tr}(\boldsymbol\gamma_A^2)\,)$. 

\paragraph{The reduced Lagrange-Poincar\'e action.}
By transforming to the rotational strain $\boldsymbol\gamma_m$ and frequency
$\boldsymbol\nu$, the action $\mathcal{S}$ for liquid crystals in
(\ref{liqxtal-action-director}) may thus be {\bfi reduced}, that is, rewritten
in fewer variables, by defining,
\begin{equation}\label{liqxtal-action-nu/gamma}
\mathcal{S} = \int dt\int d^3X \
\mathcal{L}(\boldsymbol{\dot{\mathbf{x}}}, J, \boldsymbol\nu,
\boldsymbol\gamma_m)
\,,
\end{equation} 
where the reduced variables $\boldsymbol\nu$ and $\boldsymbol\gamma_m$ are
perpendicular to the director $\mathbf{n}$.

A calculation using their definitions shows that the variations of
$J$ and $\boldsymbol\nu$ satisfy
\begin{equation}\label{del-nu}
\delta{J} 
=
J \frac{\partial X_A}{\partial x_i} \frac{\partial \delta x_i}{\partial X_A}
\,,\qquad
\delta\boldsymbol\nu
=
\big(\mathbf{n}\times\delta\mathbf{n}\big)\boldsymbol{\dot{\,}}
-2\boldsymbol\nu\times\big(\mathbf{n}\times\delta\mathbf{n}\big)
\,,
\end{equation}
in terms of the variational quantity
$\mathbf{n}\times\delta\mathbf{n}$. Likewise, we calculate the variation of
the rotational strain $\boldsymbol\gamma_m$ as
\begin{eqnarray}\label{del-gamma}
\delta\boldsymbol\gamma_m
&=&
J^{-1}\frac{\partial}{\partial X_A}
\Big[
J\frac{\partial X_A}{\partial x_m}
\big(\mathbf{n}\times\delta\mathbf{n}\big)\Big]
\nonumber\\
&&-\
2\boldsymbol\gamma_m\times
\big(\mathbf{n}\times\delta\mathbf{n}\big)
-
\boldsymbol\gamma_p\, J^{-1}
\frac{\partial}{\partial X_A}
\Big[
J\frac{\partial X_A}{\partial x_m}
\delta x_p
\Big]
\,.
\end{eqnarray}
This variational expression for $\boldsymbol\gamma_m$ may be rearranged into
the more suggestive form,
\begin{equation}\label{del-gamma-2}
\delta\boldsymbol\gamma_m
+
\boldsymbol\gamma_p \nabla_m\, \delta x_p
=
2\big(\mathbf{n}\times\delta\mathbf{n}\big)
\times\boldsymbol\gamma_m
+
\nabla_m
\big(\mathbf{n}\times\delta\mathbf{n}\big)
\,,
\end{equation}
in which its similarity with the kinematic relation for $\gamma$
in (\ref{LPgamma-dot}) is more readily apparent, especially when the kinematic
equations (\ref{J-dot-eqn}) and (\ref{LPgamma-dot}) are rewritten in the
notation of equation (\ref{director-grad}) as
\begin{equation}\label{LP-kinemat-rel}
\dot{J}
=
J \nabla_i\, \dot{x}_i
\quad\hbox{and}\quad
\boldsymbol{\dot{\gamma}}_m
+
\boldsymbol\gamma_p\nabla_m\, \dot{x}_p
=
2\boldsymbol\nu\times\boldsymbol\gamma_m
+
\nabla_m\,\boldsymbol\nu
\,.
\end{equation}
The variation of the Lagrange-Poincar\'e action $\mathcal{S}$ in
(\ref{liqxtal-action-nu/gamma}) for liquid crystals
in the Lagrangian fluid description may now be rewritten as
\begin{eqnarray}\label{x-nu-gamma-act-var}
\delta\mathcal{S}
&=&
-\int dt\int d^3X
\bigg\{\delta x_p\bigg[ \bigg(
\frac{\partial \mathcal{L}}{\partial\dot{x}_p}
\bigg)^{\!\!\boldsymbol{\dot{\,}}}
 +
\,J\frac{\partial}{\partial x_p}
\frac{\partial \mathcal{L}}{\partial J}
-
J\frac{\partial}{\partial x_m}
\bigg(  J^{-1}
\frac{{\partial \mathcal{L}}}{\partial
\boldsymbol\gamma_{m}}\boldsymbol{\cdot\,\gamma}_{p}^{}\bigg)
\bigg]
\nonumber\\
&&\hspace{-.5in}
+\
\big(\mathbf{n}\times\delta\mathbf{n}\big)
\!\boldsymbol{\cdot}\!
\bigg[
\bigg(\frac{{\partial \mathcal{L}}}
{\partial \boldsymbol\nu}\bigg)^{\boldsymbol{\dot{\,}}}
- 2{\boldsymbol\nu}\times
\frac{{\partial \mathcal{L}}}{\partial \boldsymbol\nu}
+\, 
J \frac{\partial}{\partial x_m}
\bigg(J^{-1}
\frac{{\partial \mathcal{L}}}{\partial \boldsymbol\gamma_{m}}\bigg)
-
2\boldsymbol\gamma_m\times
\frac{{\partial \mathcal{L}}}{\partial \boldsymbol\gamma_{m}}
\,\bigg]
\bigg\}
\nonumber\\
&&\hspace{-.5in}
+
\int dt\int d^3X
\bigg\{
\Big(\delta x_p
\frac{\partial \mathcal{L}}{\partial\dot{x}_p}
+
\big(\mathbf{n}\times\delta\mathbf{n}\big)
\boldsymbol{\cdot}
\frac{{\partial \mathcal{L}}}
{\partial \boldsymbol\nu}\
\Big)^{\!\boldsymbol{\dot{\,}}}
\\
&&\hspace{-.5in}
+\
\frac{\partial}{\partial X_A}\bigg[
\frac{\partial X_A}{\partial x_m}\
\delta x_p 
\Big(
J\frac{\partial \mathcal{L}}{\partial J}\delta_{mp}
-
\frac{{\partial \mathcal{L}}}{\partial \boldsymbol\gamma_{m}}
\boldsymbol{\cdot\,\gamma}_{p}\Big)
+
\frac{\partial X_A}{\partial x_m}\,
\big(\mathbf{n}\times\delta\mathbf{n}\big) 
\boldsymbol\cdot
\frac{{\partial \mathcal{L}}}{\partial \boldsymbol\gamma_{m}}
\,\bigg]\,\bigg\}.
\nonumber
\end{eqnarray}
Consequently, the action principle $\delta \mathcal{S}=0$ yields the following
{\bfi Lagrange-Poincar\'e equations} for liquid crystals,
\begin{eqnarray}\label{x-nu-gamma-mot}
&&
\delta x_p:\quad
\bigg(\frac{\partial \mathcal{L}}{\partial
\dot{x}_p}\bigg)^{\!\!\boldsymbol{\dot{\,}}}
 +
\,J\frac{\partial}{\partial x_p}\frac{\partial \mathcal{L}}{\partial J}
-
J\frac{\partial}{\partial x_m}
\bigg(  J^{-1}
\frac{{\partial \mathcal{L}}}{\partial
\boldsymbol\gamma_{m}}\boldsymbol{\cdot\,\gamma}_{p}^{}\bigg)
=
0\,,
\\
&&\mathbf{n}\times\delta\mathbf{n}:\quad
\bigg(\frac{{\partial \mathcal{L}}}
{\partial \boldsymbol\nu}\bigg)^{\boldsymbol{\dot{\,}}}
- 2{\boldsymbol\nu}\times
\frac{{\partial \mathcal{L}}}{\partial \boldsymbol\nu}
+\, 
J \frac{\partial}{\partial x_m}
\bigg(J^{-1}
\frac{{\partial \mathcal{L}}}{\partial \boldsymbol\gamma_{m}}\bigg)
-
2\boldsymbol\gamma_m\times
\frac{{\partial \mathcal{L}}}{\partial \boldsymbol\gamma_{m}}
=0
\,,
\nonumber
\end{eqnarray}
with natural boundary conditions, cf. equation (\ref{liqxtal-bc}),
\begin{equation}\label{nu-gamma-bc}
\partial \mathcal{L}/\partial J = 0
\quad\hbox{and}\quad
\hat{n}_m\,
\frac{{\partial \mathcal{L}}}{\partial \boldsymbol\gamma_{m}}
=0\,,
\end{equation} 
on a boundary, or material interface, whose unit normal
has spatial Cartesian components $\hat{n}_m$. These are the {\bfi dynamical
boundary conditions} for liquid crystal motion. These conditions ensure that
the fluid pressure $\partial \mathcal{L}/\partial J$ and the normal stress are
continuous across a fluid interface. The term  
$\boldsymbol\gamma_p\boldsymbol\cdot 
\partial\mathcal{L}/\partial\boldsymbol\gamma_m$ contributes to the stress
tensor of the complex fluid and arises from the dependence of its free
energy upon the rotational strain $\boldsymbol\gamma_i$. Again $J=1$ for
incompressible flow, as imposed by the Lagrange multiplier $p$, the fluid
pressure. (The Lagrange multiplier $q$ is no longer necessary.)

Upon specializing to the Oseen-Frank-type Lagrangian
(\ref{liqxtal-OF-Lag}), stationarity of the action 
$\mathcal{S}(\dot\mathbf{x},J,\boldsymbol\nu,\boldsymbol\gamma)$
in equation (\ref{liqxtal-action-nu/gamma}) recovers the Ericksen-Leslie
equations (\ref{Ericksen-Leslie eqns}) written in these variables.
This is an example of {\bfi Lagrangian reduction} -- from the variables 
$(\mathbf{n},\boldsymbol{\dot\mathbf{n}},\nabla\mathbf{n})$ to
$(\boldsymbol\nu,\boldsymbol\gamma)$. The equations in fewer variables that
result from the first stage of Lagrangian reduction were named the
Lagrange-Poincar\'e equations in {Marsden and Scheurle [1995]}, {Cendra,
Marsden and Ratiu [1999]}. One of the goals of this paper is to characterize
the geometrical properties of such equations for PCFs,
using the mathematical framework of Lagrangian reduction by stages established
in {Cendra, Marsden and Ratiu [1999]}. From this viewpoint, the first stage of
the Lagrangian reduction  for liquid crystals is finished.

\paragraph{Action principle \#3 -- Euler-Poincar\'e equations.}
The Lagrange-Poincar\'e equations (\ref{x-nu-gamma-mot}) and their kinematic
relations (\ref{LP-kinemat-rel}) are still expressed in the Lagrangian fluid
description. We shall pass to the Eulerian fluid description by applying
a second stage of Lagrangian reduction, now defined by the following right
actions of the diffeomorphism group,
\begin{eqnarray}\label{diffeo-actions}
u(x,t) &=& \dot{x}(X,t) g^{-1}(t)
\,,
\nonumber\\
D (x,t) d^3x &=& d^3X g^{-1}(t)
\,,
\nonumber\\
\boldsymbol\nu (x,t) 
&=&
\boldsymbol\nu (X,t) g^{-1}(t)
\,,
\\
\boldsymbol\gamma_i (x,t) dx_i
&=&
\Big(\boldsymbol\gamma_i (X,t) 
\frac{\partial x_i\,(X,t)}{\partial X_A} dX_A\Big)  g^{-1}(t)
\,,
\nonumber
\end{eqnarray}
where the right action denoted $x(X,t) = X g(t)$ defines the fluid motion as
following   a time-dependent curve $g(t)$ in the diffeomorphism group
as it acts on the reference configuration of the fluid with coordinate $X$. In
the Eulerian fluid description, the action $S$ in Hamilton's principle $\delta
S = 0$ is written as
\begin{equation}\label{liqxtal-action-nu/gamma-Eul}
\mathcal{S} = \int dt\int d^3x\
\ell(\mathbf{u}, D, \boldsymbol\nu, \boldsymbol\gamma)
\,,
\end{equation} 
in terms of the Lagrangian density $\ell$ given by
\begin{equation}\label{liqxtal-nu/gamma-Eul/Lag}
\ell(\mathbf{u}, D, \boldsymbol\nu, \boldsymbol\gamma) d^3x
=
\mathcal{L}\Big(\dot{x}g^{-1}(t), Jg^{-1}(t), 
\boldsymbol\nu g^{-1}(t), \boldsymbol\gamma g^{-1}(t)\Big)
\Big(d^3X g^{-1}(t)\Big)
\,.
\end{equation} 
The variations of these Eulerian fluid quantities are computed from their
definitions to be,
\begin{eqnarray}\label{del-Eul-var-u}
\delta u_j &=& \frac{\partial\eta_j}{\partial t}
+
u_k\frac{\partial\eta_j}{\partial x_k}
-
\eta_k\frac{\partial u_j}{\partial x_k}
\,,
\\
\delta D &=& - \,\frac{\partial D\eta_j}{\partial x_j}
\,,\label{del-Eul-var-Dee}
\\
\delta \boldsymbol\nu
&=&
\frac{\partial\boldsymbol\Sigma}{\partial t}
+
u_m\frac{\partial\boldsymbol\Sigma}{\partial x_m}
-
2\boldsymbol\nu\times\boldsymbol\Sigma
-
\eta_m\frac{\partial\boldsymbol\nu}{\partial x_m}
\,,\label{del-Eul-var-nu}
\\
\delta \boldsymbol\gamma_m 
&=&
\frac{\partial\boldsymbol\Sigma}{\partial x_m}
-
2\boldsymbol\gamma_m\times\boldsymbol\Sigma
-
\eta_k\frac{\partial\boldsymbol\gamma_m}{\partial x_k}
-
\boldsymbol\gamma_k\frac{\partial\eta_k}{\partial x_m}
\,,\label{del-Eul-var-gamma}
\end{eqnarray}
where $\boldsymbol\Sigma(x,t) \equiv 
(\mathbf{n}\times\delta\mathbf{n})(X,t)g^{-1}(t)$
and $\eta \equiv \delta{g}g^{-1}(t)$. One may compare these Eulerian variations
with the Lagrangian variations in (\ref{del-nu}) and (\ref{del-gamma}). See
also {Holm, Marsden and Ratiu [1998]} for more discussion of such constrained
variations in Eulerian fluid dynamics. Here the quantities $D$ and
$\boldsymbol\gamma_m$ satisfy the following {\bfi Eulerian kinematic
equations}, also obtained from their definitions, cf. the Lagrangian kinematic
relations (\ref{LP-kinemat-rel}),
\begin{eqnarray}\label{D - gamma eqns}
\frac{\partial D}{\partial t}
 &=& - \,\frac{\partial Du_j}{\partial x_j}
\,,
\\
\frac{\partial \boldsymbol\gamma_m }{\partial t}
&=&
\frac{\partial\boldsymbol\nu}{\partial x_m}
\,+\,
2\,\boldsymbol\nu\times\boldsymbol\gamma_m
-
u_k\frac{\partial\boldsymbol\gamma_m}{\partial x_k}
-
\boldsymbol\gamma_k\frac{\partial u_k}{\partial x_m}
\,.\label{D - gamma eqns1}
\end{eqnarray}
Note the similarity in form between the Eulerian variations of $D$ and
$\boldsymbol\gamma_m$, and their corresponding kinematic equations. This
similarity arises because both the variations and the evolution equations for
these quantities are obtained as infinitesimal Lie group actions.

\paragraph{Remark.}
The kinematic formula (\ref{D - gamma eqns1}) for the
evolution of $\boldsymbol\gamma_m$ immediately implies the following {\bfi
$\boldsymbol\gamma-$circulation theorem} for the spatial rotational strain,
\begin{equation}\label{gamma-circ-thm}
\frac{d}{dt}\oint_{c(u)}\!\!\boldsymbol\gamma_m\, dx_m
= 
\oint_{c(u)}\!\!
2\,\boldsymbol\nu\times\boldsymbol\gamma_m\, dx_m
\,.
\end{equation}
Thus, the circulation of $\boldsymbol\gamma$ around a loop $c(u)$ moving with
the fluid is only conserved when $\boldsymbol\gamma_m\times\boldsymbol\nu$ is a
gradient. Otherwise, the curl of
$\boldsymbol\gamma_m\times\boldsymbol\nu$ generates circulation of
$\boldsymbol\gamma$ around fluid loops. By Stokes' theorem and equation
(\ref{Hopf-vector}) we have,
\begin{equation}\label{gamma-circ-thm-Stokes}
\oint_{c(u)} \gamma^a_m dx_m
=
\int_{S(u)} \frac{\partial\gamma^a_m}{\partial x_j}\,dx_j\wedge dx_m
=
\int_{S(u)} \epsilon^{abc} dn^b\wedge dn^c
\,,
\end{equation}
for a surface $S(u)$ whose boundary is the fluid loop $c(u)$.
Consequently, the $\boldsymbol\gamma-$circulation theorem
(\ref{gamma-circ-thm}) implies that a nonvanishing curl of
$\boldsymbol\gamma_m\times\boldsymbol\nu$ generates a time-changing flux of
$\epsilon^{abc} \nabla n^b \times \nabla n^c$ for
$a=1,2,3,$ through those surfaces whose boundaries move with the fluid. 

\paragraph{The nonabelian 2-cocycle terms create withe and linkages.}
We use the kinematic equation (\ref{D - gamma eqns1}) for gamma and the
definition of $\boldsymbol\nu$ to calculate the evolution of
curl$\mathbf{v}\cdot d\mathbf{S} = n^a\epsilon^{abc}dn^b\wedge dn^c$ as
\[
\big(\partial_t + \pounds_u\big) 
\big({\rm curl\,}\mathbf{v}\cdot d\mathbf{S}\big)
=
2\,d\nu^b\wedge dn^b
=
d\big(\mathbf{n}\boldsymbol\cdot 2\boldsymbol\nu\times \boldsymbol\gamma_m\
dx_m\big)
=
d\big(\mathbf{n}\boldsymbol\cdot {\rm ad}_{\nu}\!
\boldsymbol\gamma_m\,
dx_m\big)
\]
This means the nonabelian aspect of the generalized 2-cocycle (the
$\nabla\boldsymbol\nu$ in the gamma equation) {\bfi creates writhe} in the
boundary of a surface moving with the fluid under the gamma-evolution. That
is, by Stokes law,
\[
\frac{d}{dt}\oint_{c(u)}
\mathbf{v}\cdot d\mathbf{x}
=
\oint_{c(u)}
\mathbf{n}\boldsymbol\cdot 2\boldsymbol\nu\times \boldsymbol\gamma_m\
dx_m
=
\oint_{c(u)}
\mathbf{n}\boldsymbol\cdot {\rm ad}_{\nu}\!
\boldsymbol\gamma_m\,
dx_m
\,,
\]
and the writhe of a fluid loop is not preserved. Likewise, we calculate 
\[
\big(\partial_t + \pounds_u\big) 
\big(\mathbf{v}\cdot d\mathbf{x}\wedge d(\mathbf{v}\cdot d\mathbf{x})\big)
\ne
{\rm exact \ form}
\,.
\]
Therefore, we find
\[
\frac{d}{dt} \int
\mathbf{v}\cdot {\rm curl\,}\mathbf{v}\  d^3x
\ne
0
\,.
\]
Thus, perhaps surprisingly, the linkage number, the Hopf degree on mapping in
equation (\ref{Hopf-index}) is also not preserved by the
gamma-evolution. So the nonabelian generalized 2-cocycle term in the gamma
equation also {\bfi creates linkages} in the Mermin-Ho quantity,
curl$\mathbf{v}$. Conversely, the preservation of these quantities is an
Abelian characteristic.
%

\paragraph{Euler-Poincar\'e action variations.}
We compute the variation of the action
(\ref{liqxtal-action-nu/gamma-Eul}) in Eulerian variables at fixed time $t$
and spatial position $\mathbf{x}$ as,
\begin{eqnarray}\label{u-nu-gamma-act-var-Eul}
\delta\mathcal{S}
\!\!\!&=&\!\!\!\!\!
\int \!\! dt \!\!\int\!\! d^3x\bigg[
\frac{\delta \ell}{\delta u_j}\delta u_j
+
\frac{\delta \ell}{\delta D} \delta D
+
\frac{\delta \ell}{\delta
\boldsymbol\nu} \boldsymbol\cdot \delta\boldsymbol\nu
+
\frac{\delta \ell}{\delta
\boldsymbol\gamma_m} \boldsymbol\cdot \delta\boldsymbol\gamma_m
\bigg]
\nonumber\\
\!\!\!&=&\!\!\!\!\!
\int \!\! dt \!\!\int\!\! d^3x\Bigg\{
\eta_j\bigg[
-\,\frac{\partial}{\partial t}\frac{\delta \ell}{\delta u_j}
-\,\frac{\delta \ell}{\delta u_k}
\frac{\partial u_k}{\partial x_j}
-\,
\frac{\partial}{\partial x_k}
\Big(\frac{\delta \ell}{\delta u_j}u_k\Big)
+\,
D\frac{\partial}{\partial x_j}
\frac{\delta \ell}{\delta D}
\nonumber\\&&\hspace{.75in}
-\
\frac{\delta \ell}{\delta
\boldsymbol\nu} \boldsymbol\cdot \frac{\partial\boldsymbol\nu}{\partial x_j}
\,-\,
\frac{\delta \ell}{\delta \boldsymbol\gamma_m}
\boldsymbol\cdot 
\frac{\partial\boldsymbol\gamma_m}{\partial x_j}
+
\frac{\partial}{\partial x_m}
\bigg( \boldsymbol\gamma_j \boldsymbol\cdot
\frac{{\delta\ell}}{\delta
\boldsymbol\gamma_{m}}\bigg)
\bigg]
\label{liqxtal-Euler-act-var}\\&&
\nonumber\\
&&
\hspace{-.4in}
+\
\boldsymbol{\Sigma\,\,\cdot}
\bigg[
-\,
\frac{\partial}{\partial t}
\frac{\delta \ell}{\delta\boldsymbol\nu}
\,-\,
\frac{\partial}{\partial x_m}
\Big(
\frac{{\delta\ell}}
{\delta \boldsymbol\nu}u_m
+
\frac{{\delta\ell}}
{\delta \boldsymbol\gamma_m}
\Big)
+ \ 
2{\boldsymbol\nu}\times
\frac{{\delta \ell}}{\delta \boldsymbol\nu}
+ \, 
2{\boldsymbol\gamma_m}\times
\frac{{\delta \ell}}{\delta \boldsymbol\gamma_m}
\bigg]
\nonumber\\
&&
\hspace{1.4in}
+\
\frac{\partial}{\partial t}
\bigg[
\eta_j\frac{\delta \ell}{\delta u_j}
+
\boldsymbol{\Sigma\,\cdot\,}\frac{\delta \ell}{\delta\boldsymbol\nu}
\bigg]
\nonumber\\
&&
\hspace{-.4in}
+\
\frac{\partial}{\partial x_m}
\bigg[\eta_j \Big(\frac{\delta \ell}{\delta u_j}\, u_m
-
D \frac{\delta \ell}{\delta D}\, \delta_{jm}
\,-\,
\boldsymbol{\gamma}_j\,\boldsymbol\cdot\,
\frac{\delta \ell}{\delta \boldsymbol\gamma_m}
\Big)
+
\boldsymbol{\Sigma\,\cdot\,}\Big(
\frac{\delta \ell}{\delta\boldsymbol\nu}\, u_m
+
\frac{\delta \ell}{\delta \boldsymbol\gamma_m}
\Big)\bigg]
\Bigg\}\,,
\nonumber
\end{eqnarray}
where we have substituted the variational expressions
(\ref{del-Eul-var-u})-(\ref{del-Eul-var-gamma}) and integrated by parts.

The dynamical equations resulting from Hamilton's principle
$\delta\mathcal{S}=0$ are obtained by requiring the coefficients of the
arbitrary variations $\eta_j$ and $\boldsymbol\Sigma$ to vanish. We assume
these variations themselves vanish at the temporal endpoints and we defer
discussing the boundary terms for a moment. Hence, we obtain the {\bfi
Euler-Poincar\'e equations} for liquid crystals,
\begin{eqnarray}
\eta_j:\quad
\frac{\partial}{\partial t}\frac{\delta \ell}{\delta u_j}
&=&
-\,\frac{\delta \ell}{\delta u_k}
\frac{\partial u_k}{\partial x_j}
-\,
\frac{\partial}{\partial x_k}
\Big(\frac{\delta \ell}{\delta u_j}u_k\Big)
+\,
D\frac{\partial}{\partial x_j}
\frac{\delta \ell}{\delta D}
\label{liqxtal-mot-Eul}
\\&&\hspace{.75in}
-\
\frac{\delta \ell}{\delta \boldsymbol\nu} 
\boldsymbol\cdot\frac{\partial\boldsymbol\nu}{\partial x_j}
-\,
\frac{\delta \ell}{\delta
\boldsymbol\gamma_m}
\boldsymbol\cdot \frac{\partial\boldsymbol\gamma_m}{\partial x_j}
+
\frac{\partial}{\partial x_m}
\bigg( \boldsymbol\gamma_j
\boldsymbol\cdot
\frac{{\delta\ell}}{\delta
\boldsymbol\gamma_{m}}\bigg)
,
\nonumber\\
\boldsymbol\Sigma:\quad
\frac{\partial}{\partial t}
\frac{\delta \ell}{\delta\boldsymbol\nu}
&=&
-\,\frac{\partial}{\partial x_m}
\Big(
\frac{{\delta\ell}}
{\delta \boldsymbol\nu}u_m
+
\frac{{\delta\ell}}
{\delta \boldsymbol\gamma_m}
\Big)
\nonumber\\
&&\hspace{.75in}
+ \ 2{\boldsymbol\nu}\times
\frac{{\delta \ell}}{\delta \boldsymbol\nu}
+ \, 2{\boldsymbol\gamma_m}\times
\frac{{\delta \ell}}{\delta \boldsymbol\gamma_m}
\,.
\label{liqxtal-micromot-Eul}
\end{eqnarray}
Equation (\ref{liqxtal-micromot-Eul}) agrees with the Lagrangian version of the
micromotion equation governing $\boldsymbol\nu$ in (\ref{x-nu-gamma-mot}). To
see this, it is helpful to recall that
$\delta\ell/\delta\boldsymbol\nu$ is an Eulerian density, so a Jacobian is
involved in the transformation to the Lagrangian version. 

\paragraph{Momentum conservation.}
In momentum conservation form, the liquid crystal motion equation
(\ref{liqxtal-mot-Eul}) in the Eulerian fluid description becomes
\begin{equation}\label{liqxtal-momentum-eqn}
\frac{\partial}{\partial t}\frac{\partial \ell}{\partial u_j}
=
-\
\frac{\partial}{\partial x_m}
\bigg(\frac{\partial \ell}{\partial u_j}\, u_m 
+
\Big(\ell - D\frac{\partial \ell}{\partial D}\Big)\delta_{mj}
-
\, \boldsymbol\gamma_j
\boldsymbol\cdot
\frac{{\partial\ell}}{\partial
\boldsymbol\gamma_{m}\,}
\bigg)
\,,
\end{equation}
for simple algebraic dependence of the Lagrangian $\ell$ on
$\mathbf{u}$, $D$, $\boldsymbol\nu$ and $\boldsymbol\gamma_m$.  This
momentum conservation law is in agreement with the direct passage to Eulerian
coordinates of the motion equation (\ref{x-nu-gamma-mot}) in the Lagrangian
fluid description. For this transformation, it is helpful to recognize from
equation (\ref{liqxtal-nu/gamma-Eul/Lag}) that
$\ell=(\mathcal{L}J^{-1})g(t)^{-1}$ implies, by the chain rule, 
\begin{equation}\label{L-to-ell}
\frac{\partial \mathcal{L}}{\partial J}g(t)^{-1}
=
\ell - D\frac{\partial \ell}{\partial D}
\,,\quad\hbox{since}\quad
D(x,t)=J^{-1}(X,t)g(t)^{-1}\,.
\end{equation}
The momentum conservation law (\ref{liqxtal-momentum-eqn}) acquires
additional terms, if the Lagrangian $\ell$ also depends on gradients
of $\mathbf{u}$, $D$, $\boldsymbol\nu$ and $\boldsymbol\gamma_m$. 

\paragraph{Noether's theorem.} Noether's theorem associates conservation laws
to continuous symmetries of Hamilton's principle. See, e.g.,
{Olver [1993]} for a clear discussion of the classical
theory and {Jackiw and Manton [1980]} for its applications
in gauge theories. The momentum conservation equation
(\ref{liqxtal-momentum-eqn}) also emerges from Noether's theorem, since the
action $\mathcal{S}$ in equation (\ref{liqxtal-action-nu/gamma-Eul}) admits
spatial translations, that is, since this action is invariant under the
transformations,
\begin{equation}
x_j\to x_j^\prime = x_j + \eta_j(\mathbf{x},t)
\quad\hbox{with}\quad 
\eta_j = c_j\,,
\end{equation}
for constants $c_j$, with $j=1,2,3$. To see how equation 
(\ref{liqxtal-momentum-eqn}) emerges from Noether's theorem, simply add the
term $\partial (\ell\eta_j)/\partial x_j$ (arising from transformations of the
spatial coordinate) to the endpoint and boundary terms in the variational
formula (\ref{u-nu-gamma-act-var-Eul}) arising from variations at {\it
fixed} time $t$ and spatial position $\mathbf{x}$, then specialize to $\eta_j
= c_j$. 

\paragraph{Kelvin-Noether circulation theorem for liquid crystals.}
Rearranging the motion equation (\ref{liqxtal-mot-Eul}) and using the
continuity equation for $D$ in (\ref{D - gamma eqns}) gives the
{\bfi Kelvin-Noether circulation theorem}, 
cf. {Holm, Marsden and Ratiu [1998]}, 
\begin{equation}\label{Eul-vel-circ-thm}
\frac{d}{dt}\oint_{c(u)}
\frac{1}{D}
\frac{\delta \ell}{\delta u_j} dx_j
=
- 
\oint_{c(u)}\frac{1}{D}
\bigg[
\frac{\delta \ell}{\delta\boldsymbol\nu}
\boldsymbol\cdot d\boldsymbol\nu
+\,
\frac{\delta \ell}{\delta \boldsymbol\gamma_m}
\boldsymbol\cdot d\boldsymbol\gamma_m
-
\frac{\partial}{\partial x_m}\bigg( \boldsymbol\gamma_j
\boldsymbol\cdot
\frac{{\delta \ell}}{\delta
\boldsymbol\gamma_{m}}\bigg)dx_j
\bigg]
.
\end{equation}
Hence, stresses in the director field of a liquid crystal and gradients in
its angular frequency $\boldsymbol\nu$ and rotational strain
$\boldsymbol\gamma$ can generate fluid circulation. Equivalently, by Stokes'
theorem, these gradients of liquid crystal properties can generate vorticity,
defined as 
$\boldsymbol\omega\equiv{\rm curl}\,(D^{-1}\delta \ell/\delta\mathbf{u})$.
For incompressible flows, we may set $D=1$ in these equations and write the
vorticity dynamics as,
\begin{eqnarray}\label{Eul-vort-dyn}
\frac{\partial\omega_i}{\partial t}
+
(\mathbf{u}\cdot\nabla)\omega_i
-
(\boldsymbol\omega\cdot\nabla)u_i
&=&
\bigg(\nabla\nu^a\times\nabla\frac{\delta \ell}{\delta \nu^a}\bigg)_i
+
\bigg(\nabla\gamma^a_m\times
\nabla\frac{\delta \ell}{\delta\gamma^a_m}\bigg)_i
\nonumber\\
&&+\
\epsilon_{ijk}\,\frac{\partial}{\partial x_j}
\frac{\partial}{\partial x_m}
\bigg( \boldsymbol\gamma_k
\boldsymbol\cdot
\frac{{\delta\ell}}{\delta
\boldsymbol\gamma_{m}}\bigg)
\,.
\end{eqnarray}
Thus, spatial gradients in the director angular frequency $\boldsymbol\nu$ and
rotational strain $\boldsymbol\gamma$ are sources of the fluid vorticity in a
liquid crystal.

\subsection{Hamiltonian dynamics of perfect liquid crystals}

The Euler-Lagrange-Poincar\'e formulation of liquid crystal dynamics obtained
so far allows passage to the corresponding Hamiltonian formulation via the
following {\bfi Legendre transformation} of the reduced Lagrangian $\ell$ in
the velocities $\mathbf{u}$ and $\boldsymbol\nu$, in the Eulerian fluid
description,
\begin{equation}\label{liqxtal-legendre-xform}
m_i = \frac{\delta \ell}{\delta u_i}\,, \
\boldsymbol\sigma = \frac{\delta \ell}{\delta \boldsymbol\nu}\,, \quad
h(\mathbf{m}, D, \boldsymbol\sigma, \boldsymbol\gamma_m)
 = m_i u_i + \boldsymbol{\sigma\cdot\nu}
- \ell(\mathbf{u}, D, \boldsymbol\nu, \boldsymbol\gamma_m).
\end{equation}
Accordingly, one computes the derivatives of $h$ as
\begin{equation}\label{liqxtal-dual-var-derivs}
\frac{\delta h}{\delta m_i} 
=
 u_i \,,
\quad
\frac{\delta h}{\delta \boldsymbol\sigma} 
=
 \boldsymbol\nu\,,
\quad
\frac{\delta h}{\delta D} 
=
-\, \frac{\delta \ell}{\delta D}
\,,\quad
\frac{\delta h}{\delta \boldsymbol\gamma_m} 
=
- \,\frac{\delta \ell}{\delta \boldsymbol\gamma_m}\,.
\end{equation}

Consequently, the Euler-Poincar\'e equations (\ref{liqxtal-mot-Eul}) -
(\ref{liqxtal-micromot-Eul}) and the auxiliary kinematic equations 
(\ref{D - gamma eqns}) - (\ref{D - gamma eqns1}) for liquid
crystal dynamics in the Eulerian description imply the following equations,
for the Legendre-transformed variables, $(\mathbf{m}, D, \boldsymbol\sigma,
\boldsymbol\gamma)$,
\begin{eqnarray}
\frac{\partial m_i}{\partial t} 
 &=& 
-\,m_j \frac{\partial}{\partial x_i}
\frac{\delta h}{\delta m_j}
\,
 - 
\frac{\partial}{\partial x_j}
\bigg(m_i\frac{\delta h}{\delta m_j}\bigg)
-
D\,\frac{\partial}{\partial x_i}\frac{\delta h }{ \delta D}
\nonumber\\
&&
\ +\
\bigg(\frac{\partial\boldsymbol\gamma_j}{\partial x_i}\bigg)
\!\boldsymbol\cdot
\frac{\delta h }{ \delta \boldsymbol\gamma_j}
\ -\
\frac{\partial}{\partial x_j}\bigg(
\boldsymbol\gamma_i
\boldsymbol\cdot
\frac{\delta h }{ \delta \boldsymbol\gamma_j}\bigg)
-\
\boldsymbol\sigma\boldsymbol\cdot
\frac{\partial}{\partial x_i}
\frac{\delta h }{ \delta \boldsymbol\sigma}
\,,
\label{m-eqn-liqxtal}\\
\frac{\partial D}{\partial t}
 &=& - \,\frac{\partial }{\partial x_j}
\bigg(D\frac{\delta h}{\delta m_j}\bigg)
\,,
\label{D-eqn-liqxtal}\\
\frac{\partial\boldsymbol\sigma}{\partial t}  
 &=& 
-\,
\frac{\partial}{\partial x_j}
\bigg(\boldsymbol\sigma\,\frac{\delta h}{\delta m_j}
-\frac{\delta h}{\delta \boldsymbol\gamma_j}\bigg)
-
2\boldsymbol\sigma\times
\frac{\delta h}{\delta \boldsymbol\sigma}
- 
2 \boldsymbol\gamma_j\times \frac{ \delta h }{ \delta
\boldsymbol\gamma_j}
\,,
\label{sigma-eqn-liqxtal}\\
\frac{\partial \boldsymbol\gamma_i }{\partial t}
&=&
-\,
\boldsymbol\gamma_j\frac{\partial}{\partial x_i}
\frac{\delta h}{\delta m_j}
-
\bigg(\frac{\partial\boldsymbol\gamma_i}{\partial x_j}\bigg)
\frac{\delta h}{\delta m_j}
+\,\frac{\partial}{\partial x_i}
\frac{\delta h}{\delta\boldsymbol\sigma} 
-
2\,\boldsymbol\gamma_i\times\frac{\delta h}{\delta\boldsymbol\sigma}
\,.\qquad\label{gamma-eqn-liqxtal}
\end{eqnarray}
These equations are {\bfi Hamiltonian}. That is, they may be expressed in the
form 
\begin{equation}\label{Ham-form-liqxtals}
\frac{\partial \mathbf{z} }{\partial t}
=
\{\mathbf{z},h\} = \hbox{\sffamily b}\,\cdot
\frac{\delta h }{\delta \mathbf{z}}\,,
\end{equation}
where $\mathbf{z}\in (\mathbf{m}, D, \boldsymbol\sigma,
\boldsymbol\gamma)$ and the Hamiltonian matrix
{\sffamily b} defines the Poisson bracket
\begin{equation} \label{PB-def-liqxtals}
\{f,h\}= \int d^{\,n\,}x \,
\frac{\delta f }{\delta \mathbf{z}}
\cdot\,\hbox{\sffamily b}\,\cdot
\frac{\delta h }{\delta \mathbf{z}}
\,,
\end{equation}
which is bilinear, skew symmetric and satisfies the {\bfi Jacobi identity},
\[\{f,\{g,h\}\}+ c.p.(f,g,h) = 0.\] 
Assembling the liquid crystal equations (\ref{m-eqn-liqxtal}) -
(\ref{gamma-eqn-liqxtal}) into the Hamiltonian form (\ref{Ham-form-liqxtals})
gives, {\small
\begin{equation}\label{Ham-matrix-diff-liqxtal}
\frac{\partial}{\partial t}
\left[ \begin{array}{c} 
m_i \\ D \\ \boldsymbol\gamma_i \\ \boldsymbol\sigma
\end{array}\right]
= -
\left[ \begin{array}{cccc} 
m_j\partial_i + \partial_j m_i & 
D\partial_i &
 (\partial_j\boldsymbol\gamma_i
 - \boldsymbol\gamma_{j\,,\,i})\,\boldsymbol\cdot & 
\boldsymbol\sigma\boldsymbol\cdot\partial_i
\\ 
\partial_jD & 0 & 0 & 0
\\
\boldsymbol\gamma_j\partial_i + \boldsymbol\gamma_{i\,,\,j}& 0 & 0 & 
- \partial_i 
+ 2\boldsymbol\gamma_i\times
\\
\partial_j\boldsymbol\sigma & 0 & 
- \partial_j 
+ 2\boldsymbol\gamma_j\times
& 2\boldsymbol\sigma\,\times
\end{array} \right]
\left[ \begin{array}{c} 
{\delta h/\delta m_j} \\ 
{\delta h/\delta D} \\ 
{\delta h/\delta\boldsymbol\gamma_j} \\ 
{\delta h/\delta\boldsymbol\sigma}
\end{array}\right]
\end{equation}
}
$\!\!$In components, this Hamiltonian matrix expression becomes,
{\small
\begin{equation}\label{Ham-matrix-diff-liqxtal-compon}
\frac{\partial}{\partial t}
\left[ \begin{array}{c} 
m_i \\ D \\ \gamma_i^{\,\alpha} \\ \sigma_\alpha
\end{array}\right]
= -
\left[ \begin{array}{cccc} 
m_j\partial_i + \partial_j m_i & 
D\partial_i &
 \partial_j\gamma_i^{\,\beta} - \gamma_{j\,,\,i}^{\,\beta} & 
\sigma_\beta\partial_i
\\ 
\partial_jD & 0 & 0 & 0
\\
\gamma_j^{\,\alpha}\partial_i + \gamma_{i\,,\,j}^{\,\alpha}& 0 & 0 & 
-\delta_\beta^{\,\alpha}\partial_i 
- t_{\beta\kappa}^{\,\alpha}\gamma_i^{\,\kappa}
\\
\partial_j\sigma_\alpha & 0 & 
-\delta_\alpha^{\,\beta}\partial_j 
+ t_{\alpha\kappa}^{\,\beta} \gamma_j^{\,\kappa}
& -t_{\alpha\beta}^{\,\kappa} \sigma_\kappa
\end{array} \right]
\left[ \begin{array}{c} 
{\delta h/\delta m_j} \\ 
{\delta h/\delta D} \\ 
{\delta h/\delta\gamma_j^{\,\beta}} \\ 
{\delta h/\delta\sigma_\beta}
\end{array}\right]
\end{equation}
}
$\!\!$where $t^{\,\alpha}_{\ \beta\kappa} = 2\epsilon_{\alpha\beta\kappa}$
represents (twice) the vector cross product for liquid crystals. In this
expression, the operators act to the right on all terms in a product by the
chain rule and, as usual, the summation convention is enforced on repeated
indices. At this point we have switched to using both lower and upper Greek
indices for the internal degrees of freedom, so that we will agree later with
the more general theory, in which upper Greek indices refer to a basis set in
a Lie algebra, and lower Greek indices refer to the corresponding dual basis.
Lower Latin indices still denote spatial components.

\paragraph{Remarks about the Hamiltonian matrix.}
The Hamiltonian matrix in equation (\ref{Ham-matrix-diff-liqxtal-compon}) was
discovered some time ago in the context of spin-glasses and Yang-Mills
magnetohydrodynamics (YM-MHD) by using the Hamiltonian approach in 
{Holm and Kupershmidt [1988]}. There, it was shown to be a valid Hamiltonian
matrix by associating its Poisson bracket as defined in equation
(\ref{PB-def-liqxtals}) with the dual space of a certain Lie algebra of
semidirect-product type that has a generalized two-cocycle on it. The
mathematical discussion of this Lie algebra and its generalized two-cocycle is
given in {Holm and Kupershmidt [1988]}. A related Poisson bracket for spin
glass fluids is given in {Volovik and Dotsenko [1980]}. A different Poisson
bracket for nematic liquid crystals is given in {Kats and Lebedev [1994]}, who
discuss a constrained Poisson bracket that in general does not satisfy the
Jacobi identity. The liquid crystal equations in Kats and Lebedev [1994] also
differ from the Ericksen-Leslie equations by being first order in the time
derivatives of the director, rather than second order, as in the
Ericksen-Leslie theory. See also Isaev et al. [1995] for a discussion of
this first order theory using the Poisson bracket approach. The present work
ignores the first order (kinematic) theory in what follows and concentrates on
second order (dynamic) theory.

Being dual to a Lie algebra, our matrix in equation
(\ref{Ham-matrix-diff-liqxtal-compon}) is in fact a {\bfi Lie-Poisson
Hamiltonian matrix}. See, e.g., {Marsden and Ratiu [1999]} and references
therein for more discussions of such Hamiltonian matrices. For our present
purposes, its rediscovery in the PCF context links the
physical and mathematical interpretations of the variables in the theory of
PCFs with earlier work in the gauge theory approach to
condensed matter, see, e.g., {Kleinert [1989]}. These gauge theory aspects
emerge upon rewriting the Lie-Poisson Hamiltonian equations in terms of
{\bfi covariant derivatives} with respect to the space-time connection
one-form given by $\boldsymbol\nu dt + \boldsymbol\gamma_m dx_m$, as done in
{Holm and Kupershmidt [1988]}. The gauge theory approach to liquid crystal
physics is reviewed in, e.g., {Trebin [1982]}, {Kleman [1983]},
and {Kleman [1989]}.

The generalized two-cocycle in the Hamiltonian matrix 
(\ref{Ham-matrix-diff-liqxtal-compon}) is somewhat exotic for a classical 
fluid. This generalized two-cocycle consists of the partial derivatives in
equation (\ref{Ham-matrix-diff-liqxtal-compon}) appearing with the Kronecker
deltas in the $\boldsymbol\sigma-\boldsymbol\gamma$ cross terms. The first hint
of these terms comes from the exterior derivative $d\boldsymbol\nu$ appearing
in the kinematic equation (\ref{D - gamma eqns1}) for $\boldsymbol\gamma_m$. 
Finding such a feature in the continuum theory of liquid crystals may provide
a bridge for transfering ideas and technology between the classical and
quantum fluid theories.%
%
\footnote{In quantum field theory, these partial derivative operators in the
Poisson bracket (or commutator relations) are called non-ultralocal terms, or
Schwinger terms, after {Schwinger [1951, 1959]}. These terms
lead to the so-called ``quantum anomalies.'' Of course, no quantum effects are
considered here. However, the Poisson bracket
(\ref{Ham-matrix-diff-liqxtal-compon}) still contains {\it classical}
Schwinger terms.} See {Volovick [1992]} and {Volovick and Vachaspati [1996]}
for discussions of similar opportunities for technology transfer in the  
theory of superfluid Helium. The implications of the generalized two-cocycle
for the solutions of the liquid crystal equations can be seen by considering
two special cases: static solutions with one-dimensional spatial dependence;
and spatially homogeneous, but time-dependent, liquid crystal dynamics.

\paragraph{Static perfect liquid crystal solutions with $z$-variation.}
Static (steady, zero-velocity) solutions, with one-dimensional spatial
variations in, say, the $z$-direction obey equations (\ref{m-eqn-liqxtal}) -
(\ref{gamma-eqn-liqxtal})  specialized to
\begin{eqnarray}
-D\,\frac{d}{dz}\,\frac{\delta h }{ \delta D}
&=&
\boldsymbol\gamma_3\boldsymbol\cdot\frac{d}{dz}\,
\frac{\delta h }{ \delta \boldsymbol\gamma_3}
+\
\boldsymbol{\sigma\,\cdot\,}
\frac{d}{dz}\,
\frac{\delta h }{ \delta \boldsymbol\sigma}
\
=\
0
\,,
\nonumber\label{m-eqn-liqxtal-stat}\\
\frac{d}{dz}\,
\frac{\delta h}{\delta \boldsymbol\gamma_3}
&=&
2 \boldsymbol\gamma_3\times \frac{ \delta h }{ \delta
\boldsymbol\gamma_3}
+
2\boldsymbol\sigma\times
\frac{\delta h}{\delta \boldsymbol\sigma} 
\,,
\label{sigma-eqn-liqxtal-stat}\\
\frac{d}{dz}\,
\frac{\delta h}{\delta\boldsymbol\sigma} 
&=&
2\,\boldsymbol\gamma_3\times\frac{\delta h}{\delta\boldsymbol\sigma}
\,.\qquad
\nonumber\label{gamma-eqn-liqxtal-stat}
\end{eqnarray}

The sum of terms in the first equation of the set
(\ref{sigma-eqn-liqxtal-stat}) vanishes to give zero pressure gradient, as a
consequence of the latter two equations.  We compare the latter two equations
with the {\bfi $E(3)$ Lie-Poisson Hamiltonian systems}, given by
\begin{eqnarray}
-\,\frac{d\boldsymbol\Pi}{dt}\,
&=&
\frac{ \partial H }{ \partial
\boldsymbol\Pi}\times\boldsymbol\Pi
+ 
\frac{\partial H}{\partial \boldsymbol\Gamma}
\times\boldsymbol\Gamma
\,,
\label{heavy-top-Pi}\\
-\,\frac{d\boldsymbol\Gamma}{dt}
&=&
\frac{\partial H}{\partial\boldsymbol\Pi}
\times\boldsymbol\Gamma
\,.
\nonumber\label{heavy-top-Gamma}
\end{eqnarray}
When the Hamiltonian $H(\boldsymbol\Pi,\boldsymbol\Gamma)$ in these
equations is specialized to 
\begin{equation} \label{heavy-top-Ham}
H = \bigg(\,\sum_{i\,=\,1}^3\,\frac{(\Pi_i)^2}{2I_i}\bigg)
+ Mg\boldsymbol{\chi\cdot\Gamma}
\,,
\end{equation}
where $I_i$, $i=1,2,3$, and $Mg\boldsymbol\chi$ are constants, then the
$E(3)$ Lie-Poisson equations (\ref{heavy-top-Pi}) specialize to the classical
Euler-Poisson equations for a heavy top, discussed in, e.g., 
{Marsden and Ratiu [1999]}.

By comparing these two equation sets, one observes that (at least for
algebraic $h$) the static perfect liquid crystal flows with $z$-variation in
equations (\ref{sigma-eqn-liqxtal-stat}) and the
$E(3)$ Lie-Poisson tops governed by equations (\ref{heavy-top-Pi}) are
{\bfi Legendre duals} to each other under the map,
\begin{equation} \label{liqxtals-heavy-top-map1}
\frac{d}{dz} = -2\,\frac{d}{dt}
\,,\quad
h(\boldsymbol\gamma_3,\boldsymbol\sigma)
= 
\boldsymbol{\Pi\cdot\gamma}_3
+
\boldsymbol{\Gamma\cdot\sigma}
- 
H(\boldsymbol\Pi,\boldsymbol\Gamma) 
\,,
\end{equation}
so that
\begin{equation} \label{liqxtals-heavy-top-map2}
\frac{ \partial h }{ \partial
\boldsymbol\gamma_3} = 
\boldsymbol\Pi
\,,\quad
\frac{\partial h}{\partial\boldsymbol\sigma}
=\boldsymbol\Gamma  
\,,\quad 
\boldsymbol\gamma_3 = \frac{\partial H}{\partial\boldsymbol\Pi}
\,,\quad
\boldsymbol\sigma= 
\frac{\partial H}{\partial\boldsymbol\Gamma}
\,.
\end{equation}
Hence, we arrive at the result:
\begin{quote}
{\it The class of static one dimensional flows of a perfect liquid crystal is
Legendre-isomorphic to the class of $E(3)$ Lie-Poisson tops.}
\end{quote}
These tops conserve $\boldsymbol{\Pi\cdot\Gamma}$ and
$|\boldsymbol\Gamma|^2$, but in general they are not integrable.

\paragraph{Spatially homogeneous, time-dependent perfect liquid crystal flows.}
Spatially homogeneous solutions of equations (\ref{m-eqn-liqxtal}) -
(\ref{gamma-eqn-liqxtal}) obey the dynamical equations,
\begin{eqnarray}
\frac{1}{2}\,\frac{d\boldsymbol\sigma}{dt}  
 &=& 
\frac{\delta h}{\delta \boldsymbol\sigma}
\times\boldsymbol\sigma
+
\frac{ \delta h }{ \delta \boldsymbol\gamma_m}
\times\boldsymbol\gamma_m
\,,
\label{sigma-eqn-liqxtal-tee}\\
\frac{1}{2}\,\frac{d\boldsymbol\gamma_m }{dt}
&=&
\frac{\delta h}{\delta\boldsymbol\sigma}
\times\boldsymbol\gamma_m
\,.
\label{gamma-eqn-liqxtal-tee}
\end{eqnarray}
For a single value of the spatial index, say $m=3$, these are nothing more than
the $E(3)$ top equations (\ref{heavy-top-Pi}) with time re-parameterized by
$t\to-2t$. Hence, in this case, the Hamiltonian internal dynamics of a
spatially homogeneous liquid crystal is essentially identical to the $E(3)$
Lie-Poisson dynamics of a top. In the multi-component case, one sums
over $m=1,2,3$, in the second term of equation (\ref{sigma-eqn-liqxtal-tee})
and, thus, the resulting dynamics is more complex than the simple top.
Hence, we have:
\begin{quote}
{\it The class of spatially homogeneous, time-dependent perfect liquid crystal
flows is isomorphic to the generalization (\ref{sigma-eqn-liqxtal-tee}) -
(\ref{gamma-eqn-liqxtal-tee}) of the $E(3)$ Lie-Poisson tops.}
\end{quote}
%

\paragraph{Action principle \#4 -- Clebsch representation.}
Another representation of Hamilton's principle for liquid crystals in the
Eulerian fluid description may be found by constraining the Eulerian action
$S$ in equation (\ref{liqxtal-action-nu/gamma-Eul}) by using Lagrange
multipliers to enforce the kinematic equations (\ref{D - gamma eqns})
and (\ref{D - gamma eqns1}). The constrained action is, thus,
\begin{eqnarray}\label{liqxtal-action-Clebsch}
\mathcal{S} &=& \int dt\int d^3x\
\ell(\mathbf{u}, D, \boldsymbol\nu, \boldsymbol\gamma)
+\
\phi\Big(\frac{\partial D}{\partial t}
 + \,\frac{\partial Du_j}{\partial x_j}
\Big)
\nonumber\\
&&+\
\boldsymbol\beta_m\cdot\Big(\frac{\partial \boldsymbol\gamma_m }{\partial t}
-
\frac{\partial\boldsymbol\nu}{\partial x_m}
+
2\,\boldsymbol\gamma_m\times\boldsymbol\nu
+
u_k\frac{\partial\boldsymbol\gamma_m}{\partial x_k}
+
\boldsymbol\gamma_k\frac{\partial u_k}{\partial x_m}
\Big)
\,,
\end{eqnarray} 
with Lagrange multipliers $\phi$ and $\boldsymbol\beta_m$.
Stationarity of $S$ under variations in $u_k$ and $\boldsymbol\nu$ implies
the relations
\begin{eqnarray}\label{liqxtal-Clebsch-rep-mom}
\delta u_k: 
&&
\frac{\delta \ell }{\delta u_k}
\,-\,
D\,\frac{\partial \phi }{\partial x_k}
\,+\,
\boldsymbol\beta_m\cdot
\frac{\partial\boldsymbol\gamma_m}{\partial x_k}
\,-\,
\frac{\partial}{\partial x_m}
\Big(
\boldsymbol\gamma_k\cdot\boldsymbol\beta_m
\Big)
=
0
\,,
\\
\delta\boldsymbol\nu:
&&
\frac{\delta \ell }{\delta \boldsymbol\nu}
\,+\,
\frac{\partial\boldsymbol\beta_m}{\partial x_m}
\,+\,
2\,\boldsymbol\beta_m\times\boldsymbol\gamma_m
=
0
\,.\label{liqxtal-Clebsch-rep-sigma}
\end{eqnarray} 
These are the {\bfi Clebsch relations} for the momentum of the motion
$m_k=\delta \ell/\delta u_k$ and the director angular momentum of the
micromotion $\boldsymbol\sigma = \delta \ell /\delta \boldsymbol\nu$.
Stationary variations of $S$ in $D$ and $\boldsymbol\gamma_m$ give,
respectively, the dynamical equations for the {\bfi canonical momenta},
$\pi_D=\phi$ and
$\boldsymbol\pi_{\gamma_m}=\boldsymbol\beta_m$, as
\begin{eqnarray}\label{liqxtal-Clebsch-dyn-phi}
\delta D: 
&&
\frac{\partial \phi }{\partial t}
+\,
u_k\,\frac{\partial \phi }{\partial x_k}
-
\frac{\delta\ell}{\delta D}
=
0
\,,
\\
\delta\boldsymbol\gamma_m:
&&
\frac{\partial \boldsymbol\beta_m }{\partial t}
+
\frac{\partial}{\partial x_k}
\Big(
\boldsymbol\beta_m u_k
\Big)
-
\boldsymbol\beta_k\frac{\partial u_m}{\partial x_k}
-
2\,\boldsymbol\nu\times\boldsymbol\beta_m
-
\frac{\delta \ell }{\delta \boldsymbol\gamma_m}
=
0
\,.\qquad\label{liqxtal-Clebsch-dyn-beta}
\end{eqnarray} 
Finally, variations in the Lagrange multipliers $\phi$ and
$\boldsymbol\beta_m$ imply the kinematic equations (\ref{D - gamma eqns}) and
(\ref{D - gamma eqns1}), respectively. These two kinematic equations combine
with the four variational equations (\ref{liqxtal-Clebsch-rep-mom}) through
(\ref{liqxtal-Clebsch-dyn-beta}) to recover the Eulerian motion and
micromotion equations after a calculation using the Clebsch relations, the
Kelvin-Noether form of the motion equation, and the dynamical equations for
the Clebsch potentials. 

At the end of in Section \ref{Ham-Princ-Lag-Red}, we shall systematize this
type of calculation and, thus, clarify its meaning as a Poisson map. For now,
we simply remark that the evolutionary Clebsch relations are Hamilton's
canonical equations for the Hamiltonian obtained from the constrained action in
(\ref{liqxtal-action-Clebsch}) by the usual Legendre transformation. Perhaps
not unexpectedly, this Hamiltonian agrees exactly with that in equation
(\ref{liqxtal-legendre-xform}) obtained from the Legendre transformation in
$\mathbf{u}$ and $\boldsymbol\nu$ alone.

As we shall discuss more generally in the next section, the Clebsch
representations for the momentum and director angular momentum provide a
Poisson map from the canonical Poisson bracket in the Clebsch variables to the
Lie-Poisson bracket for the Hamiltonian matrix with generalized two-cocycle
found in equation (\ref{Ham-matrix-diff-liqxtal-compon}). 

Historically, the Hamiltonian approach has been very fruitful in modeling the
hydrodynamics of complex fluids and quantum liquids, including superfluids,
going back to the seminal work of {Khalatnikov and Lebedev [1978]},
{Khalatnikov and Lebedev [1980]} and {Dzyaloshinskii and Volovick [1980]}. The
Clebsch approach has provided a series of physical examples of
Lie-Poisson brackets: for superfluids in {Holm and Kupershmidt [1982]}; 
superconductors in {Holm and Kupershmidt [1983a]}; 
Yang-Mills plasmas (chromohydrodynamics) in 
{Gibbons, Holm and Kupershmidt [1982]}, and 
{Gibbons, Holm and Kupershmidt [1983]};
magnetohydrodynamics, multifluid  plasmas, and elasticity, in 
{Holm and Kupershmidt [1983b]};
Yang-Mills magnetohydrodynamics in {Holm and Kupershmidt [1984]}; and
its relation to superfluid plasmas in {Holm and Kupershmidt [1987]} and
spin-glasses in {Holm and Kupershmidt [1988]}. Many, but not all, of these
Lie-Poisson brackets fit into the present Euler-Poincar\'e framework for
PCFs. 

The Euler-Poincar\'e framework also accomodates many of the various types of
Poisson brackets (such as ``rigid body fluids'') studied over the years by
Grmela, Edwards, Beris, and others, as summarized in {Beris and Edwards
[1994]}. For liquid crystals, these authors develop a bracket description both
for Ericksen-Leslie equations and the Doi-Edwards theory based on the
conformation tensor
$\mathbf{C}$, which is related to the director theory by
$\mathbf{C}=\mathbf{n}\otimes\mathbf{n}$. The extension of the present results
to this case may be accomplished, e.g., by following the Clebsch approach of
{Holm and Kupershmidt [1983b]} who treated the corresponding case of
Lie-Poisson brackets for nonlinear elasticity. The treatment in 
{Beris and Edwards [1994]} ignores the geometrical content of the Lie-Poisson
formulation in preference for its tensor properties alone.

\subsection{Summary for perfect liquid crystals}

We now recapitulate the steps in the procedure we have followed in 
deriving the Euler-Lagrange-Poincar\'e-Clebsch equations and the
Lie-Poisson Hamiltonian formulations of the dynamics of perfect liquid
crystals.

\vspace{-.25in}

{\singlespace
\begin{enumerate}
\item Define the order parameter group and its coset space.
\item Write Hamilton's principle in the Lagrangian fluid
description.
\item Make 2 stages of reduction: 
\begin{description} 
\item [$1^{st},$] to introduce the
reduced set of variables in the
Lagrangian fluid description; and 
\item [$2^{nd},$] to pass to the
Eulerian fluid description.
\end{description}
\item Legendre transform to obtain the Hamiltonian formulation. 
\end{enumerate}
}

The alternative Clebsch procedure starts directly with an action for
Hamilton's principle that is defined in the Eulerian fluid description and
constrained by the Eulerian kinematic equations. Its Hamiltonian formulation
is canonical and passes to a Lie-Poisson formulation via the Poisson map that
is defined by the Clebsch representations of the momentum and internal angular
momentum in equations (\ref{liqxtal-Clebsch-rep-mom}) and
(\ref{liqxtal-Clebsch-rep-sigma}), respectively. 
 
Many physical extensions of these results for perfect liquid crystals are
available, e.g., to include MHD, compressibility, anisotropic dielectric and
diamagnetic effects, linear wave excitation properties, etc. However, we wish
to spend the most of the rest of this paper setting the formulations we have
established here for perfect liquid crystal dynamics into the geometrical
framework of Lagrangian reduction by stages developed in {Cendra, Marsden and
Ratiu [1999]}. This geometrical setting will take advantage of the unifying
interpretation of order parameters as coset spaces of broken symmetry groups.
(The coset interpretation of order parameters for liquid crystals, superfluids
and spin glasses is reviewed, e.g., in  {Mermin [1979]}.) The present
formulations are  geometrical variants of the Ericksen-Leslie equations for
liquid crystal dynamics that illuminate some of their mathematical
features from the viewpoint of Lagrangian reduction.

\section{Action principles and Lagrangian reduction}
\label{Ham-Princ-Lag-Red}

As we have seen, the passage to reduced variables $\boldsymbol\nu$ and
$\boldsymbol\gamma_m$ for liquid crystals restricts the variables $\mathbf{n},
\boldsymbol{\dot\mathbf{n}}, \nabla\mathbf{n}$ to the coset space
$SO(3)/O(2)$ of rotations that properly affect the director $\mathbf{n}$ and
imposes invariance of the theory under the reflections
$\mathbf{n}\to-\mathbf{n}$. The reduced variables transform properly under
$SO(3)$, because the $O(2)$ isotropy subgroup of $\mathbf{n}$ has been
{\sl factored out} of them. Thus, we ``mod out'' or ``reduce'' the
symmetry-associated degrees of freedom by passing to variables that transform
properly under rotations in $SO(3)$ and admit the
$Z_2$ reflections $\mathbf{n}\to-\mathbf{n}$. The removal of degrees of
freedom associated with symmetries is the essential idea behind
Marsden-Weinstein group reduction in {Marsden and Weinstein [1974]}. 
{\bfi Marsden-Weinstein reduction} first appeared in
the Hamiltonian setting. However, this sort of reduction by symmetry groups
has been recently extended to the Lagrangian setting, see 
{Cendra, Marsden and Ratiu [1999]} and {Marsden, Ratiu and Scheurle [1999]}.
The remainder of the paper applies the mathematical framework of 
{\bfi Lagrangian reduction by stages} due to {Cendra, Marsden and Ratiu [1999]}
to express some of the properties of PCF dynamics in the Eulerian description
for an arbitrary order parameter group.

A synthesis of the nonlinear dynamics for the motion and micromotion of
various perfect complex fluid models is possible, due to their common
mathematical basis. The mathematical basis common to all ideal fluid motion --
both classical and complex fluids -- is {\bfi Hamilton's principle}, see, e.g.,
{Serrin [1959]},
\begin{equation}\label{Ham-princ}
\delta S=\delta\int L\,dt=0
\,.
\end{equation}
In the Lagrangian (or material) representation for fluids, the motion is
described by the Euler-Lagrange equations for this action principle.

In the Eulerian (or spatial) representation for fluids, the Euler-Lagrange
equations for the dynamics are replaced by the Euler-Poincar\'e
equation. The distinction between Euler-Lagrange equations and
Euler-Poincar\'e equations is exemplified by the distinction between rigid
body motion expressed in terms of the Euler angles and their time derivatives
on the tangent space $TSO(3)$ of the Lie group of proper rotations $SO(3)$, and
that same motion expressed in body angular velocity variables in its Lie
algebra $so(3)$. {Poincar\'e [1901]} was the first to write the latter
equations on an arbitrary Lie algebra; hence, the name {\bfi Euler-Poincar\'e
equations}. 

Euler-Poincar\'e equations may be understood and derived via the theory of
Lagrangian reduction as in 
{Cendra et al. [1999]}, and {Cendra, Marsden and Ratiu [1999]}. 
Euler-Poincar\'e equations arise when Euler-Lagrange equations and their
corresponding Hamilton principles are mapped from a velocity phase space
$TQ$ to the quotient space $TQ/G$ (a vector bundle) by a Lie-group action of a
symmetry group $G$ on the configuration space $Q$. If $L$ is a $G$-invariant
Lagrangian on $TQ$, this process maps it to a reduced Lagrangian and a
corresponding reduced variational principle for the Euler-Poincar\'e dynamics
on $TQ/G$ in which the variations are constrained. See 
{Weinstein [1996]} and {Cendra, Marsden and Ratiu [1999]}, for expositions
of the mathematical framework that underlies Lagrangian reduction by stages
and {Holm, Marsden and Ratiu [1998]} for a discussion of Euler-Poincar\'e
equations and their many applications in classical ideal fluid dynamics from
the viewpoint of the present paper.  See {Marsden, Ratiu and Scheurle [1999]}
for additional insight and recent results in Lagrangian reduction.

The order parameters of PCFs are material variables. The
Lagrangian in Hamilton's principle (\ref{Ham-princ}) for PCFs is the map,
\begin{equation}\label{micrpol-Lag}
L:TG\times V^{\ast}\times T\mathcal{O}
\longmapsto\mathbb{R}
\,.
\end{equation}
That is, the velocity phase space for the PCF Lagrangian $L$ in
material variables is the Cartesian product of three spaces:
\begin{enumerate}
\item [$TG$,]  the tangent space of the Lie group $G$ of fluid motions (the
diffeomorphisms that take the fluid parcels from their reference configuration
to their current positions in the domain of flow), 
\item [$V^\ast$,]  the vector space of advected quantities carried with the
fluid motion, and 
\item [$T\mathcal{O}$,]  the tangent space of the Lie group $\mathcal{O}$ of
fluid micromotions ($\mathcal{O}$ is the order parameter Lie group). 
\end{enumerate}
The advected quantities in $V^\ast$ include the volume element or mass density
and whatever else is carried along with the fluid parcels, such as the
magnetic field intensity in the case of magnetohydrodynamics. The new
feature of PCFs relative to the simple fluids with
advected parameters treated in {Holm, Marsden and Ratiu [1998]} is the
dependence of their Lagrangian on $T\mathcal{O}$. The order parameter coset
space at each material point is acted upon by the order parameter Lie group.
(We choose the convention of group action from the right.) Since the order
parameter is a material property, the {\bfi diffeomorphism group $G$ also acts
on the order parameter group}, as $\mathcal{O}\times G\to\mathcal{O}$, denoting
action from the right.

In this Section, we shall use Hamilton's principle (\ref{Ham-princ}) with
Lagrangian (\ref{micrpol-Lag}) to obtain the dynamical equations for the
motion and micromotion of PCFs whose order parameters are
defined as coset spaces of Lie groups. In doing so, we shall begin by
{\it assuming} this Lagrangian is invariant under the right action of the
order parameter Lie group $\mathcal{O}$ on its tangent space $T\mathcal{O}$.
(This right action on the space of internal variables leaves the other
components of the configuration space $TG$ and $V^\ast$ fixed.) We shall 
assume this Lagrangian is also invariant under the right action of the
diffeomorphisms $G$, which relabel the fluid parcels. (This action of $G$ {\it
does indeed} affect the material variables defined on $T\mathcal{O}$ and
$V^\ast$.) Under these symmetry assumptions we shall perform the following
two group reductions 
\[
\big(TG\times V^{\ast}\times (T\mathcal{O}/\mathcal{O})\big)/G
\simeq
\mathfrak{g}\times ( V^{\ast} \times \mathfrak{o})g^{-1}(t)
\,,
\]
with respect to the right actions of first $\mathcal{O}$ and then $G$, 
by applying group reduction to the velocity phase space of this
Lagrangian in two stages,
\begin{eqnarray}\label{1st-micrpol-Lag-red}
\hbox{1st stage,}\quad
(TG\times V^{\ast}\times T\mathcal{O})/\mathcal{O}
&\simeq&
TG \times V^{\ast} \times \mathfrak{o}\,,
\\
\hbox{2nd stage,}\quad
(TG \times V^{\ast} \times \mathfrak{o})/G
&\simeq&
\mathfrak{g}\times ( V^{\ast} \times \mathfrak{o})g^{-1}(t)
\,.\label{2nd-micrpol-Lag-red}
\end{eqnarray}
Here we denote isomorphisms as, e.g., $\mathfrak{o}\simeq T\mathcal{O}/O$
and $\mathfrak{g}\simeq TG/G$, where the Lie algebras $\mathfrak{o}$ and
$\mathfrak{g}$ correspond, respectively, to the Lie groups $\mathcal{O}$ and
$G$. The first stage is Lagrangian reduction by the right action of 
$\mathcal{O}$, the order parameter group.%
%
\footnote{Variants exist. In particular, for liquid crystals, only a part of
the Lie algebra $\mathfrak{o}$ is required; namely, $so(3)/O(2)$, the part of
the Lie algebra $so(3)$ that is invariant under the $O(2)$ isotropy group of
the director, $\mathbf{n}$.} The second stage is Lagrangian reduction of
the first result by the right action of the diffeomorphisms $G$ in the first
factor and by composition of functions in the second factor. Because of the
assumed invariances of our Lagrangian, these two stages of reduction of the
velocity phase spaces will each yield a reduced Lagrangian and a corresponding
reduced variational principle for the dynamics. The group actions at each
stage are assumed to be free and proper, so the reduced spaces will be local
principle fiber bundles.%
%
\footnote{A natural flat connection appears on this bundle, but this bundle
picture should be made intrinsic and global, while including defect dynamics.
A strategy for obtaining equations for the defect dynamics will
be discussed in the next Section. However, the global bundle picture is
for future work.}
The mathematical formulation of the process of Lagrangian reduction
by stages and the introduction of various connections on the
{\bfi Lagrange-Poincar\'e bundles} that arise in Lagrangian reduction are
discussed in {Cendra, Marsden and Ratiu [1999]}. These
Lagrange-Poincar\'e bundles are special cases of {\bfi Lie algebroids}. See
{Weinstein [1996]} for a fundamental description of the
relation between Lagrangian mechanics and Lie algebroids.

\subsection{Lagrangian reduction by stages}

We are dealing with a Lagrangian defined by the map
\[
L(g,\dot{g},a_0,\dot\chi,d\chi):TG \times V^{\ast} \times T\mathcal{O}
\longmapsto\mathbb{R}
\,,
\]
where $G$ is the diffeomorphism group that acts on both the vector space
$V^\ast$ of advected material quantities and the order parameter group
$\mathcal{O}$. We assume that $L$ has the following invariance properties,
\begin{equation}\label{micrpol-Lag-invar}
L(g,\dot{g},a_0,\dot\chi,d\chi)
=
L(g,\dot{g},a_0,\dot\chi\psi,d\chi\psi)
=
L(gh,\dot{g}h,a_0h,\dot\chi\psi h,d\chi\psi h)
\,,
\end{equation}
for all $\psi\in\mathcal{O}$ and $h\in G$. In particular, we shall choose
$\psi={\chi^{-1}}(t)$ in the first stage and $h=g^{-1}(t)$ in the second
stage of the reduction, so that
\begin{equation}\label{micrpol-red-Lag1}
L(g,\dot{g},a_0,\dot\chi,d\chi)
=
L(g,\dot{g},a_0,\dot\chi{\chi^{-1}},d\chi{\chi^{-1}})
\,,
\end{equation}
after the first stage of reduction, and
\begin{equation}\label{micrpol-red-Lag2}
L(g,\dot{g},a_0,\dot\chi,d\chi)
=
L(e,\dot{g}g^{-1},a,\dot\chi{\chi^{-1}} g^{-1},d\chi{\chi^{-1}} g^{-1})
\equiv
l(\xi,a,\nu,\gamma)
\,,
\end{equation}
after the second stage, with $\xi \equiv \dot{g}g^{-1}$, $a\equiv a_0g^{-1}$,
$\nu \equiv (\dot\chi{\chi^{-1}})g^{-1}$ and
$\gamma\cdot dx \equiv (d\chi{\chi^{-1}})g^{-1}$. 
After the first stage of reduction, the reduced action principle yields the
Lagrange-Poincar\'e equations, and after the second stage we shall obtain the
Euler-Poincar\'e equations for a perfect complex fluid with an arbitrary
order parameter group. 

\subsubsection{Lagrange-Poincar\'e equations}

The first stage
\[
TG\times V^{\ast}\times T\mathcal{O}
\longmapsto TG \times V^{\ast} \times \mathfrak{o}
\,,
\]
of the two-stage symmetry reduction in (\ref{1st-micrpol-Lag-red}) -
(\ref{2nd-micrpol-Lag-red}) affects only the internal variables and passes
from coordinates on the order parameter Lie group, $\mathcal{O}$, to
coordinates on its Lie algebra, $\mathfrak{o}$, obtained from the tangent
vectors of the order parameter Lie group at the identity by the isomorphism
$\mathfrak{o}\simeq T\mathcal{O}/O$. The results at this first stage consist
of Euler-Lagrange equations for the fluid motion, coupled through additional
components of the stress tensor to equations of a type called
Lagrange-Poincar\'e equations in {Marsden and Scheurle [1995]}, {Cendra,
Marsden and Ratiu [1999]}. In our case, these Lagrange-Poincar\'e equations
describe the micromotion in the Lagrangian (or material) fluid description. 

The first stage of reduction results in the {\bfi Lagrange-Poincar\'e action
principle}, 
\begin{equation}\label{LP-action}
\delta\!\!\int\!\! dt\!\!\int\!\! d^3X\
 L(\dot{x}, J, \nu, \gamma) 
=
0
\,,
\end{equation}
written in the material representation and denoted as follows,
\begin{description}
\item $L(\dot{x},J,\nu,\gamma)$ is the reduced Lagrangian on
$TG\times V^{\ast} \times \mathfrak{o}$,
\item $J(X,t)={\rm det}(\partial x/\partial X)\in V^\ast$ is the
volume element,
\item $\nu(X,t) = \dot{\chi} {\chi^{-1}}(X,t)\in\mathfrak{o}$ is the
material angular frequency and
\item $\gamma\cdot dx =  d\chi {\chi^{-1}}(X,t)\in
\mathfrak{o}$ with components denoted as $\gamma_{m}$ given by
\begin{eqnarray}\label{gamma-mat-def}
\gamma_m\, dx_m (X,t) 
&=& d\chi {\chi^{-1}}(X,t)
\,,\nonumber\\
 &=&
\gamma_m  
\frac{\partial x_m}{\partial X_A} dX_A
=
\gamma^{mat}_A(X,t)  dX_A
\,,
\end{eqnarray}
where $\gamma\cdot dx$ is the Cosserat strain one-form introduced in
{Cosserat [1909]} 
and superposed ``dot" $(\,\,)\boldsymbol{\dot{\,}}$ denotes time
derivative at {\it fixed material position} $X$.  
\end{description}
These material quantities satisfy auxiliary {\bfi kinematic equations},
obtained by differentiating their definitions,
\begin{eqnarray}\label{LPvol-advection}
(J^{-1}d^{\,3\,}x)\boldsymbol{\dot{\,}} 
&=&
(d^{\,3\,}X)\boldsymbol{\dot{\,}}
=
 0\,,
\\
(\gamma\cdot dx)\boldsymbol{\dot{\,}}
&=&
(d\chi {\chi^{-1}})\boldsymbol{\dot{\,}}
=
d\nu
+
{\rm ad}_{\nu}(\gamma\cdot dx)
\,,\label{LPgamma-strain-rate-kinematics}
\end{eqnarray}
The material angular frequency $\nu=\dot{\chi} {\chi^{-1}}$ and the material
Cosserat strain one-form $\gamma\cdot dx =  d\chi {\chi^{-1}}$ take their
values in the right-invariant Lie algebra $\mathfrak{o}$ of the order
parameter Lie group
$\mathcal{O}$. The {\bfi ad-operation} appearing in equation
(\ref{LPgamma-strain-rate-kinematics}) denotes multiplication, or commutator,
in the Lie algebra $\mathfrak{o}$.

The dynamical {\bfi Lagrange-Poincar\'e equations} determine the complex
fluid's motion with fluid trajectory $\phi_t(X)=x(X,t)$ with $\phi_t\in G$ and
micromotion $\chi(X,t)\in\mathcal{O}$ in the material fluid description. These
equations take the following forms,
\begin{eqnarray}\label{LPmotion-eqn}
J^{-1}\Big(\frac{\partial L}{\partial \dot{x}_p}\Big)^{\boldsymbol{\dot{\,}}}
+\,\frac{\partial}{\partial x_p}\frac{\partial L}{\partial J}
-\frac{\partial}{\partial x_m}
\bigg\langle J^{-1}
\frac{\partial L}{\partial \gamma_{m}}\,,\gamma_{p}^{}\bigg\rangle
&=&0
\,,\\
\label{LPmicromotion-eqn}
\,\Big(\frac{\partial L}{\partial \nu}\Big)^{\boldsymbol{\dot{\,}}}
- {\rm ad}^*_{\nu}\frac{\partial L}{\partial \nu}
+\, J \frac{\partial}{\partial x_m}
\bigg(J^{-1}\frac{\partial L}{\partial \gamma_{m}}\bigg)
-\,{\rm ad}^*_{\gamma_{m}^{}}\frac{\partial L}{\partial \gamma_{m}}
&=&0
\,.
\end{eqnarray}
The {\bfi ad$^*$-operation} appearing in (\ref{LPmicromotion-eqn})
is defined in terms of the ad-operation and the symmetric pairing
$\langle\Dot,\Dot\rangle:\mathfrak{o}^*\times\mathfrak{o}\rightarrow\mathbb{R}$
between elements of the right Lie algebra $\mathfrak{o}$ and its dual
$\mathfrak{o}^*$ as, e.g.,
\begin{equation}\label{ad*-def}
-\,\bigg\langle{\rm ad}^*_{\nu^{}}
\frac{\partial L}{\partial \nu}\,,\Sigma\bigg\rangle
= \bigg\langle\frac{\partial L}{\partial \nu}\,,\,
{\rm ad}_{\nu^{}}\Sigma\bigg\rangle  
=
\bigg\langle
\frac{\partial L}{\partial \nu}\,,\,[\nu^{}\,,\Sigma\,]
\bigg\rangle\,.
\end{equation}
In a Lie algebra basis satisfying
$[e_\alpha,e_\beta] = t^{\,\iota}_{\alpha\beta}e_\iota$ and its dual
basis
$e^\kappa$ satisfying $\langle e^\kappa,
e_\iota\rangle = 
\delta^{\,\kappa}_\iota$, we may write this formula as
\begin{equation}\label{ad*-def-basis}
-\,\bigg\langle{\rm ad}^*_{\nu^{}}
\frac{\partial L}{\partial \nu}\,,\Sigma\bigg\rangle
= \bigg\langle\frac{\partial L}{\partial \nu}\,,\,
{\rm ad}_{\nu^{}}\Sigma\bigg\rangle  
= 
\frac{\partial L}{\partial \nu^{\,\kappa}} \, t^{\,\kappa}_{\alpha\beta}\,
\nu^{\,\alpha}\,\Sigma^{\,\beta}
\,.
\end{equation}
Thus, for the sign conventions we choose in (\ref{ad*-def}), the
ad$^*$-operation is defined as the {\bfi negative transpose} of the
ad-operation.

The dynamical equations (\ref{LPmotion-eqn}) and
(\ref{LPmicromotion-eqn}) follow from the Lagrange-Poincar\'e action principle
(\ref{LP-action}) for PCF dynamics in the material fluid
description. These Lagrange-Poincar\'e equations may be calculated directly,
as
\begin{eqnarray}\label{HP-perfect}
0&=&\delta\!\!\int\!\! dt\!\!\int\!\! d^3X\
 L(\dot{x}, J, \nu, \gamma)
\\
&=&
\!\!\int\!\! dt\!\!\int\!\!  d^3X\
\bigg[
\frac{\partial L}{\partial \dot{x}_p} {\delta \dot{x}_p}
+
\frac{\partial L}{\partial J}{\delta J}
+
\bigg\langle\frac{\partial L}{\partial \nu}\,, {\delta \nu}\bigg\rangle
+
\bigg\langle\frac{\partial L}{\partial \gamma_m}\,,
 {\delta \gamma_m}\bigg\rangle
\bigg]
\nonumber\\
&=& \!\!\int\!\! dt\!\!\int\!\! d^3X\
\bigg\{\delta x_p \bigg[
-\,\Big(\,\frac{\partial L}{\partial \dot{x}_p}
\Big)^{\boldsymbol{\dot{\,}}}
-\,J\frac{\partial}{\partial x_p}\frac{\partial L}{\partial J}
+J\frac{\partial}{\partial x_m}
\bigg\langle J^{-1}
\frac{\partial L}{\partial \gamma_{m}}\,,\gamma_{p}^{}\bigg\rangle
\bigg]
\nonumber\\
&&
\ +\
\bigg\langle \bigg[
-\,\Big(\frac{\partial L}{\partial \nu}
\Big)^{\boldsymbol{\dot{\,}}}
+ {\rm ad}^*_{\nu}\frac{\partial L}{\partial \nu}
-\, J \frac{\partial}{\partial x_m}\bigg(J^{-1}
\frac{\partial L}{\partial \gamma_{m}}\bigg)
+\,{\rm ad}^*_{\gamma_{m}^{}}\frac{\partial L}{\partial \gamma_{m}}
\bigg]\,,
\Sigma\bigg\rangle
\bigg\}\nonumber\\
&+&
\!\!\int\!\! dt\oint\! d^2S\ \
\hat{n}_A\
\frac{\partial X_A}{\partial x_m}\
\bigg(
\bigg\langle
\frac{\partial L}{\partial \gamma_{m}}
\,,
\Sigma
\bigg\rangle
-
\bigg\langle 
\frac{\partial L}{\partial \gamma_{m}}\,,\gamma_{p}^{}\bigg\rangle
\delta x_p
+ J\frac{\partial L}{\partial J}\,\delta x_m
\bigg).
\nonumber
\end{eqnarray}
Here we define $\Sigma\equiv\delta\chi\, {\chi^{-1}}$, in terms of which we
calculate,
\begin{eqnarray}\label{nu-gamma-Sigma}
\delta\nu
&=&
\dot{\Sigma}
-
\big[ \nu, \Sigma \big]
=
\dot{\Sigma}
-
{\rm ad}_\nu\, \Sigma 
\,,\\
\delta\gamma^{\,mat}_A
&=&
\frac{\partial \Sigma}{\partial X_A} 
-
\big[ \gamma^{\,mat}_A, \Sigma \big]
=
\frac{\partial \Sigma}{\partial X_A} 
-
{\rm ad}_{\gamma^{\,mat}_A} \, \Sigma 
\,.
\end{eqnarray} 
Since $\gamma_m = \gamma_A^{\,mat}(\partial X_A/\partial x_m)$, this means that
\begin{eqnarray}\label{gamma-Sigma}
\delta\gamma_m
&=&
-\,
\gamma_p
\frac{\partial X_B}{\partial x_m} 
\frac{\partial }{\partial X_B} \delta x_p
+
\frac{\partial X_A}{\partial x_m} 
\,
\delta\gamma^{\,mat}_A
\nonumber\\
&=&
-\,
\gamma_p
\frac{\partial X_B}{\partial x_m} 
\frac{\partial }{\partial X_B} \delta x_p
+
\frac{\partial X_A}{\partial x_m} 
\frac{\partial \Sigma}{\partial X_A}
-
{\rm ad}_{\gamma_m} \, \Sigma 
\,.
\end{eqnarray} 
Thus, the variations in $\gamma_m$ couple the two Lagrange-Poincar\'e
equations. We also drop endpoint terms that arise from integrating by parts in
time, upon taking $\delta{x}_p$ and $\Sigma$ to vanish at these endpoints. The
natural boundary conditions
\begin{equation}\label{nu-gamma-bc-interface}
\frac{\partial \mathcal{L}}{\partial J} = 0
\,\quad\hbox{and}\quad
\hat{n}_m\,
\frac{{\partial \mathcal{L}}}{\partial \boldsymbol\gamma_{m}}
=0\,,
\end{equation} 
ensure that the fluid pressure and the normal stress are continuous across a
fluid interface.

\subsubsection{Euler-Poincar\'e equations}

The passage next from the Lagrangian fluid description of continuum mechanics
to the Eulerian fluid description will yield the Euler-Poincar\'e
equations. We obtain these equations by applying to Hamilton's principle the
second stage,
\[
TG \times V^{\ast} \times \mathfrak{o}
\longmapsto \mathfrak{g}\times 
(V^{\ast} \times \mathfrak{o})g^{-1}(t)
\,,
\]
of the two-stage Lagrangian reduction in (\ref{1st-micrpol-Lag-red}) -
(\ref{2nd-micrpol-Lag-red}). This second stage of reduction results in the
{\bfi Euler-Poincar\'e action principle},
\begin{equation}\label{EP-var-princ}
\delta\!\int\!{l}(\xi,a,\nu,\gamma)\,dt=0
\,,
\end{equation}
with constrained variations 
\begin{eqnarray} 
\delta \xi &=& \frac{\partial\eta}{\partial t} 
+ \big[\,\xi , \eta \,\big], 
\quad
\delta a =  -a\,\eta ,
\nonumber\\
\delta\nu
&=& \frac{\partial\Sigma}{\partial t}
+ \xi\cdot\nabla\Sigma - {\rm ad}_{\nu}\Sigma - \nu\,\eta\,,
\label{variationsright1}
\\
\delta\,(\gamma\cdot dx)
&=& d\,\Sigma 
+ {\rm ad}_{\Sigma\,}(\gamma\cdot dx) 
- (\gamma\cdot dx)\,\eta\,,
\nonumber
\end{eqnarray}
where $\eta(t)= \delta g(t) g(t)^{-1} \in \mathfrak g$ 
and $\Sigma(t)= \delta \chi(t) \chi(t)^{-1} \in \mathfrak o$
both vanish at the endpoints.

The Euler-Poincar\'e action principle produces the following equations defined 
on $\mathfrak g \times (V^{\ast} \times \mathfrak{o})g^{-1}(t)$ for the
motion and micromotion, in which $\partial/\partial t$ denotes Eulerian time
derivative at {\it fixed spatial position} $x$,
\begin{eqnarray} \label{LP-eulerpoincare-motion}
\frac{\partial}{\partial t} \frac{\delta l}{\delta \xi}
&=& 
- \, \mbox {\rm ad}_{\xi}^{\ast} \frac{ \delta l }{ \delta \xi}
+ \frac{\delta l }{ \delta a}\diamond a
+ \frac{\delta l }{ \delta \nu}\diamond \nu
+ \frac{\delta l }{ \delta \gamma_m}\diamond \gamma_m
\,,
\\
\label{LP-eulerpoincare-micromotion}
\frac{\partial}{\partial t} \frac{\delta l}{\delta \nu} 
&=&-\,
{\rm div}\Big(\xi\frac{\delta l}{\delta \nu}
+\frac{\delta l}{\delta \gamma}\Big)
+ 
{\rm ad}^{\ast}_{\nu} \frac{ \delta l }{ \delta \nu}
+ {\rm ad}^{\ast}_{\gamma_m} \frac{ \delta l }{ \delta \gamma_m}\,.
\end{eqnarray}
These are the {\bfi Euler-Poincar\'e equations} for a perfect complex
fluid. In these equations, $l$ is the  reduced Lagrangian on
$\mathfrak{g}\times (V^\ast \times \mathfrak{o})g^{-1}(t)$. Also, 
\begin{equation}
{\rm ad}_{\xi}\,\eta 
\equiv 
[\,\xi\,,\,\eta\,]
\,,\quad\hbox{with}\quad
\xi\,,\eta\in\mathfrak{g}
\,,
\end{equation}
is the commutator in the Lie algebra of vector fields, $\mathfrak{g}$.
In addition, we define the two operations ad$^\ast_{\xi}$ and
$\diamond$ as
\begin{equation}
\Bigg \langle
\mbox {\rm ad}_{\xi}^{\ast}\,\frac{\delta l}{\delta \xi}\,,\,\eta
\Bigg \rangle
\
\equiv \
- \
\Bigg \langle
\frac{\delta l}{\delta \xi}\,,
\mbox {\rm ad}_{\xi}\,\eta \Bigg \rangle
\,,
\label{ad-star-def}
\end{equation}
and
\begin{equation}
\Bigg \langle \frac{\delta l }{ \delta a}\diamond a\,,\,
\eta \Bigg \rangle
\
\equiv \
-\ 
\Bigg \langle \frac{\delta l}{\delta a} \,,\, a\,\eta
\Bigg \rangle
\,.
\label{diamond-def}
\end{equation}
The concatenation $a\,\eta$ denotes the right Lie algebra action of
$\eta\in\mathfrak{g}$ on $a\in V^\ast$ (by Lie derivative).  
The pairing $\langle\boldsymbol\cdot\,,\,\boldsymbol\cdot\rangle$
now includes {\it spatial integration} and, thus, allows for integration by
parts. Similar definitions hold for 
$(\delta l /\delta \gamma_m\diamond \gamma_m)$
and $(\delta l/\delta \nu\diamond \nu )$. 

In components, these quantities are given by,
\begin{eqnarray}
\Bigg \langle \frac{\delta l }{ \delta \gamma_m}\diamond \gamma_m\,,\,
\eta \Bigg \rangle
=-\
\Bigg \langle \frac{\delta l }{ \delta \gamma_m}\,,\,
\gamma_m\,\eta \Bigg \rangle
&=&
\Bigg \langle 
\frac{\partial }{\partial x_m}
\bigg( 
\frac{\delta l}{\delta \gamma_m^\beta} 
\gamma_j^\beta
\bigg)
-
\frac{\delta l}{\delta \gamma_m^\beta} 
\frac{\partial \gamma_m^\beta}{\partial x_j}
\,,\, \eta_j
\Bigg \rangle
\,,
\label{diamond-def-gamma}
\nonumber\\
\Bigg \langle \frac{\delta l }{ \delta \nu}\diamond \nu\,,\,
\eta \Bigg \rangle
=-\
\Bigg \langle \frac{\delta l }{ \delta \nu}\,,\,
\nu\,\eta \Bigg \rangle
&=&
\Bigg \langle  
-
\frac{\delta l}{\delta \nu^\beta} 
\frac{\partial \nu^\beta}{\partial x_j}
\,,\, \eta_j
\Bigg \rangle
\,,
\label{diamond-def-nu}
\\
\Bigg \langle \frac{\delta l }{ \delta D}\diamond D\,,\,
\eta \Bigg \rangle
=-\
\Bigg \langle \frac{\delta l }{ \delta D}\,,\,
D\,\eta \Bigg \rangle
&=&
\Bigg \langle D \frac{\partial}{\partial x_j}
\frac{\delta l}{\delta D} \,,\, \eta_j
\Bigg \rangle
\,.
\label{diamond-def-D}
\nonumber
\end{eqnarray}
%
\paragraph{Remark.}
At this point, one might have also introduced $(3+1)$ {\bfi covariant
derivatives} acting on $\mathfrak{o}-$valued functions of space and time, as
\begin{equation}
D_m \equiv \frac{\partial}{\partial x_m} 
-
ad_{\gamma_m}\,,
\quad\hbox{and}\quad
D_t \equiv \frac{\partial}{\partial t} 
-
ad_{\nu}\,,
\label{covar-der}
\end{equation}
with associated curvature (or Yang-Mills magnetic field) given by
\begin{equation}
ad_{B_{ij}} \equiv \Big[D_i\,,\,D_j\Big]\,,
\label{B-field-def}
\end{equation}
whose components are expressed, cf. equation (\ref{curl-gamma}),
\begin{equation}
B_{ij}^{\,\alpha} 
= 
\gamma_{i,j}^{\,\alpha} - \gamma_{j,i}^{\,\alpha} 
+ 
t^{\,\alpha}_{\beta\kappa}\gamma_i^{\,\beta}\gamma_j^{\,\kappa}
\,,
\end{equation}
as in Holm and Kupershmidt [1988]. However, the operations ad, ad$^*$,
$\diamond$ and Lie derivative are sufficient for our present purposes.

\paragraph{Eulerian kinematic equations.}
By definition, an {\bfi Eulerian advected quantity} $a\in V^\ast
g(t)^{-1}$ satisfies 
$$\frac{\partial a}{\partial t} + a\,\xi = 0\,.$$ 
This advection relation may be written equivalently as 
$$\frac{\partial a}{\partial t} + \pounds_\xi\, a = 0\,,$$ 
where $\pounds_\xi$ is the Lie derivative with respect to
$\xi=\dot{g}(t)g^{-1}(t)$, the Eulerian fluid velocity, often denoted also as
${u}(x,t)$.

The Eulerian versions of the Lagrangian {\bfi kinematic equations}
(\ref{LPvol-advection}) and (\ref{LPgamma-strain-rate-kinematics}) are given
in terms of the Lie derivative by
\begin{eqnarray}\label{cont-eqn}
&&\Big(\frac{\partial }{\partial t} + \pounds_\xi\Big) 
\Big(D\,d^{\,3\,}x\Big)
=0
\,,
\\
\label{LP-micro-kinematics}
&&\Big(\frac{\partial }{\partial t} + \pounds_\xi\Big) 
(\gamma\cdot dx) 
= d\nu + {\rm ad}_{\nu}(\gamma\cdot dx)
\,.
\end{eqnarray}
In these equations, the quantity $D(x,t)=J^{-1}(X,t)g(t)^{-1}$ is the
Eulerian mass density and the quantities $\nu(x,t)=\nu(X,t)g(t)^{-1}$ and
$$
\gamma(x,t)\cdot dx
=
\Big(\gamma(X,t)\cdot\frac{\partial x}{\partial X}(X,t)\cdot
dX\Big)g(t)^{-1}
$$
are the Eulerian counterparts of the right-invariant material quantities
$\nu(X,t)$ and $\gamma(X,t)$ in equations
(\ref{LPgamma-strain-rate-kinematics}).

\paragraph{Remark.} We note that equation (\ref{LP-micro-kinematics}) implies
the {\bfi $\gamma-$circulation theorem} for PCFs, cf.
equation (\ref{gamma-circ-thm}), 
\begin{equation}\label{gamma-circ-thm-perfect}
\frac{d}{dt}\oint_{c(\xi)}\gamma\cdot dx
=
\oint_{c(\xi)}
{\rm ad}_{\nu}(\gamma\cdot dx)
\,.
\end{equation}
Thus, the circulation of $\gamma$ around a loop $c(\xi)$ moving with
the fluid is conserved when ${\rm ad}_{\nu}\,\gamma$ is a
gradient. Otherwise, the curl of this quantity generates circulation of
$\gamma$ around fluid loops.\bigskip

The Euler-Poincar\'e equations (\ref{LP-eulerpoincare-motion}) and
(\ref{LP-eulerpoincare-micromotion}) may be obtained directly from the
Euler-Poincar\'e action principle  (\ref{EP-var-princ}), as follows.

\paragraph{Euler-Poincar\'e action variations.}
We compute the variation of the action
(\ref{EP-var-princ}) in Eulerian variables at fixed time $t$
and spatial position $\mathbf{x}$ as,
\begin{eqnarray}\label{xi-a-nu-gamma-act-var-Eul}
\delta\mathcal{S}
\!\!\!&=&\!\!\!\!\!
\int \!\! dt\
\bigg[
\bigg\langle\frac{\delta l}{\delta \xi}\,,\delta \xi \bigg\rangle
+
\bigg\langle\frac{\delta l}{\delta a}\,,\delta a \bigg\rangle
+
\bigg\langle\frac{\delta l}{\delta \nu}
\,,\delta \nu \bigg\rangle
+
\bigg\langle\frac{\delta l}{\delta \gamma_m}
\,,\delta \gamma_m \bigg\rangle
\bigg]
\nonumber\\
\!\!\!&=&\!\!\!\!\!
\int \!\! dt\
\Bigg\{
\bigg\langle\!\!
-\,\frac{\partial}{\partial t}\frac{\delta l}{\delta \xi}
-\,{\rm ad}^*_\xi \frac{\delta l}{\delta \xi}
+ a\diamond \frac{\delta l}{\delta a}
+ \nu\diamond \frac{\delta l}{\delta \nu}
+ \gamma_m\diamond \frac{\delta l}{\delta \gamma_m}
\,,\,\eta
\bigg\rangle
\label{perfect-mot-Eul}
\\
&&
\hspace{-.4in}
+\
\bigg\langle
-\,
\frac{\partial}{\partial t}
\frac{\delta l}{\delta\nu}
\,-\,
\frac{\partial}{\partial x_m}
\Big(
\frac{{\delta l}}{\delta \nu}\,\xi_m
+
\frac{{\delta l}}{\delta \gamma_m}
\Big)
+ 
{\rm ad}^*_\nu \frac{{\delta l}}{\delta \nu}
+ 
{\rm ad}^*_{\gamma_m} \frac{{\delta l}}{\delta \gamma_m}
\,,\,{\Sigma}
\bigg\rangle
\Bigg\}
\label{perfect-micromot-Eul}
\nonumber\\
&&
\hspace{-.4in}
+
\int \!\! dt \!\!\int\!\! d^3x\
\Bigg\{
\,\frac{\partial}{\partial t}
\bigg[
\frac{\delta l}{\delta \xi_j}\,\eta_j 
+
\frac{\delta l}{\delta\nu^{\,\beta}}
\,\Sigma^{\,\beta}
\bigg]
-
d\,\bigg(
\eta\,\rfloor a\,,\,\frac{\delta l}{\delta a}
\bigg)
\nonumber\\
&&
+
\,\frac{\partial}{\partial x_m}
\bigg[\,
\eta_j \bigg(\xi_m\,\frac{\delta l}{\delta \xi_j}\, 
\,-\,
\frac{\delta l}{\delta \gamma_m^{\,\beta}}
\,\gamma_j^{\,\beta}
\bigg)
+
\bigg(
\frac{\delta l}{\delta\nu^{\,\beta}}\, \xi_m
+
\frac{\delta l}{\delta \gamma_m^{\,\beta}}\
\bigg)
\,\Sigma^{\,\beta}
\bigg]
\Bigg\}\,,
\nonumber
\end{eqnarray}
where we have used the variational expressions in 
(\ref{variationsright1}) and integrated by parts.
Here $\eta \, \rfloor a$ in the first boundary term denotes substitution of
the vector field $\eta$ into the tensor differential form $a$ and
$(\boldsymbol\cdot\,,\,\boldsymbol\cdot)$ denotes the natural pairing between
a tensor field and its dual. 

The dynamical Euler-Poincar\'e equations (\ref{LP-eulerpoincare-motion}) and
(\ref{LP-eulerpoincare-micromotion}) are thus obtained from the
Euler-Poincar\'e action principle  (\ref{EP-var-princ}), by requiring the
coefficients of the arbitrary variations $\eta$ and $\Sigma$ to
vanish in the variational formula (\ref{perfect-mot-Eul}). The remaining
terms in (\ref{perfect-mot-Eul}) yield {\bfi Noether's theorem} for this
system, which assigns a conservation law to each symmetry of the
Euler-Poincar\'e variational principle.

\paragraph{Momentum conservation.}
In momentum conservation form, the PCF motion equation
in the Eulerian fluid description (\ref{LP-eulerpoincare-motion}) becomes, for
{\it algebraic} dependence of the Lagrangian density $l$ on $(\xi, a, \nu,
\gamma_m)$,
\begin{equation}\label{perfect-momentum-eqn}
\hspace{-.1in}
\frac{\partial}{\partial t}\frac{\partial  l}{\partial \xi_j}
=
-\
\frac{\partial}{\partial x_m}
\bigg(\xi_m\,\frac{\partial l}{\partial \xi_j}\, 
+
l\, \delta_{mj}
-
\frac{\partial l}{\partial \gamma_m^{\,\beta}}
\, \gamma_j^{\,\beta}
\bigg)
\,+\
d\,\bigg(
\frac{\partial}{\partial x_j}
\,\rfloor\, a\,,\,\frac{\partial l}{\partial a}
\bigg)
\,.
\end{equation}
In this equation, expressed in Cartesian coordinates, there is an implied sum
over the various types of advected tensor quantities, $a$. This momentum
conservation law also arises from Noether's theorem, as a consequence of
the invariance of the variational principle (\ref{EP-var-princ}) under spatial
translations. In fact, the simplest derivation of this equation is obtained by
evaluating the variational formula (\ref{xi-a-nu-gamma-act-var-Eul}) on 
the equations of motion and using the translational symmetry of the
Lagrangian in Noether's theorem with $\eta_j=\partial/\partial x_j$. 
For an algebraic Lagrangian density $l(\xi, D, \nu, \gamma_m)$, this momentum
conservation law becomes, cf. equation (\ref{liqxtal-momentum-eqn}), 
\begin{equation}\label{perfect-momentum-eqn-1}
\frac{\partial}{\partial t}\frac{\partial  l}{\partial \xi_j}
=
-\
\frac{\partial}{\partial x_m}
\bigg(\xi_m\,\frac{\partial l}{\partial \xi_j}\, 
+
\Big(l - D\frac{\partial l}{\partial D}\Big)\, \delta_{mj}
-
\frac{\partial l}{\partial \gamma_m^{\,\beta}}
\, \gamma_j^{\,\beta}
\bigg)
\,.
\end{equation}
%

\paragraph{Kelvin-Noether circulation theorem for PCFs.}
Rearranging the motion equation (\ref{LP-eulerpoincare-motion}) and using the
continuity equation for $D$ in (\ref{cont-eqn}) gives 
\begin{equation}\label{Eul-vel-circ-thm-perfect}
\frac{d}{dt}\oint_{c(\xi)}
\frac{1}{D}
\frac{\delta  l}{\delta \xi_j} dx_j
=
\oint_{c(\xi)}\frac{1}{D}
\bigg[
\frac{\delta l }{ \delta a}\diamond a
+ \frac{\delta l }{ \delta \nu}\diamond \nu
+ \frac{\delta l }{ \delta \gamma_m}\diamond \gamma_m
\bigg]
,
\end{equation}
where the circulation loop $c(\xi)$ moves with the fluid velocity
$\xi$ and we have used the following relation, valid for one-form densities,
\begin{equation}\label{Eul-vel-circ-thm-perfect-1}
{\rm ad}^*_\xi\, \frac{\delta l}{\delta \xi}
=
\pounds_\xi\, \frac{\delta l}{\delta \xi}
\,,\quad\hbox{with}\quad
\frac{\delta l}{\delta \xi}
=
\bigg(\frac{\delta l}{\delta \xi_j}dx_j\,\otimes\,d^3x\bigg)
.
\end{equation}
This relation may be checked explicitly in Cartesian coordinates, as follows,
\begin{eqnarray}\label{Lie-der/ad-star-rel}
\Big\langle  
\pounds_\xi \, \frac{\delta l}{\delta \xi}\,,\,\eta 
\Big\rangle
&=&
\int
\bigg(
\frac{\partial}{\partial x^j}
\Big( \xi^j
\frac{\delta l}{\delta \xi^i}
\Big)
+
\frac{\delta l}{\delta \xi^j}
\frac{\partial\xi^j}{\partial x^i}
\bigg)\ \eta^i\
d^3x
\nonumber\\
&=&
-\,
\int
\frac{\delta l}{\delta \xi^j}
\bigg(
\xi^i 
\frac{\partial\eta^j}{\partial x^i}
-
\eta^i 
\frac{\partial\xi^j}{\partial x^i}
\bigg)\
d^3x
\nonumber\\
&=&
-\,
\Big\langle 
\frac{\delta l}{\delta \xi}\,,\,{\rm ad}_\xi\, \eta 
\Big\rangle
=
\Big\langle
{\rm ad}^*_\xi\, \frac{\delta l}{\delta \xi}\,,\,\eta 
\Big\rangle
\,.
\end{eqnarray}
See {Holm, Marsden and Ratiu [1998]} 
for more explanation and discussion of the Kelvin-Noether circulation theorem
for Euler-Poincar\'e systems. 

In components, the Kelvin-Noether circulation theorem
(\ref{Eul-vel-circ-thm-perfect}) for PCFs with
Lagrangian $l(\xi, D, \nu, \gamma_m)$ may be written using equation
(\ref{diamond-def-nu}) as,
\begin{equation}\label{Eul-vel-circ-thm-perfect-example}
\frac{d}{dt}\oint_{\!c(\xi)\,}\!\!
\frac{1}{D}
\frac{\delta  l}{\delta \xi_j} dx_j
=
\oint_{\!c(\xi)\,}\!\!
\frac{1}{D}
\bigg[
D \frac{\partial}{\partial x_j}\frac{\delta l}{\delta D}
-
\frac{\delta l}{\delta \nu^\beta} 
\frac{\partial \nu^\beta}{\partial x_j}
+
\frac{\partial }{\partial x_m}
\bigg( 
\frac{\delta l}{\delta \gamma_m^\beta} 
\gamma_j^\beta
\bigg)
-
\frac{\delta l}{\delta \gamma_m^\beta} 
\frac{\partial \gamma_m^\beta}{\partial x_j}
\bigg]
\,dx_j
.
\end{equation}
Thus, gradients of angular
frequency and order parameter strain may cause fluid circulation.

\subsection{Hamiltonian dynamics of PCFs}

\paragraph{The Legendre Transformation.} One passes from
Euler-Poincar\'e equations on a Lie algebra $\mathfrak{g}$ to Lie--Poisson
equations on the dual $\mathfrak{g}^\ast$ by means of the Legendre
transformation, see, e.g., {Holm, Marsden and Ratiu [1998]}. In our case, we
start with the reduced Lagrangian $l$ on
$\mathfrak{g}\times (V^\ast \times
\mathfrak{o})g(t)^{-1}$ and perform a Legendre transformation in the
variables $ \xi $ and $\nu$ only, by writing
\begin{equation}\label{legendre-xform}
\mu = \frac{\delta l}{\delta \xi}\,, \quad
\sigma = \frac{\delta l}{\delta \nu}\,, \quad
h(\mu, a, \sigma, \gamma)
 = \langle \mu, \xi\rangle + \langle \sigma, \nu \rangle
- l(\xi, a, \nu, \gamma).
\end{equation}
One then computes the variational derivatives of $h$ as
\begin{equation}\label{dual-var-derivs}
\frac{\delta h}{\delta \mu} 
=
 \xi \,,
\quad
\frac{\delta h}{\delta \sigma} 
=
 \nu\,,
\quad
\frac{\delta h}{\delta a} 
=
-\, \frac{\delta l}{\delta a}
\,,\quad
\frac{\delta h}{\delta \gamma} 
=
- \,\frac{\delta l}{\delta \gamma}\,.
\end{equation}
Consequently, the Euler-Poincar\'e equations (\ref{LP-eulerpoincare-motion}) -
(\ref{LP-micro-kinematics}) for PCF dynamics in the Eulerian
description imply the following equations, for the
Legendre-transformed variables, $(\mu, a, \sigma, \gamma)$,
cf. equations (\ref{m-eqn-liqxtal}) -- (\ref{gamma-eqn-liqxtal})
\begin{eqnarray} \label{Ham-perfect-eqns}
\frac{\partial\mu}{\partial t} 
 &=& 
-\,\mbox {\rm ad}_{\delta h / \delta \mu}^{\ast}\, \mu
- \frac{\delta h }{ \delta a}\diamond a
- \frac{\delta h }{ \delta \gamma_m}\diamond \gamma_m
- \frac{\delta h }{ \delta \sigma}\diamond \sigma
\,,
\nonumber\\
\frac{\partial a}{\partial t} 
 &=& -\,\pounds_{\delta h/\delta \mu}a\,
\,,
\nonumber\\
\frac{\partial \gamma}{\partial t}\cdot dx
 &=& -\,\pounds_{\delta h/\delta \mu} (\gamma\cdot dx)
+ d\Big(\frac{\delta h}{\delta \sigma}\Big)
+ {\rm ad}_{\delta h/\delta \sigma}(\gamma\cdot dx)\,,
\nonumber\\
\frac{\partial\sigma}{\partial t}  
 &=& 
-\,
{\rm div}\bigg(\frac{\delta h}{\delta \mu}\,\sigma
-\frac{\delta h}{\delta \gamma}\bigg)
 +
{\rm ad}^{\ast}_{\delta h/\delta \sigma}\, \sigma
- {\rm ad}^{\ast}_{\gamma_m} \frac{ \delta h }{ \delta
\gamma_m}\,.
\end{eqnarray}
As for the case of liquid crystals discussed earier, these equations are {\bfi
Hamiltonian} and may be expressed in terms of a Lie-Poisson bracket.

\paragraph{Lie-Poisson bracket for PCFs.}
Assembling the PCF equations (\ref{Ham-perfect-eqns}) into
Hamiltonian form gives, symbolically,
{\small
\begin{equation} \label{LP-Ham-struct-symbol}
\frac{\partial}{\partial t}
\left[ \begin{array}{c} 
\mu \\ a \\ \gamma \\ \sigma
\end{array}\right]
= -
\left[ \begin{array}{cccc} 
{\rm ad}^\ast_\Box\,\mu & 
\Box\diamond a &
\Box\diamond\gamma & 
\Box\diamond\sigma
\\ 
\pounds_\Box\,a & 0 & 0 & 0
\\
\pounds_\Box\,\gamma& 0 & 0
&  -\,({\rm grad} - {\rm ad}_\gamma)\Box
\\
\pounds_\Box\,\sigma & 0 & 
  -\,({\rm div} - {\rm ad}^\ast_\gamma)\Box
& -\,{\rm ad}^\ast_\Box\, \sigma
\end{array} \right]
\left[ \begin{array}{c} 
\delta h/\delta\mu \\ 
\delta h/\delta a \\ 
\delta h/\delta\gamma \\ 
\delta h/\delta\sigma
\end{array}\right]
\end{equation}
}
$\!\!$with boxes $\Box$ indicating where the matrix operations occur. In the
$\gamma-\sigma$ entry of the Hamiltonian matrix (\ref{LP-Ham-struct-symbol}),
one recognizes the {\bfi covariant spatial derivative} defined in
equation (\ref{covar-der}), and finds its adjoint operator in the
$\sigma-\gamma$ entry.  More explicitly, in terms of indices and differential
operators, and for $a=D$, the mass density, this Hamiltonian matrix form
becomes 
{\small
\begin{equation}\label{Ham-matrix-diff}
\frac{\partial}{\partial t}
\left[ \begin{array}{c} 
\mu_i \\ D \\ \gamma_i^{\,\alpha} \\ \sigma_\alpha
\end{array}\right]
= -
\left[ \begin{array}{cccc} 
\mu_j\partial_i + \partial_j\mu_i & 
D\partial_i &
 \partial_j\gamma_i^{\,\beta} - \gamma_{j\,,\,i}^{\,\beta} & 
\sigma_\beta\partial_i
\\ 
\partial_jD & 0 & 0 & 0
\\
\gamma_j^{\,\alpha}\partial_i + \gamma_{i\,,\,j}^{\,\alpha}& 0 & 0 & 
-\delta_\beta^{\,\alpha}\partial_i 
- t_{\beta\kappa}^{\,\alpha}\gamma_i^{\,\kappa}
\\
\partial_j\sigma_\alpha & 0 & 
-\delta_\alpha^{\,\beta}\partial_j 
+ t_{\alpha\kappa}^{\,\beta} \gamma_j^{\,\kappa}
& -\,t_{\alpha\beta}^{\,\kappa} \sigma_\kappa
\end{array} \right]
\left[ \begin{array}{c} 
{\delta h/\delta\mu_j} \\ 
{\delta h/\delta D} \\ 
{\delta h/\delta\gamma_j^{\,\beta}} \\ 
{\delta h/\delta\sigma_\beta}
\end{array}\right]
\end{equation}
}
$\!\!$Here, the summation convention is enforced on repeated indices. Upper
Greek indices refer to the Lie algebraic basis set, lower Greek indices refer
to the dual basis and Latin indices refer to the spatial reference frame. The
partial derivative $\partial_j=\partial/\partial x_j$, say, acts to the right
on all terms in a product by the chain rule. For the case that  
$t^{\,\alpha}_{\ \beta\kappa}$ are structure constants
$\epsilon_{\alpha\beta\kappa}$ for the Lie algebra $so(3)$, the Lie-Poisson
Hamiltonian matrix (\ref{Ham-matrix-diff-liqxtal-compon}) for liquid crystals
is recovered, modulo an inessential factor of 2.

\paragraph{Remark.}
As mentioned earlier in our discussion of Hamiltonian dynamics of
liquid crystals, the Hamiltonian matrix in equation (\ref{Ham-matrix-diff})
was discovered some time ago in the context of investigating the
relation between spin-glasses and Yang-Mills magnetohydrodynamics (YM-MHD) by
using the Hamiltonian approach in {Holm and Kupershmidt [1988]}. There, it was
shown to be a valid Hamiltonian matrix by associating its Poisson bracket with
the dual space of a certain Lie algebra of semidirect-product type that has a
{\bfi generalized two-cocycle} on it. This generalized
two-cocycle contributes the grad and div terms appearing in the more symbolic
expression of this Hamiltonian matrix in equation (\ref{LP-Ham-struct-symbol}).

The mathematical discussion of this Lie algebra and its generalized
two-cocycle, as well as the corresponding Lie-Poisson Hamiltonian equations
for spin-glass fluids and YM-MHD, are given in {Holm and Kupershmidt [1988]}.
The present work provides a rationale for the derivation of such Lie-Poisson
brackets from the Lagrangian side.

\paragraph{Spatially one-dimensional static solutions with $z$-variation.}
Static (steady, zero-velocity) solutions for PCFs, with
constant pressure and one-dimensional spatial variations in, say, the
$z$-direction obey equations (\ref{Ham-matrix-diff}) for $i=3=j$, rewritten as
\begin{eqnarray}
-D\,\frac{d}{dz}\,\frac{\delta h }{ \delta D}
&=&
\gamma_3^{\,\alpha}\frac{d}{dz}\,
\frac{\delta h }{ \delta \gamma_3^{\,\alpha}}
+\
\sigma_\alpha
\frac{d}{dz}\,
\frac{\delta h }{ \delta \sigma_\alpha}
\
=\
0
\,,
\nonumber\label{m-eqn-perfect-stat}\\
\frac{d}{dz}\,
\frac{\delta h}{\delta \gamma_3^{\,\alpha} }
&=&
t^\beta_{\alpha\kappa}\gamma_3^{\,\kappa} 
\frac{ \delta h }{ \delta \gamma_3^\beta}
-\,
t^\kappa_{\alpha\beta}\sigma_\kappa
\frac{\delta h}{\delta \sigma_\beta} 
\,,
\label{sigma-eqn-perfect-stat}\\
\frac{d}{dz}\,
\frac{\delta h}{\delta\sigma_\alpha} 
&=&
-\, t^\alpha_{\beta\kappa}\gamma_3^{\,\kappa} 
\frac{\delta h}{\delta \sigma_\beta}
\,.\qquad
\nonumber\label{gamma-eqn-perfect-stat}
\end{eqnarray}
As for the case of liquid crystals, the sum of terms in the first equation of
the set (\ref{sigma-eqn-perfect-stat}) vanishes to give zero pressure
gradient, as a consequence of the latter two equations.

Under the Legendre transformation 
\begin{equation} \label{perfect-Legendre-heavy-top-map}
h(\gamma_3,\sigma)
= 
\Pi_\alpha \gamma_3^{\,\alpha}
+
\Gamma^{\,\alpha} \sigma_\alpha
- 
H(\Pi,\Gamma) 
\,,
\end{equation}
these equations become,
\begin{eqnarray}
\frac{d}{dz}\,\Pi_\alpha
&=&
t^\beta_{\alpha\kappa}
\frac{ \delta H}{ \delta \Pi_\kappa}\Pi_\beta
-\,
t^\kappa_{\alpha\beta}
\frac{ \delta H}{ \delta \Gamma^\kappa}\Gamma^\beta
\,,
\label{Pi-eqn-perfect-stat}\\
\frac{d}{dz}\,\Gamma^\alpha
&=&
-\, t^\alpha_{\beta\kappa}
\frac{ \delta H}{ \delta \Pi_\kappa}\Gamma^\beta
\,.
\nonumber\label{gamma-eqn-perfect-stat-1}
\end{eqnarray}
These Legendre-transformed equations are Poincar\'e's [1901] generalization of
Euler's equations for a heavy top, expressing them on an arbitrary Lie algebra
with structure constants $t^\alpha_{\beta\kappa}$. Thus, 
\begin{quote}
{\it The steady, spatially one-dimensional solutions for all PCFs have the
underlying Lie algebra structure discovered in {Poincar\'e [1901]}.}
\end{quote}

\paragraph{Spatially homogeneous, time-dependent PCF flows.}
Spatially homogeneous solutions of equations (\ref{Ham-matrix-diff}) obey the
dynamical equations,
\begin{eqnarray}
\frac{d\sigma_\alpha}{dt}\,
&=&
-\, t^\beta_{\alpha\kappa}\gamma_j^{\,\kappa} 
\frac{ \delta h }{ \delta \gamma_j^\beta}
+\,
t^\kappa_{\alpha\beta}\sigma_\kappa
\frac{\delta h}{\delta \sigma_\beta} 
\,,
\label{sigma-eqn-perfect-homog}\\
\frac{d\gamma_i^\alpha}{dt}\,
&=&
t^\alpha_{\beta\kappa}\gamma_i^{\,\kappa} 
\frac{\delta h}{\delta \sigma_\beta}
\,.
\nonumber\label{gamma-eqn-perfect-homog}
\end{eqnarray}
For a single spatial index, say $\gamma_i$, these are again Poincar\'e's
[1901] equations generalizing Euler's equations for a heavy top to an
arbitrary Lie algebra. Of course, the corresponding Hamiltonian matrix for
this system is the lower right corner of the matrix in equation
(\ref{Ham-matrix-diff}). 

When $t^\alpha_{\beta\kappa}=\epsilon_{\alpha\beta\kappa}$ for the Lie algebra
$so(3)$, Poincar\'e's equations (\ref{sigma-eqn-perfect-homog}) correspond
to the {\bfi Leggett equations} for $^3He-A$ with spin density $\sigma_\alpha$
and spin anisotropy vector $\gamma_i^{\,\alpha}$, see {Leggett [1975]}. For
special solutions of these and other related equations in the context of
$^3He-A$, see {Golo and Monastyrskii [1977]}, {Golo and Monastyrskii [1978]}, 
{Golo et al. [1979]}.

\paragraph{Evolution of the disclination density.}
{Holm and Kupershmidt [1988]} use the chain rule
and the defining relation for the disclination density, cf. equation
(\ref{curl-gamma}),
\begin{equation}
B_{ij}^{\,\alpha} 
\equiv 
\gamma_{i,j}^{\,\alpha} - \gamma_{j,i}^{\,\alpha} 
+ 
t^{\,\alpha}_{\beta\kappa}\gamma_i^{\,\beta}\gamma_j^{\,\kappa}
\,,
\end{equation}
to transform the Hamiltonian matrix (\ref{Ham-matrix-diff}) to a
new Hamiltonian matrix, whose Lie-Poisson Hamiltonian dynamics may be written
as 
 {\small
\begin{eqnarray}\label{Ham-matrix-diff-B_ij}
\frac{\partial}{\partial t}
\left[ \begin{array}{c} 
\mu_i \\ D \\ B_{ij}^{\,\alpha} \\ \sigma_\alpha
\end{array}\right]
&&\\
&&\hspace{-1in}
= -
\left[ \begin{array}{cccc} 
\mu_k\partial_i + \partial_k\mu_i & 
D\partial_i &
-
B_{lm,i}^{\,\beta}
+
 \partial_m B_{li}^{\,\beta} 
- 
 \partial_l B_{mi}^{\,\beta}  & 
\sigma_\beta\partial_i
\\ 
\partial_kD & 0 & 0 & 0
\\
B_{ij,k}^{\,\alpha}
+
 B_{i\,k}^{\,\alpha}  \partial_j
- 
 B_{jk}^{\,\alpha}  \partial_i  & 0 & 0 & 
- 
t_{\beta\kappa}^{\,\alpha}B_{ij}^{\,\kappa}
\\
\partial_k\sigma_\alpha & 0 & 
 t_{\alpha\kappa}^{\,\beta} B_{lm}^{\,\kappa}
& -\,t_{\alpha\beta}^{\,\kappa} \sigma_\kappa
\end{array} \right]
\left[ \begin{array}{c} 
{\delta h/\delta\mu_k} \\ 
{\delta h/\delta D} \\ 
{\delta h/\delta B_{lm}^{\,\beta}} \\ 
{\delta h/\delta\sigma_\beta}
\end{array}\right]
\nonumber
\end{eqnarray}
}
$\!\!$The corresponding {\bfi PCF dynamics for the disclination density}
emerges as 
\begin{equation}
\frac{\partial B_{ij}^{\,\alpha}}{\partial t} 
= 
-\,
\Big( B_{ij,k}^{\,\alpha}
+
 B_{i\,k}^{\,\alpha}  \partial_j
- 
 B_{jk}^{\,\alpha}  \partial_i \Big)\frac{\delta h}{\delta\mu_k}
\
+\
t_{\beta\kappa}^{\,\alpha}B_{ij}^{\,\kappa}
\frac{\delta h}{\delta\sigma_\beta}
\,.
\end{equation}
As expected, this equation preserves the trivial solution
$B_{ij}^{\,\alpha}=0$, which is the case when the $\gamma_m-$strain field is
continuous and the complex fluid has no defects. We shall mention a strategy
for dealing with defects in Section \ref{strat-defect}.

\subsection{Clebsch approach for PCF dynamics}

Following {Serrin [1959]}, we call the auxiliary
constraints imposed by the Eulerian kinematic equations the {\bfi Lin
constraints}. As we shall see, the diamond operation $\diamond$ defined
in equation (\ref{diamond-def}) arises naturally in imposing the Lin
constraints. Taking variations of the constrained Eulerian action,
\begin{equation}\label{constrain-Eul-act}
\mathcal{S} = \int\!\!dt\
\bigg\{ l(\xi, a, \nu, \gamma)
+
\Big\langle v, \frac{\partial a}{\partial t}
+ \pounds_\xi a \Big\rangle
+
\Big\langle \beta, \frac{\partial \gamma}{\partial t}
+ \pounds_\xi \gamma
-d\nu
-
{\rm ad}_\nu\gamma \Big\rangle
\bigg\}
\end{equation}
yields the following {\bfi PCF Clebsch relations},
\begin{eqnarray}\label{perfect-Clebsch-relations}
\delta \xi: 
&&
\frac{\delta l}{\delta \xi}
-
v\diamond a
-
\beta\diamond \gamma
=
0
\,,
\nonumber\\
\delta \nu:
&&
\frac{\delta l}{\delta \nu}
+
d\beta
-
{\rm ad}^*_\gamma\,\beta
=
0
\,,
\nonumber\\
\delta a:
&&
\frac{\delta l}{\delta a}
-
\frac{\partial v }{\partial t}
-
\pounds_\xi\,v
=
0
\,,
\\
\delta v:
&&
\frac{\partial a }{\partial t}
+ 
\pounds_\xi\,a
=
0
\,,
\nonumber\\
\delta \gamma:
&&
\frac{\delta l}{\delta \gamma}
-
\frac{\partial \beta }{\partial t}
-
\pounds_\xi\,\beta
+
{\rm ad}^*_\nu\,\beta
=
0
\,,
\nonumber\\
\delta \beta:
&&
\frac{\partial \gamma }{\partial t}
+
\pounds_\xi\,\gamma
-
d\nu
-
{\rm ad}_\nu\,\gamma
=
0
\,.\nonumber
\end{eqnarray} 
We shall show that these Clebsch relations recover the Euler-Poincar\'e
equations (\ref{LP-eulerpoincare-motion}) -
(\ref{LP-eulerpoincare-micromotion}). (In what follows, we shall ignore
boundary and endpoint terms that arise from integrating by parts.)

The {\bfi diamond operation} $\diamond$ is defined by 
\begin{equation}
\langle v \diamond a \,,\, \eta \rangle
\equiv
-\,
\langle v \,,\, \pounds_\eta\, a \rangle
=
-\,
\langle v \,,\, a\, \eta \rangle
\,.
\end{equation}
This operation is {\bfi antisymmetric},
\begin{equation}
\langle v \diamond a \,,\, \eta \rangle
=
-\,
\langle a \diamond v\,,\, \eta \rangle
\,,
\end{equation}
as obtained from,
\begin{equation}\label{antisym-rel}
\langle v \,,\, \pounds_\eta\, a \rangle
+
\langle \pounds_\eta\, v \,,\, a \rangle
=
0
\,,\quad\hbox{or,}\quad
\langle v \,,\, a\,\eta \rangle
+
\langle v\,\eta \,,\, a \rangle
=
0
\,,
\end{equation}
and the symmetry of the pairing
$\langle\boldsymbol\cdot\,,\,\boldsymbol\cdot\rangle$. The diamond operation
also satisfies the {\bfi chain rule under the Lie derivative},
\begin{equation}
\langle \pounds_\xi\,(v \diamond a)\,,\, \eta \rangle
=
\langle (\pounds_\xi\,v) \diamond a\,,\, \eta \rangle
+
\langle v \diamond (\pounds_\xi\,a)\,,\, \eta \rangle
\,.
\end{equation}
This property can be verified, as follows,
\begin{eqnarray}\hspace{-.4in}
\langle \pounds_\xi\, v \diamond a\,,\, \eta \rangle
+
\langle v \diamond \pounds_\xi\, a\,,\, \eta \rangle
&=&
\langle v\,\xi\,\eta \,,\, a \rangle
-
\langle v\,\eta\,\xi \,,\, a \rangle
\nonumber\\
&=&
\langle a \,,\, v\,({\rm ad}_\xi\,\eta) \rangle
=
-\, \langle a \diamond v\,,\,({\rm ad}_\xi\,\eta) \rangle
\nonumber\\
&=&
    \langle {\rm ad}^*_\xi (a \diamond v)\,,\,\eta \rangle
=
    \langle \pounds_\xi (a \diamond v)\,,\,\eta \rangle
\,,
\end{eqnarray}
where we have used $\langle v\,\xi \,,\, a\,\eta \rangle
+
\langle v\,\xi\eta \,,\, a \rangle = 0$, implied by (\ref{antisym-rel}),
in the first step. Finally, we have the {\bfi useful identity},
\begin{equation}
\langle \, \beta \diamond d\nu\,,\, \eta \rangle
=
-\,\langle d\beta \diamond \nu\,,\, \eta \rangle
\,,
\end{equation}
as obtained from $(d\nu)\eta = d(\nu\eta)$ and 
\begin{equation}
\langle \,\beta \,,\, d(\nu\eta)\, \rangle
+
\langle \,d\beta \,,\, \nu\eta\, \rangle
=
0
\,.
\end{equation}
These three properties of the $\diamond$
operation and the PCF Clebsch relations
(\ref{perfect-Clebsch-relations}) together imply 
\begin{eqnarray}
\Big( \frac{\partial }{\partial t}
+
\pounds_\xi\,\Big)
\Big( v\diamond a
+
\beta\diamond \gamma\Big)
&=&
\frac{\delta l }{ \delta a}\diamond a
+ \frac{\delta l }{ \delta \nu}\diamond \nu
+ \frac{\delta l }{ \delta \gamma_m}\diamond \gamma_m
\\
&&+\
\Big[
({\rm ad}^*_\nu\,\beta) \diamond \gamma
+
\beta \diamond {\rm ad}_\nu\,\gamma
+
({\rm ad}^*_\gamma\,\beta) \diamond \nu
\Big]
\,.
\nonumber
\end{eqnarray}
The term in square brackets is seen to vanish, upon pairing it with a vector
field, integrating by parts and again using the properties of the $\diamond$
operation. This manipulation recovers the {\bfi PCF motion equation}
(\ref{LP-eulerpoincare-motion}) as
\begin{equation}
\Big( \frac{\partial }{\partial t}
+
\pounds_\xi\,\Big)
\frac{\delta l }{\delta \xi}
=
\frac{\delta l }{ \delta a}\diamond a
+ \frac{\delta l }{ \delta \nu}\diamond \nu
+ \frac{\delta l }{ \delta \gamma_m}\diamond \gamma_m
\,,
\end{equation}
since, as we have seen,
\begin{equation}
\pounds_\xi\,
\frac{\delta l }{\delta \xi}
=
{\rm ad}^*_\xi\,\frac{\delta l }{\delta \xi}
\,,
\end{equation}
for one-form densities such as $\delta l/\delta \xi$.

The {\bfi PCF micromotion equation} (\ref{LP-eulerpoincare-micromotion}) is
also  recovered from the Clebsch relations (\ref{perfect-Clebsch-relations}).
This is accomplished by taking the time derivative of the $\delta\nu-$formula,
substituting the $\delta\beta-$ and $\delta\gamma-$formulas, and using
linearity of ad$^*$ to find
\begin{eqnarray}
\Big( \frac{\partial }{\partial t}
+
\pounds_\xi\,\Big)
\frac{\delta l }{ \delta \nu}
\!\!\!&+&\!\!\!
d\,\frac{\delta l }{ \delta \gamma}
-
{\rm ad}^*_\gamma\,\frac{\delta l }{ \delta \gamma}
\\
\!\!\!&=&\!\!\!
-\,\Big[
d({\rm ad}^*_\nu\,\beta) 
+
{\rm ad}^*_{d\nu}\,\beta
-
{\rm ad}^*_\gamma\,({\rm ad}^*_\nu\,\beta)
-
{\rm ad}^*_{({ad}_\nu\gamma)}\,\beta
\Big]
\nonumber\\
\!\!\!&=&\!\!\!
-\,\Big[
{\rm ad}^*_\nu\,d\beta
-
{\rm ad}^*_\nu\,({\rm ad}^*_\gamma\,\beta)
\Big]
\nonumber\\
\!\!\!&=&\!\!\!
{\rm ad}^*_\nu\,\frac{\delta l }{ \delta \nu}
\,.
\nonumber
\end{eqnarray}
Hence, the Clebsch relations (\ref{perfect-Clebsch-relations}) also  
recover the micromotion equation (\ref{LP-eulerpoincare-micromotion}).

\paragraph{Remarks.}
From the Hamiltonian viewpoint, the pairs $(v,a)$ and $(\beta,\gamma)$ are
canonically conjugate variables and the {\bfi Clebsch map} $(v,a,\beta,\gamma)
\to (\mu,\sigma)$, with
\begin{eqnarray}
\frac{\delta l }{\delta \xi}
\ \equiv\
\mu
\!\!&=&\!\!
v \diamond a
+
\beta \diamond \gamma
\,,\\
\frac{\delta l }{\delta \nu}
\ \equiv\
\sigma
\!\!&=&\!\!
-\,d\beta
+
{\rm ad}^*_\gamma\,\beta
\,,
\end{eqnarray}
is a {\bfi Poisson map} from the canonical Poisson bracket to the Lie-Poisson
Hamiltonian structure given in equation (\ref{Ham-matrix-diff}), in which 
$a=D$. Of course, there is no obstruction against allowing $a$ to be any
advected quantity, as discussed in {Holm, Marsden and Ratiu [1998]}. \bigskip

The generalized two-cocycle associated with the Hamiltonian matrix in
(\ref{Ham-matrix-diff}) arises from the term $d\beta$ in the $\sigma-$part of
this Poisson map. \bigskip

Various other applications of the Lin constraint and Clebsch representation
approach in formulating and analyzing ideal fluid and plasma dynamics as
Hamiltonian systems appear in  {Holm and Kupershmidt [1983b]}, {Marsden and
Weinstein [1983]},  {Zakharov et al. [1985]}, {Zakharov and Kusnetsov [1997]}.

\subsection{Conclusions for PCFs}\label{Conclus}
Perfect complex fluids (PCFs) have internal variables whose micromotion is
coupled to the fluid's motion. Examples of PCFs include spin-glass fluids,
superfluids and liquid crystals. PCF internal variables are
materially advected order parameters that may be represented equivalently as
either geometrical objects, or as coset spaces of Lie groups.  The new feature
of PCFs relative to simple fluids with advected parameters treated in {Holm,
Marsden and Ratiu [1998]} is the dependence of their Lagrangian
$$
L:TG\times V^{\ast}\times T\mathcal{O}
\longmapsto\mathbb{R}\,,
$$
on $T\mathcal{O}$, the tangent space of their order parameter group.
Moreover, the diffeomorphisms $G$ act on $T\mathcal{O}$.
We treat  Lagrangians that are invariant under the right actions of both  
the order parameter group $\mathcal{O}$ and the diffeomorphisms $G$. In this
case, reaching the Euler-Poincar\'e fluid description requires {\it two stages}
of Lagrangian reduction,
$$
\big(TG\times V^{\ast}\times (T\mathcal{O}/\mathcal{O})\big)/G
\simeq
\mathfrak{g}\times ( V^{\ast} \times \mathfrak{o})g^{-1}(t)
\,,
$$
rather than the single stage of Lagrangian reduction (with respect to the
``relabeling transformations'' of $G$) employed for simple fluids in {Holm,
Marsden and Ratiu [1998]}.

After studying the example of nematics in Section \ref{liqxtal}, we derived 
the Euler-Poincar\'e dynamics of PCFs in two stages of
Lagrangian reduction in Section \ref{Ham-Princ-Lag-Red}.  The first stage 
produced the Lagrange-Poincar\'e equations derived from an action principle
defined on the right invariant Lie algebra of the order parameter group in the
Lagrangian (or material) fluid description. The second stage of Lagrangian
reduction passed from the material fluid description to the Eulerian (or
spatial) fluid description and produced the Euler-Poincar\'e equations for
PCFs. We also derived these Euler-Poincar\'e equations
using the Clebsch approach. 

In addition, we used a Legendre transformation to obtain the
Lie-Poisson Hamiltonian formulation of PCF dynamics
in the Eulerian fluid description. The Lie-Poisson Hamiltonian formulation of
these equations agreed with that found earlier in
{Holm and Kupershmidt [1987]}, {Holm and Kupershmidt [1988]},  who treated
spin-glass fluids, Yang-Mills magnetohydrodynamics and superfluid $^4He$ and
$^3He$. Thus, we found that Lagrangian reduction by stages provides a
rationale for deriving this Lie-Poisson Hamiltonian formulation from the
Lagrangian side. This approach also fits well with some gauge theoretical
descriptions of condensed matter physics.

\section{A strategy for introducing defect dynamics}\label{strat-defect}

Many other potential applications of the Euler-Poincar\'e framework abound
in the physics of condensed matter. For example, besides the perfect liquid
crystal dynamics treated here explicitly, the superfluid hydrodynamics of the
various phases of $^3He$ may be treated similarly. In particular, the
geometrical framework of Lagrangian reduction by stages is well-adapted to the
standard identification of the phases of $^3He$ with the independent cosets of
the order parameter group $SO(3)\times SO(3)\times U(1)$, as discussed, e.g.,
in {Mineev [1980]} and {Volovick [1992]}. Magnetic materials may also be
treated this way. The seminal papers on the geometrical properties of magnetic
materials are {Dzyaloshinskii [1977]}, 
{Volovik and Dotsenko [1980]} and {Dzyaloshinskii and Volovick [1980]}. 
Other recent studies of the dynamics of magnetic materials and superfluid
$^3He$ in directions relevant to the present paper appear, e.g., in  
{Holm and Kupershmidt [1988], {Balatskii [1990]}, 
{Isayev and Peletminsky [1997]}, {Isayev, Kovalevsky and Peletminsky [1997]}.

Most of these physical applications involve defects (imperfections,
or ``glitches'' that appear as discontinuities in the order parameter) and
one must describe their dynamics, as well. This is an area of intense
investigation in many contexts in condensed matter physics. Our approach is
based on an analogy with the theory of the Hall effect in a neutral
multicomponent fluid plasma when inertia is negligible in one of the
components (the electrons).

The Hamiltonian context for Hall effects in a neutral ion-electron plasma 
was considered by {Holm [1987]} for a normal-fluid plasma 
and by {Holm and Kupershmidt [1987]} for a multicomponent
electromagnetically charged superfluid plasma.  In this
context, the introduction of an independent gauge field associated with
the momentum of a distribution of superfluid vortex lines generalizes to
apply for defects or vortices in any continuous medium
possessing an order parameter description that arises from spontaneous
symmetry breaking. In the remainder of this paper, we shall use the idea of
reactive forces arising via the Hall effect to discuss the case of quantum
vortices in superfluid Helium-II. In particular, we shall apply
the Hall effect analogy to describe the hydrodynamics of the quantum
vortex tangle in superfluid turbulence as an additional ``third fluid'' in
Landau's two-fluid fluid model. The third fluid associated with the vortex
tangle carries momentum and moves with its own independent velocity in
superfluid flows.

\subsection{Vortices in superfluid $^4$He}

In superfluid $^4$He the order parameter group is $U(1)$ and the defects are
called vortices. These are {\bfi quantum vortices}, since their circulation
comes in integer multiples of $\kappa=h/m\simeq10^{-3}cm^2/sec$.  Conservation
of the number of quantum vortices moving through superfluid
$^4$He (and across the streamlines of the normal fluid component) is
expressed by
\begin{equation}
\frac{d}{dt} \int_{S}
\boldsymbol{\omega\cdot\hat n}\,dS
=
0
\,,\label{vort-cons}
\end{equation}
where the superfluid vorticity $\boldsymbol{\omega}$ is the areal density
of vortices and $\boldsymbol{\hat n}$ is the unit vector normal to the
surface $S$ whose boundary $\partial S$ moves with the vortex line
velocity $\mathbf{v}_\ell$. When 
$\boldsymbol{\omega}={\rm curl}\,\mathbf{v}_s$ this is
equivalent to a {\bfi vortex Kelvin theorem}
\begin{equation}
\frac{d}{dt} \oint_{\partial S(\mathbf{v}_\ell)}
\mathbf{v}_s\cdot d\mathbf{x}
=
0
\,,\label{vort-Kel}
\end{equation}
which in turn implies the fundamental relation
\begin{equation}
\frac{\partial\mathbf{v}_s}{\partial{t}} 
-
\mathbf{v}_\ell\times\boldsymbol{\omega}
=
\nabla\mu
\,.\label{phase-slip}
\end{equation}

The superfluid velocity naturally splits into 
$\mathbf{v}_s = \mathbf{u} - \mathbf{A}$, where $\mathbf{u} = \nabla\phi$
and (minus) the curl of $\mathbf{A}$ yields the superfluid vorticity
$\boldsymbol{\omega}$. The phase $\phi$ is then a regular function without
singularities.  This splitting will reveal that the Hamiltonian dynamics
of superfluid $^4$He with vortices may be expressed as an invariant
subsystem of a larger Hamiltonian system in which $\mathbf{u}$ and
$\mathbf{A}$ have independent evolution equations. We begin by defining a
phase frequency in the normal velocity frame as 
\begin{equation}
\frac{\partial\phi}{\partial{t}} + \mathbf{v}_n\cdot\nabla \phi
=
\nu
\,.
\end{equation}
The mass density $\rho$ and the phase $\phi$ are canonically
conjugate in the Hamiltonian formulation of the Landau two-fluid model.
Therefore, $\nu = - \,\delta h/\delta \rho$ for a given Hamiltonian $h$ and
$\mathbf{u}=\nabla\phi$ satisfies
\begin{equation}
\frac{\partial\mathbf{u}}{\partial{t}} + \mathbf{v}_n\cdot\nabla\mathbf{u}
+ (\nabla\mathbf{v}_n)^T\cdot \mathbf{u}
=
- \,\nabla\frac{\delta h}{\delta \rho}
\,.
\end{equation}
The mass density $\rho$ satisfies the dual equation
\begin{equation}
\frac{\partial\rho}{\partial{t}} + \nabla \cdot(\rho\mathbf{v}_n)
=
- \,\nabla\cdot \frac{\delta h}{\delta \mathbf{u}}
\,.
\end{equation}
In the last two equations, we see the expected two-cocycle terms for Landau's
perfect superfluid, in which $\boldsymbol{\omega}=0$. Here $\mathbf{u}$ and
$\rho$ in the superfluid play the roles of ${\gamma}$ and ${\sigma}$ for the
PCFs. The curvature $\boldsymbol{\omega}$ is nonvanishing now, because of
the vortices (defects) that are represented by $\mathbf{A}$.

Perhaps not surprisingly, the rotational and potential components of the
superfluid velocity must satisfy similar equations, but the rotational
component must be advected by another velocity -- the vortex line velocity
$\mathbf{v}_\ell$ -- instead of the normal velocity $\mathbf{v}_n$ that
advects $\mathbf{u}$. Absorbing all gradients into $\mathbf{u}$ yields the
form of the equation we should expect for $\mathbf{A}$, 
\begin{equation}
\frac{\partial\mathbf{A}}{\partial{t}} 
+
\mathbf{v}_\ell\times\boldsymbol{\omega}
=
0
\,.\label{A-eqn}
\end{equation}
Taking the difference of the equations for $\mathbf{u}$ and
$\mathbf{A}$ then recovers equation (\ref{phase-slip}) as
\begin{equation}
\frac{\partial\mathbf{v}_s}{\partial{t}} 
-
\mathbf{v}_\ell\times\boldsymbol{\omega}
=
- \,\nabla\Big(\mathbf{v}_n\cdot \mathbf{u} + \frac{\delta h}{\delta
\rho}\Big)
\quad\hbox{with}\quad
\mathbf{v}_s = \mathbf{u} - \mathbf{A}
\label{vee-super-eqn}\,,
\end{equation}
in which regularity of the phase $\phi$ allows one to set
curl$\,\mathbf{u}=0$. It remains to determine $\mathbf{v}_\ell$ from
the Euler-Poincar\'e formulation. Including the additional
degree of freedom $\mathbf{A}$ allows the vortex lines to move relative to
both the normal and super components of the fluid, and thereby introduces 
additional reactive forces associated with the momentum of the vortex lines.

For superfluid $^4$He with vortices, the momenta conjugate to the velocities
$\mathbf{v}_n$ and $\mathbf{v}_\ell$ shall be our basic dynamical variables.
To develop the Euler-Poincar\'e formulation of this problem, we must consider
a Lagrangian that first of all is invariant under the order parameter
group $\mathcal{O}=U(1)$. The Lagrangian must also be invariant under 
{\bfi two types of diffeomorphisms:} one corresponding to the material motion
of the normal fluid $G_n$ and another corresponding to the motion of the
vortices $G_\ell$. Thus we consider a Lagrangian that allows the following
{\bfi direct product} of group reductions,
\begin{eqnarray}
\Big(\Big(TG_n\times V^{\ast} \times (T\mathcal{O}/\mathcal{O})\Big)/G_n\Big)
&\times&
\Big(\big(TG_\ell\times V^{\ast}\big)/G_\ell\Big)
\nonumber\\
&\simeq&
\Big(\mathfrak{g}_n\times ( V^{\ast} \times \mathfrak{o})\,g_n^{-1}(t)\Big)
\times
\Big(\mathfrak{g}_\ell\times V^{\ast}g_\ell^{-1}(t)\Big)
\,.\nonumber
\end{eqnarray}
We denote the corresponding dependence in this Lagrangian as 
\begin{equation}\label{SF-vortex-Lag}
l \,(\mathbf{v}_n, S, \nu, \mathbf{u}\,;\,\mathbf{v}_\ell,n)
\,.
\end{equation}
Here $\mathfrak{g}_n$ and $\mathfrak{g}_\ell$ denote the Lie algebras of vector
fields associated to the velocities $\mathbf{v}_n$ and $\mathbf{v}_\ell$,
respectively. Also $\mathfrak{o}$ denotes the Lie algebra of the Abelian gauge
group $U(1)$; so $\mathfrak{o}$ contains $\nu$ and $\mathbf{u}$. The $V^{\ast}$
in each factor denotes the corresponding advected densities: entropy $S$
advected by $\mathbf{v}_n$ and vortex inertial mass $n$ advected by
$\mathbf{v}_\ell$. We denote the inverse right actions of the two
diffeomorphisms as $g_n^{-1}(t)$ and $g_\ell^{-1}(t)$. At a given time
$t$, these actions separately map spatial variables back to coordinates moving
with the normal material and with the vortices, respectively.

According to the Euler-Poincar\'e action principle, the following dynamical
equations are generated by this Lagrangian, cf. (\ref{LP-eulerpoincare-motion})
and (\ref{LP-eulerpoincare-micromotion})
\begin{eqnarray} \label{SF-eulerpoincare-motion}
\frac{\partial}{\partial t} \frac{\delta l}{\delta v_n}
&=& 
- \, \mbox {\rm ad}_{v_n}^{\ast} \frac{ \delta l }{ \delta v_n}
+ \frac{\delta l }{ \delta S}\diamond S
+ \frac{\delta l }{ \delta \nu}\diamond \nu
+ \frac{\delta l }{ \delta u_m}\diamond u_m
\,,
\nonumber\\
\label{SF-eulerpoincare-micromotion}
\frac{\partial}{\partial t} \frac{\delta l}{\delta \nu} 
&=&-\,
{\rm div}\Big(\mathbf{v}_n\frac{\delta l}{\delta \nu}
+ \frac{\delta l}{\delta \mathbf{u}}\Big)
\,,\nonumber\\
\frac{\partial}{\partial t} \frac{\delta l}{\delta v_\ell}
&=& 
- \, \mbox {\rm ad}_{v_\ell}^{\ast} \frac{ \delta l }{ \delta v_\ell}
+ \frac{\delta l }{ \delta n}\diamond n
\,.
\end{eqnarray}
These are the {\bfi Euler-Poincar\'e equations} for a superfluid with
vortices. The Eulerian kinematic equations are, 
cf. (\ref{D - gamma eqns}-\ref{D - gamma eqns1})
\begin{eqnarray}\label{S-u-n eqns}
\frac{\partial S}{\partial t}
 &=& - \,{\rm div}(S\mathbf{v}_n)
\,,
\nonumber\\
\frac{\partial \mathbf{u} }{\partial t}
&=&
\mathbf{v}_n\times{\rm curl}\,\mathbf{u}
\,-\,
\nabla\Big(\mathbf{u}\cdot\mathbf{v}_n - \nu\Big)
\,,\nonumber\\
\frac{\partial n}{\partial t}
 &=& - \,{\rm div}(n\,\mathbf{v}_\ell)
\,.\nonumber
\end{eqnarray}
If $\mathbf{v}_\ell$ and $n$ are absent, these equations reduce to the
equations for a PCF with broken $U(1)$ symmetry. The momentum density conjugate
to the frequency $\nu$ is the total mass density given by 
$\rho=-\delta l/\delta \nu$, which satisfies the equation above. So far, this
is interesting, but standard in the present context. However, now a new
feature develops because of the physical description of superfluids.
Physically, nothing is known on the Lagrangian side about the relation of the
gauge frequency $\nu$ to the other variables. However, on the Hamiltonian side
we know from the Legendre transformation that $\nu=-\delta h/\delta\rho$.
Moreover, the thermodynamic energy on the Hamiltonian side is a known function
of $\rho$. This means we should leave the Lagrangian side to finish
determining the dynamics for superfluid $^4$He with vortices. The following
Hamiltonian description for this dynamics is derived in Holm [2000].
\paragraph{Proposition:} 
{\it The dynamics for superfluid $^4$He with vortices follows from a
Lie-Poisson bracket whose Hamiltonian matrix separates into two pieces given
by
\footnote{The first of these is the Hamiltonian matrix for a
PCF with broken $U(1)$ symmetry. The other Hamiltonian matrix gives the
standard semidirect-product Lie-Poisson bracket, without two-cocycles. This
combination of Hamiltonian matrices was first introduced in Holm and
Kupershmidt [1987] for superfluid plasmas.}
%
\begin{equation}\label{Ham-matrix-SF-1}
\frac{\partial}{\partial t}
\left[ \begin{array}{c} 
M_i \\ S \\ \rho \\ u_i 
\end{array}\right]
= -
\left[ \begin{array}{cccc} 
M_j\partial_i + \partial_j M_i & 
S  \partial_i & \rho\partial_i &
 \partial_j u_i - u_{j\,,\,i}  
\\ 
\partial_j S & 0 & 0 & 0
\\
\partial_j\rho & 0 & 0 & \partial_j 
\\
u_j\partial_i + u_{i\,,\,j}& 0 & \partial_i & 0 
\end{array} \right]
\left[ \begin{array}{c} 
{\delta h/\delta M_j} \\ 
{\delta h/\delta S} \\ 
{\delta h/\delta \rho} \\ 
{\delta h/\delta u_{\,j}} 
\end{array}\right],
\end{equation}
and 
\begin{equation}\label{Ham-matrix-SF-2}
\frac{\partial}{\partial t}
\left[ \begin{array}{c} 
N_i \\ n  
\end{array}\right]
= -
\left[ \begin{array}{cc} 
N_j\partial_i + \partial_j N_i & 
n\partial_i 
\\ 
\partial_jn & 0 
\end{array} \right]
\left[ \begin{array}{c} 
{\delta h/\delta N_j} \\ 
{\delta h/\delta n} 
\end{array}\right],
\end{equation}
where $M=\delta l/\delta v_n$, $N=\delta l/\delta v_\ell$, 
$\rho =-\delta l/\delta \nu$ and the Hamiltonian $h$ is the Legendre transform
of the Lagrangian $l$ in (\ref{SF-vortex-Lag}) with respect to $v_n$, $v_\ell$
and $\nu$.  }
\paragraph{Corollary:} 
{\it If the Hamiltonian has no explicit spatial dependence, then total momentum
conservation holds as, 
\begin{equation}
\frac{\partial}{\partial t} \big(M_j + N_j \big)
=
\{M_j + N_j,h\}
=
-\,
\frac{\partial}{\partial x^k} \, T_j^{\,k}
\,.\nonumber
\end{equation}
Suppose the Hamiltonian density has dependence
$h(\mathbf{M},\rho,S,n, \mathbf{v}_s,\boldsymbol{\omega},\mathbf{A})$,
in which %
\begin{equation}
\mathbf{v}_s=\mathbf{u}-\mathbf{A}
\,,\quad
\mathbf{A}=-\mathbf{N}/n
\quad\hbox{and}\quad
\boldsymbol{\omega}={\rm curl}\,\mathbf{v}_s
\,.\nonumber
\end{equation}
Then, the stress tensor $T_j^{\,k}$ is expressed in terms of derivatives of the
Hamiltonian as
\begin{equation} \label{2nd-stress-tensor}
T_j^{\,k} 
= 
M_j\frac{\partial h}{\partial M_k}
+
v_{s\,j}\bigg(\frac{\partial h}{\partial v_{s\,k}}
+
\Big({\rm curl}\,\frac{\partial h}{\partial \boldsymbol{\omega}}\Big)_k
\bigg)
-
v_{s\,l,\,j}\epsilon_{mlk}\,\frac{\partial h}{\partial \omega_m}
+\
\delta_j^{\,k} P
-
A_j\,\frac{\partial h}{\partial A_k}\bigg|_{\mathbf{v}_s}
\,.\nonumber
\end{equation}
where the pressure $P$ is defined as
\begin{equation}
P 
= 
M_l\frac{\partial h}{\partial M_l}
+
\rho\frac{\partial h}{\partial \rho}
+
S\frac{\partial h}{\partial S}
+
n\frac{\partial h}{\partial n}
-
h
\,.\nonumber
\end{equation}
Thus, the dependence of the Hamiltonian
$h$ on  the vorticity $\boldsymbol{\omega}$ introduces reactive stresses due
to the motion of the vortices.}\bigskip

Details of the choice of Hamiltonian, as well as the derivation and
interpretation of the explicit equations are given in Holm [2000]. Other
results include the transformation of these equations to a rotating reference
frame and the resulting Taylor-Proudman theorem for superfluid $^4$He with
vortices. The generalization of this idea to complex fluids with nonabelian
broken symmetries will be discussed elsewhere.

\section*{Acknowledgements} I am grateful to A. Balatskii, H. Cendra, J. Hinch,
P. Hjorth, J. Louck, J. Marsden, T. Mullin, M. Perlmutter, T. Ratiu, J. Toner
and A. Weinstein for constructive comments and enlightening discussions during
the course of this work. I am also grateful for hospitality at the Isaac
Newton Institute for Mathematical Sciences where part of this work was
completed. This research was supported by the U.S. Department of Energy under
contracts W-7405-ENG-36 and the Applied Mathematical Sciences Program
KC-07-01-01.

\section*{Appendix. External torques and partial Lagrangian reduction.} The
anisotropic dielectric and diamagnetic effects on the director angular
momentum due to external electric and magnetic fields can be restored by
adding the torques from equation (\ref{EM-torques-def}) to the right
hand side of the second equation in (\ref{x-nu-gamma-mot}). Knowing they can be
restored this way, one could simply ignore the external torques. However, the
major applications of liquid crystals involve these torques and their
restoration also provides an example of {\bfi partial Lagrangian reduction}.
This example also briefly recapitulates the procedure of Lagrangian reduction
by stages used in the remainder of the paper.

\paragraph{Partially reduced Lagrange-Poincar\'e equations.}
Restoring the effects of external torques requires that we consider Hamilton's
principle for a Lagrangian that still retains its dependence on the director
$\mathbf{n}$,
\begin{equation}\label{liqxtal-action-director-nu}
\mathcal{S}' = \int dt\int d^3X \
\mathcal{L}(\boldsymbol{\dot{\mathbf{x}}}, J, \mathbf{n},
\boldsymbol\nu,\nabla\mathbf{n})
\,.
\end{equation} 
In this case, the partial reduction of the Lagrangian dependence from
$\boldsymbol{\dot{\mathbf{n}}}$ to
$\boldsymbol\nu = \mathbf{n}\times\boldsymbol{\dot{\mathbf{n}}}$
proceeds as follows. The variation of
$\boldsymbol\nu = \mathbf{n}\times\boldsymbol{\dot{\mathbf{n}}}$
gives
\begin{equation}\label{var-nu}
\delta\boldsymbol{\nu} 
=
(\mathbf{n}\times \delta\mathbf{n})\boldsymbol{\dot{\,}}
-
2\boldsymbol{\dot{\mathbf{n}}}\times\delta\mathbf{n}
=
(\mathbf{n}\times \delta\mathbf{n})\boldsymbol{\dot{\,}}
-
2\boldsymbol{\nu}\times(\mathbf{n}\times \delta\mathbf{n})
\,.
\end{equation}
Moreover, we have the relation
\begin{equation}\label{var-ndotA}
\delta\mathbf{n}\boldsymbol\cdot\mathbf{A}
=
(\mathbf{n}\times \delta\mathbf{n})
\boldsymbol\cdot
(\mathbf{n}\times \mathbf{A})
=
|\mathbf{n}|^2 \delta\mathbf{n}\boldsymbol\cdot\mathbf{A}
-
(\mathbf{n}\boldsymbol\cdot\delta\mathbf{n})
(\mathbf{n}\boldsymbol\cdot\mathbf{A})
\,,
\end{equation}
for any vector $\mathbf{A}$, since  $|\mathbf{n}|^2=1$, which implies that
$\mathbf{n}\cdot\delta\mathbf{n} = 0$. Substituting the identity
(\ref{var-nu}) for $\delta\boldsymbol{\nu}$ and the relation (\ref{var-ndotA})
into Hamilton's principle implies the following replacements,
\begin{eqnarray}
\delta\mathbf{n}\boldsymbol{\cdot}
\,\Big(
\frac{\partial \mathcal{L}}
{\partial\boldsymbol{\dot{\mathbf{n}}}}\Big)^{\boldsymbol{\dot{\,}}} 
&\Longrightarrow &
(\mathbf{n}\times \delta\mathbf{n})
\boldsymbol\cdot
\,\Big[\Big(
\frac{\partial \mathcal{L}}
{\partial\boldsymbol{\nu}}\Big)^{\boldsymbol{\dot{\,}}} 
-
2\boldsymbol{\nu}
\times
\frac{\partial \mathcal{L}}
{\partial\boldsymbol{\nu}}
\Big]
\\
\delta\mathbf{n}\boldsymbol{\cdot}
\,\Big(
\frac{\partial \mathcal{L}}
{\partial\mathbf{n}}\Big)
&\Longrightarrow &
(\mathbf{n}\times \delta\mathbf{n})
\boldsymbol\cdot
\,\Big(\mathbf{n}\times
\frac{\partial \mathcal{L}}
{\partial \mathbf{n}} 
\Big)
\end{eqnarray}

Varying the action $\mathcal{S}'$ in the fields $\mathbf{x}$,
$\mathbf{n}$ and $\boldsymbol\nu$ at fixed material position $\mathbf{X}$ and
time $t$ now gives
\begin{eqnarray}
\delta\mathcal{S}'
\!\!\!&=&\!\!\!
-\int dt\int d^3X
\bigg\{\delta x_p\bigg[ \bigg(
\frac{\partial \mathcal{L}}{\partial
\dot{x}_p}\bigg)^{\!\!\boldsymbol{\dot{\,}}}
 +
\,J\frac{\partial}{\partial x_p}\frac{\partial \mathcal{L}}{\partial J}
-
J\frac{\partial}{\partial x_m}
\bigg( J^{-1}
\frac{\partial \mathcal{L}}{\partial \mathbf{n}_{,m}}
\boldsymbol{\cdot}\mathbf{n}_{,p}\bigg)
\bigg]
\nonumber\\
\nonumber\\
&&\hspace{1in}
+\
(\mathbf{n}\times \delta\mathbf{n})
\boldsymbol\cdot
\,\bigg[\Big(
\frac{\partial \mathcal{L}}
{\partial\boldsymbol{\nu}}\Big)^{\boldsymbol{\dot{\,}}} 
-
2\boldsymbol{\nu}
\times
\frac{\partial \mathcal{L}}
{\partial\boldsymbol{\nu}}
\bigg]
\label{stat-act-director-nu}\\
&&\hspace{1in}
-\
(\mathbf{n}\times \delta\mathbf{n})
\boldsymbol\cdot
\,\bigg[
\mathbf{n}\times \bigg(
\frac{\partial \mathcal{L}}{\partial\mathbf{n}}
-
\, J \frac{\partial}{\partial x_m}
\Big(J^{-1}
\frac{\partial \mathcal{L}}{\partial \mathbf{n}_{,m\,}}\Big)
\bigg)
\bigg]\bigg\}
\,,\nonumber
\end{eqnarray}
with the same natural (homogeneous) boundary conditions as before.
Consequently, the action principle $\delta \mathcal{S}'=0$ yields the
following {\bfi partially reduced Lagrange-Poincar\'e equations} for liquid
crystals,
\begin{eqnarray}\label{x-nu-n-mot}
&&
\delta x_p:\quad
\bigg(\frac{\partial \mathcal{L}}{\partial
\dot{x}_p}\bigg)^{\!\!\boldsymbol{\dot{\,}}}
 +
\,J\frac{\partial}{\partial x_p}\frac{\partial \mathcal{L}}{\partial J}
-
J\frac{\partial}{\partial x_m}
\bigg( J^{-1}
\frac{\partial \mathcal{L}}{\partial \mathbf{n}_{,m}}
\boldsymbol{\cdot}\mathbf{n}_{,p}\bigg)
=
0\,,
\\
&&\mathbf{n}\times\delta\mathbf{n}:\quad
\bigg(\frac{{\partial \mathcal{L}}}
{\partial \boldsymbol\nu}\bigg)^{\boldsymbol{\dot{\,}}}
- 2{\boldsymbol\nu}\times
\frac{{\partial \mathcal{L}}}{\partial \boldsymbol\nu}
-\, 
\mathbf{n}\times \bigg(
\frac{\partial \mathcal{L}}{\partial\mathbf{n}}
-
\, J \frac{\partial}{\partial x_m}
\Big(J^{-1}
\frac{\partial \mathcal{L}}{\partial \mathbf{n}_{,m\,}}\Big)
\bigg)
=0
\,.
\nonumber
\end{eqnarray}
One may compare these with the more completely reduced Lagrange-Poincar\'e
equations for liquid crystals in equations (\ref{x-nu-gamma-mot}). The
explicit dependence of the Lagrangian on the director $\mathbf{n}$ introduces
torques and stresses not seen for Lagrangians depending only on
$\boldsymbol{\nu}$ and $\boldsymbol{\gamma}= \mathbf{n}\times d\mathbf{n}$.

\paragraph{Partially reduced Euler-Poincar\'e equations.}

We are dealing with Hamilton's principle for a Lagrangian in the class 
\begin{equation}\label{liqxtal-action-nu-n-Eul}
\mathcal{S}' = \int dt\int d^3x\
\ell(\mathbf{u}, D, \mathbf{n}, \boldsymbol\nu, \nabla\mathbf{n})
\,,
\end{equation} 
in terms of the Lagrangian density $\ell$ given by
\begin{eqnarray}\label{liqxtal-nu-n-Eul/Lag}
\ell(\mathbf{u}, D, \mathbf{n},\boldsymbol\nu, \nabla\mathbf{n})\, d^3x
=
\\
&&\hspace{-1in}
\mathcal{L}\Big(\dot{x}g^{-1}(t), Jg^{-1}(t), \mathbf{n}\,g^{-1}(t),
\boldsymbol\nu g^{-1}(t), \nabla\mathbf{n}\,g^{-1}(t)\Big)
\Big(d^3X g^{-1}(t)\Big)
\,.
\nonumber
\end{eqnarray} 
The variations of the Eulerian fluid quantities are computed from their
definitions to be,
\begin{eqnarray}\label{del-Eul-var-vel}
\delta u_j &=& \frac{\partial\eta_j}{\partial t}
+
u_k\frac{\partial\eta_j}{\partial x_k}
-
\eta_k\frac{\partial u_j}{\partial x_k}
\,,
\\
\delta D &=& - \,\frac{\partial D\eta_j}{\partial x_j}
\,,\label{del-Eul-var-D}
\\
\delta \boldsymbol\nu
&=&
\frac{\partial\boldsymbol\Sigma}{\partial t}
+
u_m\frac{\partial\boldsymbol\Sigma}{\partial x_m}
-
2\boldsymbol\nu\times\boldsymbol\Sigma
-
\eta_m\frac{\partial\boldsymbol\nu}{\partial x_m}
\,,\label{del-Eul-var-freq}
\\
\delta \mathbf{n}
&=&
-\,\mathbf{n}\times\boldsymbol\Sigma
-
\eta_k\frac{\partial \mathbf{n}}{\partial x_k}
\,,\label{del-Eul-var-n}
\end{eqnarray}
where $\boldsymbol\Sigma(x,t) \equiv 
(\mathbf{n}\times\delta\mathbf{n})(X,t)g^{-1}(t)$
and $\eta \equiv \delta{g}g^{-1}(t)$.

We compute the variation of the action
(\ref{liqxtal-action-nu-n-Eul}) in Eulerian variables at fixed time $t$
and spatial position $\mathbf{x}$ as, cf. equation
(\ref{liqxtal-Euler-act-var}),
\begin{eqnarray}\label{u-nu-n-act-var-Eul}
\delta\mathcal{S}
\!\!\!&=&\!\!\!\!\!
\int \!\! dt \!\!\int\!\! d^3x\bigg[
\frac{\delta \ell}{\delta u_j}\delta u_j
+
\frac{\delta \ell}{\delta D} \delta D
+
\frac{\delta \ell}{\delta
\boldsymbol\nu} \boldsymbol\cdot \delta\boldsymbol\nu
+
\frac{\delta \ell}{\delta \mathbf{n}} 
\boldsymbol\cdot
\delta\mathbf{n}
\bigg]
\nonumber\\
\!\!\!&=&\!\!\!\!\!
\int \!\! dt \!\!\int\!\! d^3x\Bigg\{
\eta_j\bigg[
-\,\frac{\partial}{\partial t}\frac{\delta \ell}{\delta u_j}
-\,\frac{\delta \ell}{\delta u_k}
\frac{\partial u_k}{\partial x_j}
-\,
\frac{\partial}{\partial x_k}
\Big(\frac{\delta \ell}{\delta u_j}u_k\Big)
\nonumber\\&&
\hspace{.75in}
+\
D\frac{\partial}{\partial x_j}
\frac{\delta \ell}{\delta D}
-\
\frac{\delta \ell}{\delta
\boldsymbol\nu} \boldsymbol\cdot \frac{\partial\boldsymbol\nu}{\partial x_j}
\,-\,
\frac{\delta \ell}{\delta \mathbf{n} }
\boldsymbol\cdot 
\frac{\partial \mathbf{n} }{\partial x_j}
\bigg]
\label{liqxtal-Euler-act-var-n}\\&&
\nonumber\\
&&
+\
\boldsymbol{\Sigma\,\,\cdot}
\bigg[
-\,
\frac{\partial}{\partial t}
\frac{\delta \ell}{\delta\boldsymbol\nu}
\,-\,
\frac{\partial}{\partial x_m}
\Big(
\frac{{\delta\ell}}
{\delta \boldsymbol\nu}u_m
\Big)
+ \ 
2{\boldsymbol\nu}\times
\frac{{\delta \ell}}{\delta \boldsymbol\nu}
+ \, 
\mathbf{n}\times
\frac{{\delta \ell}}{\delta \mathbf{n}}
\bigg]
\nonumber\\
&&
\hspace{.75in}
+\
\frac{\partial}{\partial t}
\bigg[
\eta_j\frac{\delta \ell}{\delta u_j}
+
\boldsymbol{\Sigma\,\cdot\,}\frac{\delta \ell}{\delta\boldsymbol\nu}
\bigg]
\nonumber\\
&&
+\
\frac{\partial}{\partial x_m}
\bigg[\eta_j \Big(\frac{\delta \ell}{\delta u_j}\, u_m
-
D \frac{\delta \ell}{\delta D}\, \delta_{jm}
\Big)
+
\boldsymbol{\Sigma\,\cdot\,}\Big(
\frac{\delta \ell}{\delta\boldsymbol\nu}\, u_m
\Big)\bigg]
\Bigg\}\,,
\nonumber
\end{eqnarray}
where we have substituted the variational expressions
(\ref{del-Eul-var-vel})-(\ref{del-Eul-var-n}) and integrated by parts.
Hence, we obtain the {\bfi partially reduced 
Euler-Poincar\'e equations} for liquid crystals,
\begin{eqnarray}
\eta_j:\quad
\frac{\partial}{\partial t}\frac{\delta \ell}{\delta u_j}
&=&
-\,\frac{\delta \ell}{\delta u_k}
\frac{\partial u_k}{\partial x_j}
-\,
\frac{\partial}{\partial x_k}
\Big(\frac{\delta \ell}{\delta u_j}u_k\Big)
+\,
D\frac{\partial}{\partial x_j}
\frac{\delta \ell}{\delta D}
\label{liqxtal-mot-Eul-n}
\\&&\hspace{.75in}
-\
\frac{\delta \ell}{\delta
\boldsymbol\nu} \boldsymbol\cdot \frac{\partial\boldsymbol\nu}{\partial x_j}
\,-\,
\frac{\delta \ell}{\delta \mathbf{n} }
\boldsymbol\cdot 
\frac{\partial \mathbf{n} }{\partial x_j}
,
\nonumber\\
\boldsymbol\Sigma:\quad
\frac{\partial}{\partial t}
\frac{\delta \ell}{\delta\boldsymbol\nu}
&=&
-\,\frac{\partial}{\partial x_m}
\Big(
\frac{{\delta\ell}}
{\delta \boldsymbol\nu}u_m
\Big)
+ \ 2{\boldsymbol\nu}\times
\frac{{\delta \ell}}{\delta \boldsymbol\nu}
+ \, \mathbf{n}\times
\frac{{\delta \ell}}{\delta \mathbf{n}}
\,.
\label{liqxtal-micromot-Eul-n}
\end{eqnarray}
One may compare these with the more completely reduced Euler-Poincar\'e
equations for liquid crystals in equations (\ref{liqxtal-mot-Eul}) and 
(\ref{liqxtal-micromot-Eul}). Again ones sees the torques and stresses
generated by the explicit dependence of the Lagrangian on the director
$\mathbf{n}$.

\paragraph{Partially reduced Hamiltonian dynamics of liquid crystals}

The Euler-Lagrange-Poincar\'e formulation of liquid crystal dynamics obtained
so far allows passage to the corresponding Hamiltonian formulation via the
following {\bfi Legendre transformation} of the partially reduced Lagrangian
$\ell$ in the velocities $\mathbf{u}$ and $\boldsymbol\nu$, in the Eulerian
fluid description,
\begin{equation}\label{liqxtal-legendre-xform-n}
m_i = \frac{\delta \ell}{\delta u_i}\,, \
\boldsymbol\sigma = \frac{\delta \ell}{\delta \boldsymbol\nu}\,, \quad
h(\mathbf{m}, D, \boldsymbol\sigma, \mathbf{n})
 = m_i u_i + \boldsymbol{\sigma\cdot\nu}
- \ell(\mathbf{u}, D, \boldsymbol\nu, \mathbf{n}).
\end{equation}
Accordingly, one computes the derivatives of $h$ as
\begin{equation}\label{liqxtal-dual-var-derivs-n}
\frac{\delta h}{\delta m_i} 
=
 u_i \,,
\quad
\frac{\delta h}{\delta \boldsymbol\sigma} 
=
 \boldsymbol\nu\,,
\quad
\frac{\delta h}{\delta D} 
=
-\, \frac{\delta \ell}{\delta D}
\,,\quad
\frac{\delta h}{\delta \mathbf{n}} 
=
- \,\frac{\delta \ell}{\delta \mathbf{n}}\,.
\end{equation}

Consequently, the Euler-Poincar\'e equations (\ref{liqxtal-mot-Eul}) -
(\ref{liqxtal-micromot-Eul-n}) and the auxiliary kinematic equations 
(\ref{D - gamma eqns}) - (\ref{D - gamma eqns1}) for liquid
crystal dynamics in the Eulerian description imply the following equations,
for the Legendre-transformed variables, $(\mathbf{m}, D, \boldsymbol\sigma,
\mathbf{n})$,
\begin{eqnarray}
\frac{\partial m_i}{\partial t} 
 &=& 
-\,m_j \frac{\partial}{\partial x_i}
\frac{\delta h}{\delta m_j}
\,
 - 
\frac{\partial}{\partial x_j}
\bigg(m_i\frac{\delta h}{\delta m_j}\bigg)
-
D\,\frac{\partial}{\partial x_i}\frac{\delta h }{ \delta D}
\nonumber\\
&&
\ +\
\bigg(\frac{\partial \mathbf{n}}{\partial x_i}\bigg)
\!\boldsymbol\cdot
\frac{\delta h }{ \delta \mathbf{n}}
-\
\boldsymbol\sigma\boldsymbol\cdot
\frac{\partial}{\partial x_i}
\frac{\delta h }{ \delta \boldsymbol\sigma}
\,,
\label{m-eqn-liqxtal-n}\\
\frac{\partial D}{\partial t}
 &=& - \,\frac{\partial }{\partial x_j}
\bigg(D\frac{\delta h}{\delta m_j}\bigg)
\,,
\label{D-eqn-liqxtal-n}\\
\frac{\partial \mathbf{n}}{\partial t}
&=&
-
\bigg(\frac{\partial \mathbf{n}}{\partial x_j}\bigg)
\frac{\delta h}{\delta m_j}
-
\mathbf{n}\times
\frac{\delta h}{\delta\boldsymbol\sigma}
\label{sigma-eqn-liqxtal-n}\\
\frac{\partial\boldsymbol\sigma}{\partial t}  
 &=& 
-\,
\frac{\partial}{\partial x_j}
\bigg(\boldsymbol\sigma\,\frac{\delta h}{\delta m_j}
\bigg)
-
\mathbf{n}\times \frac{\delta h}{\delta \mathbf{n}}
-
2\boldsymbol\sigma\times
\frac{\delta h}{\delta \boldsymbol\sigma}
\,.\qquad\label{n-eqn-liqxtal}
\end{eqnarray}
These equations are {\bfi Hamiltonian}. Assembling the liquid crystal
equations (\ref{m-eqn-liqxtal-n}) - (\ref{n-eqn-liqxtal}) into the Hamiltonian
form (\ref{Ham-form-liqxtals}) gives, {
\begin{equation}\label{Ham-matrix-diff-liqxtal-n}
\frac{\partial}{\partial t}
\left[ \begin{array}{c} 
m_i \\ D \\ \mathbf{n} \\ \boldsymbol\sigma
\end{array}\right]
= -
\left[ \begin{array}{cccc} 
m_j\partial_i + \partial_j m_i & 
D\partial_i &
- \mathbf{n}_{,i}\boldsymbol\cdot & 
\boldsymbol\sigma\boldsymbol\cdot\partial_i
\\ 
\partial_jD & 0 & 0 & 0
\\
\mathbf{n}_{,j} & 0 & 0 & 
\mathbf{n}\times
\\
\partial_j\boldsymbol\sigma & 0 & 
\mathbf{n}\times
& 2\boldsymbol\sigma\,\times
\end{array} \right]
\left[ \begin{array}{c} 
{\delta h/\delta m_j} \\ 
{\delta h/\delta D} \\ 
{\delta h/\delta \mathbf{n} } \\ 
{\delta h/\delta\boldsymbol\sigma}
\end{array}\right]
\end{equation}
}
$\!\!$%
One may compare this with the more completely reduced Hamiltonian matrix form
for liquid crystals in equations (\ref{Ham-matrix-diff-liqxtal}). The Jacobi
identity for the Poisson bracket defined by this Hamiltonian matrix is
guaranteed by associating it to the dual of a semidirect-product Lie algebra
that, in this case, has no two-cocycles. The two-cocycles are generated by the
further  transformation to $\boldsymbol{\gamma}= \mathbf{n}\times d\mathbf{n}$.

\section*{References}

\begin{description}
\singlespace  

\item Balatskii, A. [1990]
Hydrodynamics of an antiferromagnet with fermions,
{\it Phys. Rev. B} {\bf42} 8103-8109.

\item Beris, A. N. and B. J. Edwards [1994]
{\it Thermodynamics of Flowing Systems with internal microstructure},
Oxford University Press.

\item 
Cendra, H., D. D. Holm, J. E. Marsden and T. S. Ratiu [1999]
Lagrangian Reduction, the Euler-Poincar\'e
Equations, and Semidirect Products.
{\it Arnol'd Festschrift Volume II}, 
{\bf186} Am. Math. Soc. Translations Series 2, pp. 1-25.

\item Cendra, H., J. E. Marsden and T. S. Ratiu [1999] 
Lagrangian reduction by stages, {\it Preprint}.

\item Chandrasekhar, S. [1992] 
{\it Liquid Crystals}, 2nd edn. 
Cambridge University Press, Cambridge.

\item Coquereaux, R. and A. Jadcyk [1994]
{\it Riemann Geometry Fiber Bundles Kaluza-Klein Theories and all that
$\dots$.}, World Scientific, Lecture Notes in Physics Vol. {\bf16}.

\item Cosserat, E. and F. Cosserat [1909] 
{\it Th\'eorie des corps deformable}. Hermann, Paris.

\item de Gennes, P.G. and J. Prost [1993] 
{\it The Physics of Liquid Crystals}, 
2nd edn. Oxford University Press, Oxford.

\item Dunn, J. E. and J. Serrin [1985]
On the thermodynamics of interstitial working,
{\it Arch. Rat. Mech. Anal.} {\bf88} 95-133. 

\item Dzyaloshinskii, I. E. [1977]
Magnetic structure of UO2,
{\it Commun. on Phys.} {\bf2} 69-71. 

\item Dzyaloshinskii, I. E. and G. E. Volovick  [1980]
Poisson brackets in condensed matter
physics, {\it Ann. of Phys.} {\bf125}  67-97.

\item Ericksen, J. L. [1960] 
Anisotropic fluids,
{\it Arch. Rational Mech. Anal.} {\bf4} 231-237.

\item Ericksen, J. L. [1961] 
Conservation laws for liquid crystals,
{\it Trans. Soc. Rheol.} {\bf5} 23-34.

\item Eringen, A. C. [1997]
A unified continuum theory of electrodynamics of liquid crystals,
{\it Int. J. Engng. Sci.} {\bf35} 1137-1157.

\item Flanders, H. [1989] 
{\it Differential Forms with Applications to the Physical Sciences}, 
Dover Publications: New York.

\item Fuller, F. B. [1978]
Decomposition of linking number of a closed ribbon: problem from
molecular-biology,
{\it Proc. Nat. Acad. Sci. USA} {\bf75} 3557-3561.

\item Gibbons, J., D. D. Holm and B. Kupershmidt [1982] 
Gauge-invariant Poisson brackets for chromohydrodynamics, 
{\it Phys. Lett. A} {\bf 90} 281-283.

\item Gibbons, J., D. D. Holm and B. Kupershmidt [1983] 
The Hamiltonian structure of classical chromohydrodynamics, 
{\it Physica D} {\bf 6} 179-194.

\item Goldstein, R. E., T. R. Powers and C. H. Wiggins [1998]
Viscous nonlinear dynamics of twist and writhe,
{\it Phys. Rev. Lett.} {\bf80} 5232-5235.

\item Golo, V. L. and M. I. Monastyrskii [1977]
Topology of gauge fields with several vacuums, 
{\it JETP Lett.} {\bf25} (1977) 251-254.
[{\it Pis'ma Zh. Eksp. Teor. Fiz.} {\bf25} 272-276.]

\item Golo, V. L. and M. I. Monastyrskii [1978]
Currents in superfluid $^3He$, 
{\it Lett. Math. Phys.} {\bf2} 379-383.

\item Golo, V. L., M. I. Monastyrskii and S. P. Novikov [1979]
Solutions of the Ginzburg-Landau equations for planar textures 
in superfluid $^3He$,  
{\it Comm. Math. Phys.} {\bf69} 237-246.

\item Goriely, A. and M. Tabor [1997]
Nonlinear dynamics of filaments.1. Dynamical instabilities,
{\it Phys. D} {\bf105} 20-44.

\item Hall, H. E. [1985]
Evidence for intrinsic angular momentum in superfluid $^3He-A$,
{\it Phys. Rev. Lett.} {\bf54} 205-208. 

\item Hohenberg, P.C. and B.I. Halperin [1977]
Theory of dynamical critical phenomena, 
{\it Rev. Mod. Phys.} {\bf 49} 435-479.

\item Holm, D. D. [1987]
Hall magnetohydrodynamics: conservation laws and Lyapunov stability,
{\it Phys.  Fluids} {\bf 30} 1310-1322.

\item  Holm, D. D. [2000]
Finite Temperature Modifications of HVBK dynamics for Superfluid Helium
Turbulence, submitted to Proceedings of Isaac Newton Institute workshop on
Quantum Vortex Dynamics, Cambridge, UK, August 2000.

\item Holm, D. D. and B. A. Kupershmidt [1982]
Poisson structures of superfluids,
{\it Phys. Lett. A} {\bf 91} 425-430.

\item Holm, D. D. and B. A. Kupershmidt [1983a]
Poisson structures of superconductors,
{\it Phys. Lett. A} {\bf 93} 177-181.

\item Holm, D. D. and B. A. Kupershmidt [1983b]
Poisson brackets and Clebsch representations 
for magnetohydrodynamics, multifluid  plasmas, and elasticity,
{\it Physica D} {\bf 6} 347-363.

\item Holm, D. D. and B. A. Kupershmidt [1984]
Yang-Mills magnetohydrodynamics: nonrelativistic theory,
{\it Phys. Rev. D} {\bf 30} 2557-2560.
 
\item Holm, D. D. and B. A. Kupershmidt [1986]
Hamiltonian structure and Lyapunov stability of a hyperbolic system
of two-phase flow equations including surface tension,
{\it Phys. Fluids} {\bf29} 986-991.

\item Holm, D. D. and B. A. Kupershmidt [1987]
Superfluid plasmas: Multivelocity nonlinear hydrodynamics of superfluid
solutions with charged condensates coupled electromagnetically, 
{\it Phys. Rev. A} {\bf36} 3947-3956.

\item Holm, D. D. and B. A. Kupershmidt [1988]
The analogy between spin glasses and Yang-Mills fluids,
{\it J. Math. Phys.} {\bf29} 21-30.

\item Holm,  D. D., J. E. Marsden, T. S. Ratiu [1998]
The Euler-Poincar\'e equations and semidirect
products with applications to continuum theories, 
{\it Adv. in Math.} {\bf 137} 1-81.

\item 
Isaev, A. A., M. Yu. Kovalevskii and S. V. Peletminskii [1995]
Hamiltonian appraoch to continuum dynamic,  
{\it Theoret. and Math. Phys.}, {\bf 102} 208-218.
[{\it Teoret. Math. Fiz.} {\bf 102} (1995) 283-296.]

\item 
Isayev, A. A., M. Yu. Kovalevsky and S. V. Peletminsky [1997]
Hydrodynamic theory of magnets with strong exchange interaction, 
{\it Low Temp. Phys.} {\bf 23} 522-533.

\item Isayev, A. A. and S. V. Peletminsky [1997]
On Hamiltonian formulation of hydrodynamic equations for superfluid $^3He-3$,
{\it Low Temp. Phys.} {\bf 23} 955-961.

\item Jackiw, R. and N. S. Manton [1980] 
Symmetries and conservation laws in gauge theories, 
{\it Ann. Phys.} {\bf127} 257-273.

\item Kamien, R. D. [1998]
Local writhing dynamics,
{\it Eur. Phys. J. B} {\bf1} 1-4.

\item Kats, E. I. and V. V. Lebedev [1994] 
{\it Fluctuational Effects in the Dynamics of Liquid Crystals}, 
Springer: New York.

\item Khalatnikov, I. M. and V. V Lebedev [1978]
Canonical equations of hydrodynamics of quantum liquids,
{\it J. Low Temp. Phys.} {\bf32} 789-801; 

\item Khalatnikov, I. M. and V. V Lebedev [1980]
Equation of hydrodynamics of quantum liquid in the presence of continuously
distributed singular solitons, 
{\it Prog. Theo. Phys. Suppl.} {\bf69} (1980) 269-280.

\item Klapper,  I. [1996]
Biological applications of the dynamics of twisted elastic rods,
{\it J. Comp. Phys.} {\bf125} 325-337.

\item Kleinert, H. [1989]
{\it Gauge Fields in Condensed Matter},
Vols. I, II, World Scientific.

\item Kl\'eman, M. [1983] 
{\it Points, Lines and Walls in Liquid
Crystals, Magnetic Systems and Various Ordered Media}, 
John Wiley and Sons.

\item Kl\'eman, M. [1989] 
Defects in liquid crystals,
{\it Rep. on Prog. in Phys.} {\bf52} 555-654.

\item Kuratsuji, H. and H. Yabu [1998] 
Force on a vortex in ferromagnet model and the properties of vortex
configurations,
{\it J. Phys. A} {\bf31} L61-L65.

\item Lammert, P. E., D. S. Rokhsar and J. Toner [1995]
Topological and nematic ordering.I. A gauge theory,
{\it Phys. Rev. E} {\bf52} 1778-1800.

\item Leggett, A. J. [1975]
A theoretical description of the new phases of $^3He$,
{\it Rev. Mod. Phys.} {\bf47} 331-414.

\item Leslie, F. M. [1966]  
Some constitutive equations for anisotropic fluids,
{\it Quart. J. Mech. Appl. Math.} {\bf19} (1966) 357-370.

\item Leslie, F. M. [1968]  
Some constitutive equations for liquid crystals,
{\it Arch. Rational Mech. Anal.} {\bf28} (1968) 265-283.

\item Leslie, F. M. [1979]  
Theory of flow phenomena in liquid crystals,
in {\it Advances in Liquid Crystals}, Vol. {\bf4} 
(ed. G. H. Brown) 
Academic, New York pp. 1-81.

\item Marsden, J. E. and T. S. Ratiu [1999]
{\it Introduction to Mechanics and Symmetry}, 2nd Edition,
Springer-Verlag, Texts in Applied Mathematics {\bf17}.

\item Marsden, J. E., T. S. Ratiu and J. Scheurle [1999]
Reduction theory and the Lagrange-Routh equations, {\it Preprint}.

\item Marsden, J. E. and J. Scheurle [1995] 
The Lagrange-Poincar\'e equations, 
{\it Fields Institute Commun.} {\bf1} 139-164.

\item Marsden, J. E. and A. Weinstein  [1974]
Reduction of symplectic manifolds with symmetry, {\it Rep. Math. Phys.}
{\bf5} 121-130.

\item Marsden, J.E. and A. Weinstein  [1983]
Coadjoint orbits, vortices, and Clebsch variables for incompressible fluids,
{\it Physica D} {\bf7} 305-323.

\item Mermin, N. D. [1979]
The topological theory of defects in ordered media, 
{\it Rev. Mod. Phys.} {\bf51} 591-648.

\item Mermin, N. D. and T.-L. Ho [1976]
Circulation and angular momentum in the $A$ phase of superfluid Helium-3, 
{\it Phys. Rev. Lett.} {\bf36}  594-597. 

\item Mineev, V. P. [1980] 
Topologically stable defects and solitons in ordered media,
{\it Soviet Science Reviews, Section A: Physics Reviews}, Vol. 2, Edited by I.
M. Khalatnikov (Chur, London, New York: Harwood Academic Publishers) pp.
173-246.

\item Olver, P. J. [1993],
{\it Applications of Lie groups to differential equations},
2nd Edition, Springer-Verlag, New York.

\item Poincar\'{e}, H. [1901]
Sur une forme nouvelle des \'{e}quations de la m\'{e}canique, 
{\em C.R. Acad. Sci. Paris} {\bf 132} 369-371.

\item Schwinger, J. [1951]  
On gauge invariance and vacuum polarization,
{\it Phys. Rev.} {\bf82} 664-679.

\item Schwinger, J. [1959]  
Field theory commutators,
{\it Phys. Rev. Lett.} {\bf3} 296-297.

\item Serrin, J. [1959]
in {\it Mathematical Principles of Classical Fluid Mechanics}, 
Vol. VIII/1 of Encyclopedia of Physics,
edited by S. Fl\"ugge (Springer-Verlag, Berlin),
sections 14-15, pp. 125-263.

\item Stern, A. [1999] 
Duality for coset models,
{\it Nuc. Phys. B} {\bf557} 459-479.

\item Trebin, H. R.  [1982]
The topology of non-uniform media in condensed matter physics, 
{\it Adv. in Physics} {\bf31} 195-254.

\item Tsurumaru, T. and I. Tsutsui [1999] 
On topological terms in the $O(3)$ nonlinear sigma model,
{\it Phys. Lett. B} {\bf460} 94-102.

\item Volovick, G. E. [1992]
{\it Exotic Properties of Superfluid $^3He$}, 
World-Scientific, Singapore.

\item Volovick, G. E. and T. Vachaspati [1996]
Aspects of $^3He$ and the standard electro-weak model, 
{\it Int. J. Mod. Phys. B} {\bf10} 471-521.

\item Volovik, G. E. and V. S. Dotsenko [1980] 
Hydrodynamics of defects in condensed media in the concrete cases of vortices
in rotating Helium-II and of disclinations in planar magnetic substances,
{\it Sov. Phys. JETP} {\bf58} 65-80.
[{\it Zh. Eksp. Teor. Fiz.} {\bf78} (1980) 132-148.] 

\item Weatherburn, C. E.  [1974] 
{\it Differential Geometry in Three Dimensions}, Vol. 1, 
Cambridge University Press.

\item Weinstein, A. [1996]
Lagrangian mechanics and groupoids,
{\it Fields Inst. Comm.} {\bf 7} 207-231.

\item Yabu, H. and H. Kuratsuji  [1999]
Nonlinear sigma model Lagrangian for superfluid He3-A(B), 
{\it J. Phys. A} {\bf32} 7367-7374.

\item Zakharov, V.E. and E.A. Kusnetsov [1997]
Hamiltonian formalism for nonlinear waves,
{\it Usp. Fiz. Nauk} {\bf167} 1137-1167.

\item  Zakharov, V.E., S.L. Musher and A.M. Rubenchik [1985]
Hamiltonian approach to the description of nonlinear
plasma phenomena, {\it Phys. Rep.} {\bf129} 285-366.

\end{description}

\end{document}